\documentclass[acmsmall,screen]{acmart}
\pdfoutput=1
\let\savedsection\section

\usepackage{bbold} %
\usepackage{mathtools} %
\usepackage{stmaryrd} %
\usepackage{graphicx} %
\usepackage{xcolor} %
\usepackage[ruled]{algorithm2e} %
\usepackage{nccmath} %

\AddToHook{env/proposition/begin}{\crefalias{theorem}{proposition}}
\AddToHook{env/lemma/begin}{\crefalias{theorem}{lemma}}
\AddToHook{env/remark/begin}{\crefalias{theorem}{remark}}
\AddToHook{env/definition/begin}{\crefalias{theorem}{definition}}
\AddToHook{env/example/begin}{\crefalias{theorem}{example}}
\AddToHook{env/corollary/begin}{\crefalias{theorem}{corollary}}

\usepackage{xcolor}

\definecolor{nordred}{HTML}{bf616a}
\definecolor{bordeaux}{HTML}{821529}
\definecolor{bluelink}{HTML}{003399}
\definecolor{nordred}{HTML}{bf616a}
\definecolor{nordblue}{HTML}{81a1c1}
\definecolor{norddarkblue}{HTML}{5e81ac}
\definecolor{nordgreen}{HTML}{a3be8c}
\definecolor{nordnight}{HTML}{4c566a}

\AtEndPreamble{
  \RequirePackage{hyperref}
  \RequirePackage{cleveref}
  \hypersetup{
    breaklinks = true,
    linktocpage,
    colorlinks = true,
  }
}

\usepackage[xcolor,no patch,hyperref,quotation=false,electronic]{knowledge}

\knowledgeconfigure{notion}
\knowledgestyle{notion}{color=nordnight}
\knowledgestyle{intro notion}{emphasize,color=nordnight}

\knowledge{notion}
| uniformity
| Uniformity
| uniformity axiom
| Uniformity axiom
| uniformly traced distributive multicategory
| uniformly traced distributive multicategories
| Uniformly traced distributive multicategory
| Uniformly traced distributive multicategories

\knowledge{notion}
| term
| terms
| Term
| Terms

\knowledge{notion}
| interchange
| Interchange
| interchange axiom
| Interchange axiom

\knowledge{notion}
| correctness triple
| Correctness triple
| correctness triples
| Correctness triples

\knowledge{notion}
| posetal imperative category
| Posetal imperative category
| posetal imperative categories
| Posetal imperative categories

\knowledge{notion}
| posetal imperative multicategory
| Posetal imperative multicategory
| posetal imperative multicategories
| Posetal imperative multicategories

\knowledge{notion}
| guard combinators
| Guard combinators
| guard combinator
| Guard combinator

\knowledge{notion}
| command combinators
| Command combinators
| command combinator
| Command combinator

\knowledge{notion}
| monoidal category
| monoidal categories
| Monoidal category
| Monoidal categories
| symmetric monoidal category
| symmetric monoidal categories
| Symmetric monoidal category
| Symmetric monoidal categories

\knowledge{notion}
| poset-enriched monoidal category
| poset-enriched monoidal categories
| Poset-enriched monoidal category
| Poset-enriched monoidal categories

\knowledge{notion}
| poset-enriched copy-discard category
| poset-enriched copy-discard categories
| Poset-enriched copy-discard category
| Poset-enriched copy-discard categories

\knowledge{notion}
| poset-enriched uniformly-traced monoidal category
| poset-enriched uniformly-traced monoidal categories
| Poset-enriched uniformly-traced monoidal category
| Poset-enriched uniformly-traced monoidal categories

\knowledge{notion}
| poset-enriched uniformly-traced coproduct category
| poset-enriched uniformly-traced coproduct categories
| Poset-enriched uniformly-traced coproduct category
| Poset-enriched uniformly-traced coproduct categories

\knowledge{notion}
| sesquifunctor
| Sesquifunctor
| sesquifunctors
| Sesquifunctors

\knowledge{notion}
| premonoidal category
| Premonoidal category
| premonoidal categories
| Premonoidal categories
| premonoidality
| Premonoidality
| premonoidal
| Premonoidal

\knowledge{notion}
| central
| Central
| central morphism
| Central morphism
| central morphisms
| Central morphisms

\knowledge{notion}
| Symmetric premonoidal category
| Symmetric premonoidal categories
| symmetric premonoidal category
| symmetric premonoidal categories

\knowledge{notion}
| deterministic
| Deterministic
| deterministic morphism
| Deterministic morphism
| deterministic morphisms
| Deterministic morphisms

\knowledge{notion}
| total
| total morphism
| total morphisms
| Total
| Total morphism
| Total morphisms

\knowledge{notion}
| cocartesian multicategory
| Cocartesian multicategory
| cocartesian multicategories
| Cocartesian multicategories

\knowledge{notion}
| predistributive multicategory
| Predistributive multicategory
| predistributive multicategories
| Predistributive multicategories

\knowledge{notion}
| posetal distributive copy-discard multicategory
| Posetal distributive copy-discard multicategory
| posetal distributive copy-discard multicategories
| Posetal distributive copy-discard multicategories

\knowledge{notion}
| index
| indices
| Index
| Indices

\knowledge{notion}
| state
| states
| State
| States

\knowledge{notion}
| state combinator
| state combinators
| State combinator
| State combinators

\knowledge{notion}
| multicategory
| Multicategory
| Multicategories
| multicategories

\knowledge{notion}
| distributive multicategory
| distributive multicategories
| Distributive multicategory
| Distributive multicategories

\knowledge{notion}
| substitution
| Substitution
| substitutions
| Substitutions
| substituting
| Substituting

\knowledge{notion}
| guard
| guards
| Guard
| Guards

\knowledge{notion}
| predicate
| predicates
| Predicates
| Predicate

\knowledge{notion}
| predicate combinators
| Predicate combinators

\knowledge{notion}
| command
| commands
| Commands
| Command

\knowledge{notion}
| fresh
| Fresh

\knowledge{notion}
| alpha-equivalent
| alpha-equivalence
| Alpha-equivalent
| Alpha-equivalence

\knowledge{notion}
| variable
| variables
| Variable
| Variables

\knowledge{notion}
| traced distributive copy-discard multicategory
| Traced distributive copy-discard multicategory
| traced distributive copy-discard multicategories
| Traced distributive copy-discard multicategories

\knowledge{notion}
| traced distributive multicategory
| Traced distributive multicategory
| traced distributive multicategories
| Traced distributive multicategories

\knowledge{notion}
| traced predistributive copy-discard multicategory
| Traced predistributive copy-discard multicategory
| traced predistributive copy-discard multicategories
| Traced predistributive copy-discard multicategories

\knowledge{notion}
| predistributive copy-discard multicategory
| Predistributive copy-discard multicategory
| predistributive copy-discard multicategories
| Predistributive copy-discard multicategories

\knowledge{notion}
| distributive copy-discard multicategory
| Distributive copy-discard multicategory
| distributive copy-discard multicategories
| Distributive copy-discard multicategories

\knowledge{notion}
| variable substitution
| variable substitutions
| Variable substitution
| Variable substitutions

\knowledge{notion}
| posetal distributive signature
| posetal distributive signatures
| Posetal distributive signature
| Posetal distributive signatures

\knowledge{notion}
| posetal uniformity
| Posetal uniformity

\knowledge{notion}
| uniformly-traced monoidal category
| uniformly-traced monoidal categories
| Uniformly-traced monoidal category
| Uniformly-traced monoidal categories
| uniformly-traced monoidal
| uniformly-traced monoidal
| Uniformly-traced monoidal
| Uniformly-traced monoidal

\knowledge{notion}
| uniformly-traced coproduct category
| uniformly-traced coproduct categories
| Uniformly-traced coproduct category
| Uniformly-traced coproduct categories
| uniformly-traced coproduct
| uniformly-traced coproduct
| Uniformly-traced coproduct
| Uniformly-traced coproduct
| uniform coproduct-trace
| Uniform coproduct-trace

\knowledge{notion}
| posetal uniform trace
| posetal uniform traces
| Posetal uniform traces
| Posetal uniform trace
| posetal uniform traced monoidal category
| Posetal uniform traced monoidal category
| posetal uniform traced monoidal categories
| Posetal uniform traced monoidal categories

\knowledge{notion}
| label substitution
| label substitutions
| Label substitution
| Label substitutions

\knowledge{notion}
| anchor
| anchors
| Anchor
| Anchors
| label
| Label
| Labels
| labels

\knowledge{notion}
| context
| Context
| contexts
| Contexts

\knowledge{notion}
| copy-discard category
| Copy-discard category
| copy-discard categories
| Copy-discard categories
| copy-discard
| Copy-discard

\knowledge{notion}
| copy-discard monoidal category
| Copy-discard monoidal category
| copy-discard monoidal categories
| Copy-discard monoidal categories

\knowledge{notion}
| copy-discard premonoidal category
| Copy-discard premonoidal category
| copy-discard premonoidal categories
| Copy-discard premonoidal categories
| premonoidal copy-discard category
| Premonoidal copy-discard category
| premonoidal copy-discard categories
| Premonoidal copy-discard categories

\knowledge{notion}
| commutative imperative category
| Commutative imperative category
| commutative imperative categories
| Commutative imperative categories

\knowledge{notion}
| imperative category
| Imperative category
| imperative categories
| Imperative categories

\knowledge{notion}
| imperative multicategory
| Imperative multicategory
| imperative multicategories
| Imperative multicategories

\knowledge{notion}
| bimonoidally strict
| Bimonoidally strict
| strict
| Strict

\knowledge{notion}
| distributive signature
| distributive signatures
| Distributive signature
| Distributive signatures

\knowledge{notion}
| distributive category
| distributive categories
| Distributive category
| Distributive categories
| distributive copy-discard category
| distributive copy-discard categories
| Distributive copy-discard category
| Distributive copy-discard categories

\knowledge{notion}
| basic type
| basic types
| Basic type
| Basic types

\knowledge{notion}
| generator
| generators
| Generator
| Generators

\knowledge{notion}
| effectful triple
| effectful triples
| Effectful triple
| Effectful triples

\knowledge{notion}
| double signature
| double signatures
| Double signature
| Double signatures

\knowledge{notion}
| forgetful double signature
| forgetful double signatures
| Forgetful double signature
| Forgetful double signatures

\knowledge{notion}
| double signature morphism
| double signature morphisms
| Double signature morphism
| Double signature morphisms

\knowledge{notion}
| pinwheel double category
| pinwheel double categories
| Pinwheel double category
| Pinwheel double categories
| double category with pinwheels
| double categories with pinwheels
| Double category with pinwheels
| Double categories with pinwheels

\knowledge{notion}
| bigraph
| Bigraph
| bigraphs
| Bigraphs
| 2-graph
| 2-graphs
| 2-Graph
| 2-Graphs

\knowledge{notion}
| 2-graph morphism
| 2-graph morphisms

\knowledge{notion}
| 2-graph category
| 2-graph categories

\knowledge{notion}
| double category
| double categories
| Double category
| Double categories

\knowledge{notion}
| weighted polygraph
| weighted polygraphs
| Weighted polygraph
| Weighted polygraphs
| timed polygraph
| timed polygraphs
| Timed polygraph
| Timed polygraphs

\knowledge{notion}
| morphism of timed polygraphs
| Morphism of timed polygraphs
| timed polygraph morphism
| Timed polygraph morphism

\knowledge{notion}
| tilted bicategory signature
| tilted bicategory signatures
| Tilted bicategory signature
| Tilted bicategory signatures
| tilted bigraph
| Tilted bigraph
| tilted bigraphs
| Tilted bigraphs
| tilted 2-graph
| Tilted 2-graph
| tilted 2-graphs
| Tilted 2-graphs

\knowledge{notion}
| path
| paths
| Path
| Paths
| composable path
| composable paths

\knowledge{notion}
| signature
| signatures
| Signature
| Signatures

\knowledge{notion}
| congruence
| congruences
| Congruence
| Congruences

\knowledge{notion}
| symmetry
| symmetries
| Symmetry
| Symmetries

\knowledge{notion}
| profunctor
| profunctors
| Profunctor
| Profunctors

\knowledge{notion}
| left inclusion
| Left inclusion

\knowledge{notion}
| right inclusion
| Right inclusion

\knowledge{notion}
| left whiskering
| Left whiskering

\knowledge{notion}
| right whiskering
| Right whiskering

\knowledge{notion}
| inclusion and whiskering
| Inclusion and whiskering
| whiskering interacts with inclusion
| Whiskering interacts with inclusion

\knowledge{notion}
| whiskering
| Whiskering

\knowledge{notion}
| rewiring preserves whiskering
| Rewiring preserves whiskering
| whiskering preserves rewiring
| Whiskering preserves rewiring

\knowledge{notion}
| rewiring preserves interchange
| Rewiring preserves interchange

\knowledge{notion}
| rewiring preserves composition
| Rewiring preserves composition

\knowledge{notion}
| rewiring
| Rewiring 
| definition of rewiring

\knowledge{rewiring is an action}[]{notion}

\knowledge{notion}
| composition
| Composition
| term composition
| Term composition

\knowledge{notion}
| term composition is associative
| Term composition is associative

\knowledge{notion}
| term composition is unital
| Term composition is unital

\knowledge{notion}
| tensoring
| Tensoring

\knowledge{notion}
| arrow notation preterm
| Arrow notation preterm
| arrow notation preterms
| Arrow notation preterms

\knowledge{notion}
| arrow notation term
| Arrow notation term
| arrow notation terms
| Arrow notation terms

\knowledge{notion}
| observe-arrow notation
| Observe-arrow notation
| observe-arrow notation term
| Observe-arrow notation term
| observe-arrow notation terms
| Observe-arrow notation terms

\knowledge{notion}
| subdistribution
| Subdistribution
| subdistributions
| Subdistributions
| subdistributional

\knowledge{notion}
| rescaling
| Rescaling

\knowledge{notion}
| normalisation
| normalisations
| Normalisation
| Normalisations
| normalization
| normalizations
| Normalization
| Normalizations
| normalised
| Normalised

\knowledge{notion}
| observe-arrow notation copy-discard-compare category of terms

\knowledge{notion}
| category of arrow notation terms

\knowledge{notion}
| copy-discard-compare categories
| copy-discard-compare category
| Copy-discard-compare categories
| Copy-discard-compare category

\knowledge{notion}
| Kleisli extension
| Kleisli extensions

\newcommand{\klextend}[1]{#1^{\kl[Kleisli extension]{>}}}

\knowledge{notion}
| Kleisli category
| Kleisli categories

\newcommand{\kleisli}[1]{\kl[Kleisli category]{\mathsf{kl}}(#1)}

\knowledge{notion}
| monad
| monads
| Monad
| Monads

\knowledge{notion}
| strong monad
| strong monads
| Strong monad
| Strong monads

\knowledge{notion}
| commutative monad
| commutative monads
| Commutative monad
| Commutative monads

\knowledge{notion}
| copy-discard coproducts
| copy-discard coproduct
| Copy-discard coproducts
| Copy-discard coproduct

\knowledge{notion}
| partially additive
| partially additive monad
| Partially additive
| Partially additive monad

\knowledge{notion}
| sets
| category of sets

\newcommand{\Set}{\kl[sets]{\mathsf{Set}}}

\knowledge{notion}
| powerset
| powerset monad

\newcommand{\powerset}{\kl[powerset]{\mathcal{P}}}

\knowledge{notion}
| category of relations
| Category of relations

\newcommand{\Rel}{\kl[category of relations]{\mathsf{Rel}}}

\knowledge{notion}
| maybe
| maybe monad

\newcommand{\maybe}{\kl[maybe]{\mathcal{L}}}

\knowledge{notion}
| category of partial functions
| Category of partial functions

\newcommand{\Par}{\kl[category of partial functions]{\mathsf{Par}}}

\knowledge{notion}
| discrete subdistribution
| discrete subdistributions
| discrete subdistribution monad
| countably-supported subdistribution monad

\newcommand{\subdistr}{\kl[discrete subdistribution]{\mathcal{D}}}

\knowledge{notion}
| category of discrete substochastic channels
| Category of discrete substochastic channels
| substoch

\newcommand{\subStoch}{\kl[substoch]{\ncat{Stoch}_{\leq}}}

\knowledge{notion}
| category of discrete stochastic channels
| Category of discrete stochastic channels

\newcommand{\Stoch}{\kl[category of discrete stochastic channels]{\mathsf{Stoch}}}

\knowledge{notion}
| support
\newcommand{\supp}{\kl[support]{\mathrm{supp}}}

\knowledge{notion}
| mass
| probability mass
\newcommand{\mass}{\kl[mass]{\mathrm{mass}}}

\knowledge{notion}
| Dirac distribution

\knowledge{notion}
| measurable subdistribution
| measurable subdistributions
| measurable subdistribution monad
| BorelsubStoch

\newcommand{\subgiry}{\kl[measurable subdistribution]{\mathcal{G}}}
\newcommand{\BorelsubStoch}{\kl[BorelsubStoch]{\ncat{BorelStoch}_{\leq}}}

\newcommand{\BorelStoch}{\ncat{BorelStoch}}

\knowledge{notion}
| standard Borel spaces
| Standard Borel spaces

\newcommand{\StdBorel}{\kl[standard Borel spaces]{\mathsf{StdBorel}}}

\knowledge{notion}
| quantale-valued sets

\knowledge{notion}
| assertion-correctness triple
| assertion-correctness triples
| Assertion-correctness triple
| Assertion-correctness triples

\knowledge{notion}
| state-correctness triple
| state-correctness triples
| State-correctness triple
| State-correctness triples

\knowledge{notion}
| predicate-correctness triple
| predicate-correctness triples
| Predicate-correctness triple
| Predicate-correctness triples

\knowledge{notion}
| assertion-incorrectness triple
| assertion-incorrectness triples
| Assertion-incorrectness triple
| Assertion-incorrectness triples

\knowledge{notion}
| state-incorrectness triple
| state-incorrectness triples
| State-incorrectness triple
| State-incorrectness triples

\knowledge{notion}
| predicate-incorrectness triple
| predicate-incorrectness triples
| Predicate-incorrectness triple
| Predicate-incorrectness triples

\knowledge{notion}
| relational assertion-correctness triple
| relational assertion-correctness triples
| Relational assertion-correctness triple
| Relational assertion-correctness triples

\knowledge{notion}
| relational state-correctness triple
| relational state-correctness triples
| Relational state-correctness triple
| Relational state-correctness triples

\knowledge{notion}
| relational predicate-correctness triple
| relational predicate-correctness triples
| Relational predicate-correctness triple
| Relational predicate-correctness triples

\knowledge{notion}
| relational assertion-incorrectness triple
| relational assertion-incorrectness triples
| Relational assertion-incorrectness triple
| Relational assertion-incorrectness triples

\knowledge{notion}
| relational state-incorrectness triple
| relational state-incorrectness triples
| Relational state-incorrectness triple
| Relational state-incorrectness triples

\knowledge{notion}
| relational predicate-incorrectness triple
| relational predicate-incorrectness triples
| Relational predicate-incorrectness triple
| Relational predicate-incorrectness triples

\knowledge{notion}
| branch combinator

\knowledge{notion}
| coupling synchronization

\knowledge{notion}
| given
\newcommand{\given}{\kl[given]{\mid}}
\newcommand{\naturals}{\mathbb{N}}
\newcommand{\integers}{\mathbb{Z}}
\newcommand{\union}{\cup}
\newcommand{\Union}{\bigcup}
\newcommand{\intersection}{\cap}
\newcommand{\st}{\mathrel{\mid}}
\newcommand{\ket}[1]{|#1\rangle}

\newcommand{\cat}[1]{\mathbb{#1}}
\newcommand{\ncat}[1]{\mathsf{#1}}
\newcommand{\id}{\mathrm{id}} %
\newcommand{\op}{^{\mathsf{op}}} %
\newcommand{\dcomp}{\mathbin{\fatsemi}} %
\newcommand{\defn}{\coloneqq}
\newcommand{\codefn}{\eqqcolon} %

\newcommand{\tensor}{\otimes}

\knowledge{notion}
| copy
| Copy
| copy morphism
| Copy morphism
\newcommand{\cp}{\kl[copy]{\nu}}
\knowledge{notion}
| unbiased copy
| Unbiased copy
\newcommand{\ncp}{\kl[unbiased copy]{\nu}}
\knowledge{notion}
| discard
| Discard
| discard morphism
| Discard morphism
\newcommand{\discard}{\kl[discard]{\varepsilon}}
\knowledge{notion}
| symmetry
| symmetries
| swap
| Symmetry
| Symmetries
| Swap
\newcommand{\swap}{\kl[swap]{\sigma}}
\knowledge{notion}
| projection
| projections
| Projection
| Projections
\newcommand{\proj}{\kl[projection]{\pi}}

\knowledge{notion}
| distributors
| distributor
| Distributors
| Distributor
\newcommand{\rdistr}{\kl[distributor]{\delta}^{\kl[distributor]{\mathrm{R}}}}
\newcommand{\ldistr}{\kl[distributor]{\delta}^{\kl[distributor]{\mathrm{L}}}}
\newcommand{\ridistr}{\kl[distributor]{\delta}^{\kl[distributor]{\mathrm{-R}}}}
\newcommand{\lidistr}{\kl[distributor]{\delta}^{\kl[distributor]{\mathrm{-L}}}}

\knowledge{notion}
| initial map
| initmap
\newcommand{\initmap}{\kl[initmap]{\textup{!`}}}
\knowledge{notion}
| final map
| finmap
\newcommand{\finmap}{\kl[finmap]{\textup{!}}}
\knowledge{notion}
| zero morphism
| zero morphisms
| zeromap
\newcommand{\zeromap}{\kl[zeromap]{\mathbb{0}}}
\knowledge{notion}
| join morphism
| join
\newcommand{\join}{\kl[join]{\mu}}
\knowledge{notion}
| injection
| injections
| Injection
| Injections
\newcommand{\inject}{\kl[injection]{\iota}}
\knowledge{notion}
| copairing
| copairings
| Copairing
| Copairings
\newcommand{\copairing}[2]{\kl[copairing]{[}#1, #2\kl[copairing]{]}}
\knowledge{notion}
| coproduct symmetry
| coproduct symmetries
| Coproduct symmetry
| Coproduct symmetries
| cswap
\newcommand{\cswap}{\kl[cswap]{\sigma}^{\kl[cswap]{+}}}
\knowledge{notion}
| coproduct projection
| coproduct projections
| Coproduct projection
| Coproduct projections
| cproj
\newcommand{\cproj}{\kl[cproj]{\pi}^{\kl[cproj]{+}}}
\knowledge{notion}
| uniformly traced
| uniform trace
| trace
| Uniformly traced
| Uniform trace
| Trace
\newcommand{\trace}{\kl[trace]{\mathrm{tr}}}

\newcommand{\monoidal}{\odot}
\newcommand{\unit}{I}
\newcommand{\mswap}{\sigma^{\monoidal}}

\knowledge{notion}
| left strength
| Left strength
| lstr
\newcommand{\lstr}{\kl[lstr]{\mathsf{s}}}
\knowledge{notion}
| right strength
| Right strength
| rstr
\newcommand{\rstr}{\kl[rstr]{\mathsf{t}}}

\knowledge{notion}
| poset enrichment
| imperative posetal bicategory
| imperative posetal bicategories
| Imperative posetal bicategory
| Imperative posetal bicategories
\newcommand{\pleq}{\mathrel{\kl[poset enrichment]{\leq}}}
\newcommand{\pgeq}{\mathrel{\kl[poset enrichment]{\geq}}}

\knowledge{notion}
| signature
| signatures
| Signature
| Signatures
\newcommand{\sig}{\kl[signature]{\Sigma}}

\knowledge{notion}
| monoidal signature
| monoidal signatures
| Monoidal signature
| Monoidal signatures

\knowledge{notion}
| generator
| generators
| Generator
| Generators
\newcommand{\gens}{\kl[generators]{\mathcal{G}}}

\knowledge{notion}
| arity
| Arity
\newcommand{\arity}{\kl[arity]{\mathsf{ar}}}
\knowledge{notion}
| coarity
| Coarity
\newcommand{\carity}{\kl[coarity]{\mathsf{car}}}

\newcommand{\esig}{\kl[signature]{\sig_E}}
\newcommand{\egens}{\kl[generators]{\gens_E}}
\newcommand{\gsig}{\kl[signature]{\sig_G}}
\newcommand{\ggens}{\kl[generators]{\gens_G}}
\newcommand{\psig}{\kl[signature]{\sig_P}}
\newcommand{\pgens}{\kl[generators]{\gens_P}}
\newcommand{\ssig}{\kl[signature]{\sig_S}}
\newcommand{\sgens}{\kl[generators]{\gens_S}}
\newcommand{\csig}{\kl[signature]{\sig_C}}
\newcommand{\cgens}{\kl[generators]{\gens_C}}

\newcommand{\gleft}{\mathsf{L}}
\newcommand{\gright}{\mathsf{R}}
\newcommand{\gand}{\land}
\newcommand{\gnot}{\lnot}
\newcommand{\gor}{\lor}
\newcommand{\gchoice}[3]{#1 +_{#2} #3}
\newcommand{\gmute}[3]{\textstyle{\coprod}_{#1} #3 \cdot #2}

\newcommand{\ptrue}{\top}
\newcommand{\pfalse}{\bot}
\newcommand{\pand}{\land}
\newcommand{\pchoice}[3]{#1 +_{#2} #3}
\newcommand{\pred}[1]{#1^{\tiny{\vee}}} %
\newcommand{\psub}[3]{#1[#2 \backslash #3]}
\newcommand{\pmute}[3]{\textstyle{\coprod}_{#1} #3 \cdot #2}

\newcommand{\sfalse}{\bot}
\newcommand{\smute}[3]{\textstyle{\coprod}_{#1} #2 \cdot #3}
\newcommand{\schoice}[3]{#1 +_{#2} #3}
\newcommand{\restrict}[2]{#1 \downharpoonright #2}
\newcommand{\ssub}[3]{#1(#3/ #2)}

\newcommand{\noop}{\mathop{\mathsf{skip}}}
\newcommand{\abort}{\mathop{\mathsf{abort}}}
\newcommand{\ccomp}{\mathbin{;}}
\newcommand{\cchoice}[3]{\mathop{\mathsf{if}} #2 \mathop{\mathsf{then}} #1 \mathop{\mathsf{else}} #3}
\newcommand{\while}[2]{\mathop{\mathsf{while}} #1 \mathop{\mathsf{do}} #2}
\newcommand{\assert}[1]{\mathop{\mathsf{assert}} #1}
\newcommand{\assign}[2]{#1 \leftarrow #2}
\newcommand{\samp}[2]{#1 \superimpose{\tilde{}}{\leftarrow} #2}

\newcommand{\gequiv}{=}
\newcommand{\mchoice}[3]{#1 +_{#2} #3}
\knowledge{notion}
| semantic equivalence
| semantic equivalences
| Semantic equivalence
| Semantic equivalences
| semantically equivalent
| Semantically equivalent
\newcommand{\sequiv}{\mathrel{\kl[semantic equivalence]{\equiv}}}
\knowledge{notion}
| semantic inequality
| semantic inequalities
| Semantic inequality
| Semantic inequalities
| semantically comparable
| Semantically comparable
\newcommand{\sleq}{\mathrel{\kl[semantic inequality]{\leqq}}}
\newcommand{\sgeq}{\mathrel{\kl[semantic inequality]{\geqq}}}

\newcommand{\gsemn}[1]{\mathcal{I}_0(#1)}
\newcommand{\semn}[1]{\llbracket #1 \rrbracket} %
\newcommand{\andmorph}{\mathop{\mathsf{AND}}}

\knowledge{notion}
| branch
\newcommand{\branch}[1]{\underline{#1}}

\newcommand{\triple}[3]{\{#1\}#2\{#3\}}

\usepackage{mathpartir} %
\newcommand{\nooprule}{\textsc{skip}}
\newcommand{\comprule}{\textsc{comp}}
\newcommand{\assignrule}{\textsc{assign}}
\newcommand{\samplerule}{\textsc{sample}}

\knowledge{notion}
| coupling synchronization
\newcommand{\synch}[1]{#1^{\kl[coupling synchronization]{=}}}
\knowledge{notion}
| coupling
| couplings
| Coupling
| Couplings
\newcommand{\couple}[3]{#1 \mathrel{\kl[coupling]{\triangleleft}} #2 \mathbin{\kl[coupling]{\&}} #3}

\knowledge{notion}
| coupling with precondition
| couplings with precondition
| Coupling with precondition
| Couplings with precondition
\newcommand{\pcouple}[4]{(#1 \mid #2) \mathrel{\kl[coupling with precondition]{\triangleleft}} #3 \mathbin{\kl[coupling with precondition]{\&}} #4}

\newcommand{\coupcomp}{\mathbin{\overset{\triangleleft}{,}}}
\newcommand{\idcoup}{{\id^{\triangleleft}}}
\newcommand{\couptensor}{\olessthan}

\newcommand{\freev}[1]{\mathsf{FV}(#1)}
\newcommand{\modv}[1]{\mathsf{MV}(#1)}
\newcommand{\inv}[1]{\mathsf{IV}(#1)}

\newcommand{\nondetchoice}{\blacktriangleleft}
\newcommand{\coin}[1]{c_{#1}}
\newcommand{\nondetvar}{\top}
\newcommand{\unif}[1]{\mathcal{U}_{#1}}

\newcommand{\tpre}[2]{\mathsf{tpre}(#1, #2)}
\newcommand{\tpost}[2]{\mathsf{tpost}(#1, #2)}
\newcommand{\fix}[2]{\mathsf{fix}\ #1. #2}
\IfFileExists{noappendix.token}%
{\usepackage[appendix=strip,bibliography=common]{apxproof}}%
{\usepackage[bibliography=common]{apxproof}}

\theoremstyle{acmplain}
\newtheoremrep{theorem}{Theorem}
\newtheoremrep{proposition}[theorem]{Proposition}
\newtheoremrep{lemma}[theorem]{Lemma}
\newtheoremrep{corollary}[theorem]{Corollary}
\theoremstyle{acmdefinition}
\newtheorem{remark}[theorem]{Remark}
 \usepackage{stringdiagramspics}
\newcommand{\figCouplingCompositionDef}{
  \begin{tikzpicture}[baseline=-0.1cm,color=pluscolor]
    \pic(u)at(0,0){morphism=\(u\)};
    \pic(v)at(1,0.6){morphism=\(v\)};
    \pic(g1)at(v-c|-u-c){morphism=\(g_{1}\)};
    \pic(g2)at(v-c|-,-0.6){morphism=\(g_{2}\)};
    \pic(m1)at(1.8,0.3){merge=0.3};
    \pic(m2)at(1.75,-0.075){merge=0.525};
    \coordinate[input=\(A_{1}A_{2}\)](i)at(-0.5,|-u-i);
    \coordinate[output=\(C_{1}C_{2}\)](o1)at(2.3,|-v-o1);
    \coordinate[output=\(C_{1}\)](o2)at(o1|-m1-o);
    \coordinate[output=\(C_{2}\)](o3)at(o2|-m2-o);
    \draw[-](i)to(u-i);
    \draw[-](u-o1)to(v-i);
    \draw[-](u-o)to(g1-i);
    \draw[-](u-o2)to(g2-i);
    \draw[-](v-o1)to(o1);
    \draw[-](v-o)to(m1-i1);
    \draw[-](v-o2)to(m2-i1);
    \draw[-](g1-o)to(m1-i2);
    \draw[-](g2-o)to(m2-i2);
    \draw[-](m1-o)to(o2);
    \draw[-](m2-o)to(o3);
  \end{tikzpicture}
}

\newcommand{\figCouplingCompositionWellDefProofOne}{
  \begin{tikzpicture}[baseline=-0.1cm,color=pluscolor]
    \pic(u)at(0,0){morphism=\(u\)};
    \pic(v)at(1,0.6){morphism=\(v\)};
    \pic(g1)at(v-c|-u-c){morphism=\(g_{1}\)};
    \pic(g2)at(v-c|-,-0.6){morphism=\(g_{2}\)};
    \pic(m1)at(2,0.3){merge=0.3};
    \pic(p)at(m1-c|-,0.9){morphism=\(\proj_{1}\)};
    \pic(m2)at(1.95,-0.075){merge=0.525};
    \pic(m3)at(2.7,0.6){merge=0.3};
    \pic(z)at(p-o|-m2-o){bot};
    \coordinate[input=\(A_{1}A_{2}\)](i)at(-0.5,|-u-i);
    \coordinate[output=\(C_{1}\)](o)at(3,|-m3-o);
    \draw[-](i)to(u-i);
    \draw[-](u-o1)to(v-i);
    \draw[-](u-o)to(g1-i);
    \draw[-](u-o2)to(g2-i);
    \draw[-](v-o1)to(p-i);
    \draw[-](v-o)to(m1-i1);
    \draw[-](v-o2)to(m2-i1);
    \draw[-](g1-o)to(m1-i2);
    \draw[-](g2-o)to(m2-i2);
    \draw[-](m1-o)to(m3-i2);
    \draw[-](m2-o)to(z-i);
    \draw[-](p-o)to(m3-i1);
    \draw[-](m3-o)to(o);
  \end{tikzpicture}
}

\newcommand{\figCouplingCompositionWellDefProofTwo}{
  \begin{tikzpicture}[baseline=-0.1cm,color=pluscolor]
    \pic(u)at(0,0){morphism=\(u\)};
    \pic(v)at(1,0.6){morphism=\(v\)};
    \pic(g1)at(v-c|-u-c){morphism=\(g_{1}\)};
    \pic(g2)at(v-c|-,-0.6){morphism=\(g_{2}\)};
    \pic(p)at(2,1){morphism=\(\proj_{1}\)};
    \pic(m1)at(2.6,0.8){merge=0.2};
    \pic(m2)at(3,0.4){merge=0.4};
    \pic(z1)at(1.6,|-v-o2){bot};
    \pic(z2)at(z1-c|-g2-o){bot};
    \coordinate[input=\(A_{1}A_{2}\)](i)at(-0.5,|-u-i);
    \coordinate[output=\(C_{1}\)](o)at(3.5,|-m2-o);
    \draw[-](i)to(u-i);
    \draw[-](u-o1)to(v-i);
    \draw[-](u-o)to(g1-i);
    \draw[-](u-o2)to(g2-i);
    \draw[-](v-o1)to(p-i);
    \draw[-](v-o)to(m1-i2);
    \draw[-](v-o2)to(z1-i);
    \draw[-](p-o)to(m1-i1);
    \draw[-](g1-o)to(m2-i2);
    \draw[-](g2-o)to(z2-i);
    \draw[-](m1-o)to(m2-i1);
    \draw[-](m2-o)to(o);
  \end{tikzpicture}
}

\newcommand{\figCouplingCompositionWellDefProofThree}{
  \begin{tikzpicture}[baseline=-0.1cm,color=pluscolor]
    \pic(u)at(0,0){morphism=\(u\)};
    \pic(p)at(1,0.6){morphism=\(\proj_{1}\)};
    \pic(g1)at(p-c|-u-c){morphism=\(g_{1}\)};
    \pic(g11)at(1.8,|-p-o){morphism=\(g_{1}\)};
    \pic(m1)at(2.5,0.3){merge=0.3};
    \pic(z)at(g1-i|-,-0.6){bot};
    \coordinate[input=\(A_{1}A_{2}\)](i)at(-0.5,|-u-i);
    \coordinate[output=\(C_{1}\)](o)at(3,|-m1-o);
    \draw[-](i)to(u-i);
    \draw[-](u-o1)to(p-i);
    \draw[-](u-o)to(g1-i);
    \draw[-](u-o2)to(z-i);
    \draw[-](p-o)to(g11-i);
    \draw[-](g1-o)to(m1-i2);
    \draw[-](g11-o)to(m1-i1);
    \draw[-](m1-o)to(o);
  \end{tikzpicture}
}

\newcommand{\figCouplingCompositionWellDefProofFour}{
  \begin{tikzpicture}[baseline=-0.1cm,color=pluscolor]
    \pic(u)at(0,0){morphism=\(u\)};
    \pic(p)at(1,0.4){morphism=\(\proj_{1}\)};
    \pic(m1)at(1.7,0.2){merge=0.2};
    \pic(g1)at(2.3,|-m1-o){morphism=\(g_{1}\)};
    \pic(z)at(p-i|-u-o2){bot};
    \coordinate[input=\(A_{1}A_{2}\)](i)at(-0.5,|-u-i);
    \coordinate[output=\(C_{1}\)](o)at(2.8,|-g1-o);
    \draw[-](i)to(u-i);
    \draw[-](u-o1)to(p-i);
    \draw[-](u-o)to(m1-i2);
    \draw[-](u-o2)to(z-i);
    \draw[-](p-o)to(m1-i1);
    \draw[-](m1-o)to(g1-i);
    \draw[-](g1-o)to(o);
  \end{tikzpicture}
}

\newcommand{\figCouplingCompositionWellDefProofFive}{
  \begin{tikzpicture}[baseline=-0.1cm,color=pluscolor]
    \pic(p)at(0,0){morphism=\(\proj_{1}\)};
    \pic(f1)at(1,|-p-o){morphism=\(f_{1}\)};
    \pic(g1)at(2,|-f1-o){morphism=\(g_{1}\)};
    \coordinate[input=\(A_{1}A_{2}\)](i)at(-0.5,|-p-i);
    \coordinate[output=\(C_{1}\)](o)at(2.5,|-g1-o);
    \draw[-](i)to(p-i);
    \draw[-](p-o)to(f1-i);
    \draw[-](f1-o)to(g1-i);
    \draw[-](g1-o)to(o);
  \end{tikzpicture}
}

\newcommand{\figCouplingCompositionAssociativeLeftProof}{
  \begin{tikzpicture}[baseline=0cm,color=pluscolor]
    \pic(u)at(0,0){morphism=\(u\)};
    \pic(v)at(1,0.6){morphism=\(v\)};
    \pic(g1)at(v-c|-u-c){morphism=\(g_{1}\)};
    \pic(g2)at(v-c|-,-0.6){morphism=\(g_{2}\)};
    \pic(m1)at(1.7,0.3){merge=0.3};
    \pic(m2)at(1.65,-0.075){merge=0.525};
    \pic(w)at(2.4,0.9){morphism=\(w\)};
    \pic(h1)at(w-c|-m1-o){morphism=\(h_{1}\)};
    \pic(h2)at(w-c|-,-0.3){morphism=\(h_{2}\)};
    \pic(m3)at(3.2,0.6){merge=0.3};
    \pic(m4)at(3.15,0.225){merge=0.525};
    \coordinate[input=\(A_{1}A_{2}\)](i)at(-0.5,|-u-i);
    \coordinate[output=\(D_{1}D_{2}\)](o1)at(3.5,|-w-o1);
    \coordinate[output=\(D_{1}\)](o2)at(o1|-m3-o);
    \coordinate[output=\(D_{2}\)](o3)at(o2|-m4-o);
    \draw[-](i)to(u-i);
    \draw[-](u-o1)to(v-i);
    \draw[-](u-o)to(g1-i);
    \draw[-](u-o2)to(g2-i);
    \draw[-](v-o1)to(w-i);
    \draw[-](v-o)to(m1-i1);
    \draw[-](v-o2)to(m2-i1);
    \draw[-](g1-o)to(m1-i2);
    \draw[-](g2-o)to(m2-i2);
    \draw[-](m1-o)to(h1-i);
    \draw[-](m2-o)to(h2-i);
    \draw[-](w-o1)to(o1);
    \draw[-](w-o)to(m3-i1);
    \draw[-](w-o2)to(m4-i1);
    \draw[-](h1-o)to(m3-i2);
    \draw[-](h2-o)to(m4-i2);
    \draw[-](m3-o)to(o2);
    \draw[-](m4-o)to(o3);
  \end{tikzpicture}
}

\newcommand{\figCouplingCompositionAssociativeMidProof}{
  \begin{tikzpicture}[baseline=-0.7cm,color=pluscolor]
    \pic(u)at(-1,-1.2){morphism=\(u\)};
    \pic(v)at(0,0){morphism=\(v\)};
    \pic(w)at(1,0.6){morphism=\(w\)};
    \pic(h1)at(w-c|-v-c){morphism=\(h_{1}\)};
    \pic(h2)at(w-c|-,-0.6){morphism=\(h_{2}\)};
    \pic(m1)at(1.7,-0.6){merge=0.6};
    \pic(m2)at(1.7,-1.2){merge=0.6};
    \pic(g1)at(v-c|-u-o){morphism=\(g_{1}\)};
    \pic(g2)at(g1-c|-,-1.8){morphism=\(g_{2}\)};
    \pic(h11)at(h1-c|-g1-o){morphism=\(h_{1}\)};
    \pic(h12)at(h2-c|-g2-o){morphism=\(h_{2}\)};
    \pic(m3)at(2.3,0){merge=0.6};
    \pic(m4)at(2.3,-0.6){merge=0.6};
    \coordinate[input=\(A_{1}A_{2}\)](i)at(-1.5,|-u-i);
    \coordinate[output=\(D_{1}D_{2}\)](o1)at(2.7,|-w-o1);
    \coordinate[output=\(D_{1}\)](o2)at(o1|-m3-o);
    \coordinate[output=\(D_{2}\)](o3)at(o2|-m4-o);
    \draw[-](i)to(u-i);
    \draw[-](u-o1)to(v-i);
    \draw[-](u-o)to(g1-i);
    \draw[-](u-o2)to(g2-i);
    \draw[-](v-o1)to(w-i);
    \draw[-](v-o)to(h1-i);
    \draw[-](v-o2)to(h2-i);
    \draw[-](g1-o)to(h11-i);
    \draw[-](g2-o)to(h12-i);
    \draw[-](w-o1)to(o1);
    \draw[-](w-o)to(m3-i1);
    \draw[-](w-o2)to(m4-i1);
    \draw[-](h1-o)to(m1-i1);
    \draw[-](h2-o)to(m2-i1);
    \draw[-](m1-o)to(m3-i2);
    \draw[-](m2-o)to(m4-i2);
    \draw[-](h11-o)to(m1-i2);
    \draw[-](h12-o)to(m2-i2);
    \draw[-](m3-o)to(o2);
    \draw[-](m4-o)to(o3);
  \end{tikzpicture}
}

\newcommand{\figCouplingCompositionAssociativeRightProof}{
  \begin{tikzpicture}[baseline=-0.7cm,color=pluscolor]
    \pic(u)at(-1,-1.2){morphism=\(u\)};
    \pic(v)at(0,0){morphism=\(v\)};
    \pic(w)at(1,0.6){morphism=\(w\)};
    \pic(h1)at(w-c|-v-c){morphism=\(h_{1}\)};
    \pic(h2)at(w-c|-,-0.6){morphism=\(h_{2}\)};
    \pic(m1)at(1.7,0.3){merge=0.3};
    \pic(m2)at(1.65,-0.075){merge=0.525};
    \pic(g1)at(v-c|-u-o){morphism=\(g_{1}\)};
    \pic(g2)at(g1-c|-,-1.8){morphism=\(g_{2}\)};
    \pic(h11)at(h1-c|-g1-o){morphism=\(h_{1}\)};
    \pic(h12)at(h2-c|-g2-o){morphism=\(h_{2}\)};
    \pic(m3)at(2.3,-0.1){merge=0.4};
    \pic(m4)at(2.3,-0.475){merge=0.4};
    \coordinate[input=\(A_{1}A_{2}\)](i)at(-1.5,|-u-i);
    \coordinate[output=\(D_{1}D_{2}\)](o1)at(2.7,|-w-o1);
    \coordinate[output=\(D_{1}\)](o2)at(o1|-m3-o);
    \coordinate[output=\(D_{2}\)](o3)at(o2|-m4-o);
    \draw[-](i)to(u-i);
    \draw[-](u-o1)to(v-i);
    \draw[-](u-o)to(g1-i);
    \draw[-](u-o2)to(g2-i);
    \draw[-](v-o1)to(w-i);
    \draw[-](v-o)to(h1-i);
    \draw[-](v-o2)to(h2-i);
    \draw[-](g1-o)to(h11-i);
    \draw[-](g2-o)to(h12-i);
    \draw[-](w-o1)to(o1);
    \draw[-](w-o)to(m1-i1);
    \draw[-](w-o2)to(m2-i1);
    \draw[-](h1-o)to(m1-i2);
    \draw[-](h2-o)to(m2-i2);
    \draw[-](m1-o)to(m3-i1);
    \draw[-](m2-o)to(m4-i1);
    \draw[-](h11-o)to(m3-i2);
    \draw[-](h12-o)to(m4-i2);
    \draw[-](m3-o)to(o2);
    \draw[-](m4-o)to(o3);
  \end{tikzpicture}
}

\newcommand{\figCouplingCompositionUnitalRightProof}{
  \begin{tikzpicture}[baseline=-0.1cm, color=pluscolor]
    \pic(u)at(-0.3,0){morphism=\(u\)};
    \pic(m1)at(1,0){merge=0.3};
    \pic(m2)at(m1-c|-,-0.3){merge=0.3};
    \pic(z1)at(0.5,|-m1-i1){bot};
    \pic(z2)at(z1-c|-m2-i1){bot};
    \coordinate[input=\(A_{1}A_{2}\)](i)at(-0.8,|-u-i);
    \coordinate[output=\(B_{1}B_{2}\)](o1)at(1.5,0.6);
    \coordinate[output=\(B_{1}\)](o2)at(o1|-m1-o);
    \coordinate[output=\(B_{2}\)](o3)at(o2|-m2-o);
    \draw[-](i)to(u-i);
    \draw[-](u-o1)to(m1-i1|-o1)to(o1);
    \draw[-](u-o)to(m1-i2);
    \draw[-](u-o2)to(m2-i2);
    \draw[-](z1-o)to(m1-i1);
    \draw[-](z2-o)to(m2-i1);
    \draw[-](m1-o)to(o2);
    \draw[-](m2-o)to(o3);
  \end{tikzpicture}
}

\newcommand{\figCouplingCompositionUnitalLeftProof}{
  \begin{tikzpicture}[baseline=-0.1cm,color=pluscolor]
    \pic(u)at(1,0.6){morphism=\(u\)};
    \pic(m1)at(1.8,0.3){merge=0.3};
    \pic(m2)at(1.75,-0.075){merge=0.525};
    \pic(f1)at(u-c|-m1-i2){morphism=\(f_{1}\)};
    \pic(f2)at(u-c|-m2-i2){morphism=\(f_{2}\)};
    \pic(z1)at(0.3,|-f1-i){bot};
    \pic(z2)at(z1-c|-f2-i){bot};
    \coordinate[input=\(A_{1}A_{2}\)](i)at(0,|-u-i);
    \coordinate[output=\(B_{1}B_{2}\)](o1)at(2.3,|-u-o1);
    \coordinate[output=\(B_{1}\)](o2)at(o1|-m1-o);
    \coordinate[output=\(B_{2}\)](o3)at(o2|-m2-o);
    \draw[-](i)to(u-i);
    \draw[-](u-o1)to(o1);
    \draw[-](u-o)to(m1-i1);
    \draw[-](u-o2)to(m2-i1);
    \draw[-](f1-o)to(m1-i2);
    \draw[-](f2-o)to(m2-i2);
    \draw[-](z1-o)to(f1-i);
    \draw[-](z2-o)to(f2-i);
    \draw[-](m1-o)to(o2);
    \draw[-](m2-o)to(o3);
  \end{tikzpicture}
}

\newcommand{\figCouplingCompositionUnitalLeftProofOne}{
  \begin{tikzpicture}[baseline=0.2cm,color=pluscolor]
    \pic(u)at(1,0.6){morphism=\(u\)};
    \pic(m1)at(1.8,0.4){merge=0.2};
    \pic(m2)at(1.8,0.15){merge=0.3};
    \pic(z1)at(u-c|-m1-i2){bot};
    \pic(z2)at(z1-c|-m2-i2){bot};
    \coordinate[input=\(A_{1}A_{2}\)](i)at(0,|-u-i);
    \coordinate[output=\(B_{1}B_{2}\)](o1)at(2.3,|-u-o1);
    \coordinate[output=\(B_{1}\)](o2)at(o1|-m1-o);
    \coordinate[output=\(B_{2}\)](o3)at(o2|-m2-o);
    \draw[-](i)to(u-i);
    \draw[-](u-o1)to(o1);
    \draw[-](u-o)to(m1-i1);
    \draw[-](u-o2)to(m2-i1);
    \draw[-](z1-o)to(m1-i2);
    \draw[-](z2-o)to(m2-i2);
    \draw[-](m1-o)to(o2);
    \draw[-](m2-o)to(o3);
  \end{tikzpicture}
}

\newcommand{\figUniformityZeroProofIfLeft}{
  \begin{tikzpicture}[baseline=-0.1cm,color=pluscolor]
    \pic(m)at(0,0){merge=0.3};
    \pic(f1)at(-0.8,|-m-i1){morphism=\(f\)};
    \pic(f2)at(f1-c|-m-i2){morphism=\(f\)};
    \coordinate[input=\(A\)](i1)at(-1.3,|-f1-i);
    \coordinate[input=\(A\)](i2)at(i1|-f2-i);
    \coordinate[output=\(B\)](o)at(0.5,|-m-o);
    \draw[-](i1)to(f1-i);
    \draw[-](i2)to(f2-i);
    \draw[-](f1-o)to(m-i1);
    \draw[-](f2-o)to(m-i2);
    \draw[-](m-o)to(o);
  \end{tikzpicture}
}

\newcommand{\figUniformityZeroProofIfRight}{
  \begin{tikzpicture}[baseline=-0.1cm,color=pluscolor]
    \pic(m)at(0,0){merge=0.3};
    \pic(f)at(0.8,|-m-o){morphism=\(f\)};
    \coordinate[input=\(A\)](i1)at(-0.5,|-m-i1);
    \coordinate[input=\(A\)](i2)at(i1|-m-i2);
    \coordinate[output=\(B\)](o)at(1.3,|-f-o);
    \draw[-](i1)to(m-i1);
    \draw[-](i2)to(m-i2);
    \draw[-](m-o)to(f-i);
    \draw[-](f-o)to(o);
  \end{tikzpicture}
}

\newcommand{\figUniformityZeroProofThenLeft}{
  \begin{tikzpicture}[baseline=-0.1cm,color=pluscolor]
    \pic(m)at(0,0){merge};
    \pic(f)at(-0.8,|-m-i2){morphism=\(f\)};
    \coordinate[input=\(A\)](i)at(-1.3,|-f-i);
    \draw(m-o)to[trace=0.6cm](m-i1);
    \draw[-](i)to(f-i);
    \draw[-](f-o)to(m-i2);
  \end{tikzpicture}
}

\newcommand{\figLoopForever}{
  \begin{tikzpicture}[baseline=-0.1cm,color=pluscolor]
    \pic(m)at(0,0){merge};
    \coordinate[input=\(A\)](i)at(-1,|-m-i2);
    \draw(m-o)to[trace=0.4cm](m-i1);
    \draw[-](i)to(m-i2);
  \end{tikzpicture}
}
\newcommand{\figUniformityZeroProofThenRight}{\figLoopForever}

\newcommand{\figAssignmentLeft}{
  \begin{tikzpicture}[baseline=-0.1cm,color=tensorcolor]
    \coordinate[input](i1)at(-0.3,0.5);
    \coordinate[input](i2)at(i1|-,0);
    \coordinate[input](i3)at(i1|-,-0.7);
    \pic(c1)at(0,|-i2){copy};
    \pic(e1)at(0.8,0.35){morphism=\(\semn{e_{1}}\)};
    \pic(c2)at(1.5,|-c1-o2){copy};
    \pic(e2)at(2.3,-0.55){morphism=\(\semn{e_{2}}\)};
    \coordinate[output](o1)at(2.8,|-e1-o);
    \coordinate[output](o2)at(o1|-c2-o1);
    \coordinate[output](o3)at(o1|-e2-o);
    \draw[-](i1)to(e1-i1);
    \draw[-](i2)to(c1-i);
    \draw[-](i3)to(e2-i2);
    \draw[-](c1-o1)to(e1-i2);
    \draw[-](c1-o2)to(c2-i);
    \draw[-](e1-o)to(o1);
    \draw[-](c2-o1)to(o2);
    \draw[-](c2-o2)to(e2-i1);
    \draw[-](e2-o)to(o3);
  \end{tikzpicture}
}

\newcommand{\figAssignmentRight}{
  \begin{tikzpicture}[baseline=-0.1cm,color=tensorcolor]
    \coordinate[input](i1)at(-0.3,0.5);
    \coordinate[input](i2)at(i1|-,-0.2);
    \coordinate[input](i3)at(i1|-,-0.7);
    \pic(c2)at(0,|-i2){copy};
    \pic(e2)at(0.8,-0.55){morphism=\(\semn{e_{2}}\)};
    \pic(c1)at(1.5,|-c2-o1){copy};
    \pic(e1)at(2.3,0.35){morphism=\(\semn{e_{1}}\)};
    \coordinate[output](o1)at(2.8,|-e1-o);
    \coordinate[output](o2)at(o1|-c1-o2);
    \coordinate[output](o3)at(o1|-e2-o);
    \draw[-](i1)to(e1-i1);
    \draw[-](i2)to(c2-i);
    \draw[-](i3)to(e2-i2);
    \draw[-](c1-o1)to(e1-i2);
    \draw[-](c1-o2)to(o2);
    \draw[-](e1-o)to(o1);
    \draw[-](c2-o1)to(c1-i);
    \draw[-](c2-o2)to(e2-i1);
    \draw[-](e2-o)to(o3);
  \end{tikzpicture}
}

\newcommand{\figIdentity}[1][tensorcolor]{
  \begin{tikzpicture}[baseline=-0.1cm,color={#1}]
    \draw[-](0,0)to(1,0);
  \end{tikzpicture}
}

\newcommand{\figCounitality}[1][tensorcolor]{
  \begin{tikzpicture}[baseline=-0.1cm,color={#1}]
    \pic(m)at(0,0){copy};
    \pic(u)at(0.4,|-m-o1){discard};
    \draw[-](m-o2)to(0.8,|-m-o2);
    \draw[-](m-o1)to(u-i);
    \draw[-](-0.3,|-m-i)to(m-i);
  \end{tikzpicture}
}

\newcommand{\figCoassociativityFirst}[1][tensorcolor]{
  \begin{tikzpicture}[baseline=-0.1cm,color={#1}]
    \pic(m1)at(0,0){copy};
    \pic(m2)at(0.4,|-m1-o1){copy};
    \draw[-](-0.3,|-m1-i)to(m1-i);
    \draw[-](m1-o2)to(0.8,|-m1-o2);
    \draw[-](m1-o1)to(m2-i);
    \draw[-](m2-o1)to(0.8,|-m2-o1);
    \draw[-](m2-o2)to(0.8,|-m2-o2);
  \end{tikzpicture}
}

\newcommand{\figCoassociativitySecond}[1][tensorcolor]{
  \begin{tikzpicture}[baseline=-0.1cm,color={#1}]
    \pic(m1)at(0,0.3){copy};
    \pic(m2)at(0.4,|-m1-o2){copy};
    \draw[-](m1-o1)to(0.8,|-m1-o1);
    \draw[-](-0.3,|-m1-i)to(m1-i);
    \draw[-](m1-o2)to(m2-i);
    \draw[-](m2-o1)to(0.8,|-m2-o1);
    \draw[-](m2-o2)to(0.8,|-m2-o2);
  \end{tikzpicture}
}

\newcommand{\figCocommutativity}[1][tensorcolor]{
  \begin{tikzpicture}[baseline=-0.1cm,color={#1}]
    \pic(m)at(0,0){copy};
    \draw[-](m-o1)to(0.8,|-m-o2);
    \draw[-](m-o2)to(0.8,|-m-o1);
    \draw[-](-0.3,|-m-i)to(m-i);
  \end{tikzpicture}
}

\newcommand{\figComonoid}[1][tensorcolor]{
  \begin{tikzpicture}[baseline=-0.1cm,color={#1}]
    \pic(m)at(0,0){copy};
    \draw[-](-0.3,|-m-i)to(m-i);
  \end{tikzpicture}
}

\newcommand{\figDeterministicRight}{
  \begin{tikzpicture}[baseline=-0.1cm,color=tensorcolor]
    \pic(m)at(0,0){copy=0.3};
    \pic(f1)at(0.7,|-m-o1){morphism=\(f\)};
    \pic(f2)at(f1-c|-m-o2){morphism=\(f\)};
    \draw[-](-0.3,|-m-i)to(m-i);
    \draw[-](m-o1)to(f1-i);
    \draw[-](m-o2)to(f2-i);
    \draw[-](f1-o)to(1.3,|-f1-o);
    \draw[-](f2-o)to(1.3,|-f2-o);
  \end{tikzpicture}
}

\newcommand{\figDeterministicLeft}{
  \begin{tikzpicture}[baseline=-0.1cm,color=tensorcolor]
    \pic(m)at(0,0){copy=0.3};
    \pic(f)at(-0.5,|-m-i){morphism=\(f\)};
    \draw[-](-1,|-f-i)to(f-i);
    \draw[-](f-o)to(m-i);
    \draw[-](m-o1)to(0.4,|-m-o1);
    \draw[-](m-o2)to(0.4,|-m-o2);
  \end{tikzpicture}
}

\newcommand{\figTotalLeft}{
  \begin{tikzpicture}[baseline=-0.1cm,color=tensorcolor]
    \pic(m)at(0,0){discard};
    \pic(f)at(-0.5,|-m-i){morphism=\(f\)};
    \draw[-](-1,|-f-i)to(f-i);
    \draw[-](f-o)to(m-i);
  \end{tikzpicture}
}

\newcommand{\figDiscard}{
  \begin{tikzpicture}[baseline=-0.1cm,color=tensorcolor]
    \pic(m)at(0,0){discard};
    \draw[-](-0.5,|-m-i)to(m-i);
  \end{tikzpicture}
}

\newcommand{\figAssert}{
  \begin{tikzpicture}[baseline=-0.1cm,color=tensorcolor]
    \pic(m)at(0,0){copy};
    \pic(p)at(0.7,|-m-o1){morphism=\(\semn{p}\)};
    \draw[-](m-o2)to(1.2,|-m-o2);
    \draw[-](m-o1)to(p-i);
    \draw[-](-0.3,|-m-i)to(m-i);
  \end{tikzpicture}
}

\newcommand{\figUnitality}[1][pluscolor]{
  \begin{tikzpicture}[baseline=-0.1cm,color={#1}]
    \pic(m)at(0,0){merge};
    \pic(u)at(-0.4,|-m-i1){bot};
    \draw[-](-0.8,|-m-i2)to(m-i2);
    \draw[-](u-o)to(m-i1);
    \draw[-](m-o)to(0.3,|-m-o);
  \end{tikzpicture}
}

\newcommand{\figAssociativityFirst}[1][pluscolor]{
  \begin{tikzpicture}[baseline=-0.1cm,color={#1}]
    \pic(m1)at(0,0){merge};
    \pic(m2)at(-0.4,|-m1-i1){merge};
    \draw[-](m1-o)to(0.3,|-m1-o);
    \draw[-](-0.8,|-m1-i2)to(m1-i2);
    \draw[-](m2-o)to(m1-i1);
    \draw[-](-0.8,|-m2-i1)to(m2-i1);
    \draw[-](-0.8,|-m2-i2)to(m2-i2);
  \end{tikzpicture}
}

\newcommand{\figAssociativitySecond}[1][pluscolor]{
  \begin{tikzpicture}[baseline=-0.1cm,color={#1}]
    \pic(m1)at(0,0.3){merge};
    \pic(m2)at(-0.4,|-m1-i2){merge};
    \draw[-](-0.8,|-m1-i1)to(m1-i1);
    \draw[-](m1-o)to(0.3,|-m1-o);
    \draw[-](m2-o)to(m1-i2);
    \draw[-](-0.8,|-m2-i1)to(m2-i1);
    \draw[-](-0.8,|-m2-i2)to(m2-i2);
  \end{tikzpicture}
}

\newcommand{\figCommutativity}[1][pluscolor]{
  \begin{tikzpicture}[baseline=-0.1cm,color={#1}]
    \pic(m)at(0,0){merge};
    \draw[-](-0.8,|-m-i2)to(m-i1);
    \draw[-](-0.8,|-m-i1)to(m-i2);
    \draw[-](m-o)to(0.3,|-m-o);
  \end{tikzpicture}
}

\newcommand{\figMonoid}[1][pluscolor]{
  \begin{tikzpicture}[baseline=-0.1cm,color={#1}]
    \pic(m)at(0,0){merge};
    \draw[-](m-o)to(0.3,|-m-o);
  \end{tikzpicture}
}

\newcommand{\figCodeterministicRight}{
  \begin{tikzpicture}[baseline=-0.1cm,color=pluscolor]
    \pic(m)at(0,0){merge=0.3};
    \pic(f1)at(-0.7,|-m-i1){morphism=\(f\)};
    \pic(f2)at(f1-c|-m-i2){morphism=\(f\)};
    \draw[-](m-o)to(0.3,|-m-o);
    \draw[-](f1-o)to(m-i1);
    \draw[-](f2-o)to(m-i2);
    \draw[-](-1.3,|-f1-i)to(f1-i);
    \draw[-](-1.3,|-f2-i)to(f2-i);
  \end{tikzpicture}
}

\newcommand{\figCodeterministicLeft}{
  \begin{tikzpicture}[baseline=-0.1cm,color=pluscolor]
    \pic(m)at(0,0){merge=0.3};
    \pic(f)at(0.5,|-m-o){morphism=\(f\)};
    \draw[-](f-o)to(1,|-f-o);
    \draw[-](m-o)to(f-i);
    \draw[-](-0.4,|-m-i1)to(m-i1);
    \draw[-](-0.4,|-m-i2)to(m-i2);
  \end{tikzpicture}
}

\newcommand{\figCototalLeft}{
  \begin{tikzpicture}[baseline=-0.1cm,color=pluscolor]
    \pic(m)at(0,0){bot};
    \pic(f)at(0.5,|-m-o){morphism=\(f\)};
    \draw[-](f-o)to(1,|-f-o);
    \draw[-](m-o)to(f-i);
  \end{tikzpicture}
}

\newcommand{\figZero}{
  \begin{tikzpicture}[baseline=-0.1cm,color=pluscolor]
    \pic(m)at(0,0){bot};
    \draw[-](m-o)to(0.5,|-m-o);
  \end{tikzpicture}
}

\newcommand{\figIfElseLeft}{
  \begin{tikzpicture}[baseline=-0.1cm,color=pluscolor]
    \pic(b)at(0,0){morphism=\(\branch{\semn{b}}\)};
    \pic(f)at(1,0.3){morphism=\(\semn{f}\)};
    \pic(g)at(1,-0.3){morphism=\(\semn{g}\)};
    \pic(m)at(1.8,|-b-c){merge=0.3};
    \pic(h)at(2.4,|-m-o){morphism=\(\semn{h}\)};
    \draw[-](-0.5,|-b-i)to(b-i);
    \draw[-](b-o1)to(f-i);
    \draw[-](b-o2)to(g-i);
    \draw[-](f-o)to(m-i1);
    \draw[-](g-o)to(m-i2);
    \draw[-](m-o)to(h-i);
    \draw[-](h-o)to(2.9,|-h-o);
  \end{tikzpicture}
}

\newcommand{\figIfElseRight}{
  \begin{tikzpicture}[baseline=-0.1cm,color=pluscolor]
    \pic(b)at(0,0){morphism=\(\branch{\semn{b}}\)};
    \pic(f)at(1,0.3){morphism=\(\semn{f}\)};
    \pic(g)at(1,-0.3){morphism=\(\semn{g}\)};
    \pic(m)at(2.4,|-b-c){merge=0.3};
    \pic(h1)at(1.8,|-f-o){morphism=\(\semn{h}\)};
    \pic(h2)at(1.8,|-g-o){morphism=\(\semn{h}\)};
    \draw[-](-0.5,|-b-i)to(b-i);
    \draw[-](b-o1)to(f-i);
    \draw[-](b-o2)to(g-i);
    \draw[-](f-o)to(h1-i);
    \draw[-](g-o)to(h2-i);
    \draw[-](h1-o)to(m-i1);
    \draw[-](h2-o)to(m-i2);
    \draw[-](m-o)to(2.9,|-m-o);
  \end{tikzpicture}
}

\newcommand{\figTraceUniformLeftIf}{
  \begin{tikzpicture}[baseline=-0.1cm,color=pluscolor]
    \pic(f)at(0,0){morphism=\(f\)};
    \pic(u)at(0.8,|-f-o1){smallmorphism=\(u\)};
    \coordinate[input](i1)at(-0.5,|-f-i1);
    \coordinate[input](i2)at(i1|-f-i2);
    \coordinate[output](o1)at(1.3,|-u-o);
    \coordinate[output](o2)at(o1|-f-o2);
    \draw[-](i1)to(f-i1);
    \draw[-](i2)to(f-i2);
    \draw[-](f-o1)to(u-i);
    \draw[-](u-o)to(o1);
    \draw[-](f-o2)to(o2);
  \end{tikzpicture}
}

\newcommand{\figTraceUniformRightIf}{
  \begin{tikzpicture}[baseline=-0.1cm,color=pluscolor]
    \pic(f)at(0,0){morphism=\(g\)};
    \pic(u)at(-0.8,|-f-i1){smallmorphism=\(u\)};
    \coordinate[input](i1)at(-1.3,|-f-i1);
    \coordinate[input](i2)at(i1|-f-i2);
    \coordinate[output](o1)at(0.5,|-f-o1);
    \coordinate[output](o2)at(o1|-f-o2);
    \draw[-](i1)to(u-i);
    \draw[-](u-o)to(f-i1);
    \draw[-](i2)to(f-i2);
    \draw[-](f-o1)to(o1);
    \draw[-](f-o2)to(o2);
  \end{tikzpicture}
}

\newcommand{\figTrace}[1]{
  \begin{tikzpicture}[baseline=-0.1cm,color=pluscolor]
    \pic(f)at(0,0){morphism=\(#1\)};
    \coordinate[input](i)at(-0.7,|-f-i2);
    \coordinate[output](o)at(0.7,|-f-o2);
    \coordinate(c1)at(-0.3,|-f-i1);
    \coordinate(c2)at(0.3,|-f-o1);
    \draw(c2)to[trace=0.4cm](c1);
    \draw[-](i)to(f-i2);
    \draw[-](c1)to(f-i1);
    \draw[-](f-o1)to(c2);
    \draw[-](f-o2)to(o);
  \end{tikzpicture}
}

\newcommand{\figSemanticsWhile}[2]{
  \begin{tikzpicture}[baseline=-0.1cm,color=pluscolor]
    \pic(b)at(0,0){morphism=\(#1\)};
    \pic(f)at(1,0.3){morphism=\(#2\)};
    \coordinate(a1)at(1,-0.3);
    \pic(m)at(-0.5,|-b-i){merge=0.3};
    \draw(f-o)to[trace=0.4cm](m-i1);
    \draw[-](-1.3,|-m-i2)to(m-i2);
    \draw[-](m-o)to(b-i);
    \draw[-](b-o1)to(f-i);
    \draw[-](b-o2)to(a1);
    \draw[-](a1)to(1.8,|-a1);
  \end{tikzpicture}
}

\newcommand{\figGuardChoice}{
  \begin{tikzpicture}[baseline=-0.1cm,color=pluscolor]
    \pic(b)at(0,0){morphism=\(\branch{b}\)};
    \pic(a)at(1,0.3){morphism=\(a\)};
    \pic(c)at(1,-0.3){morphism=\(c\)};
    \pic(m1)at(1.7,0.15){merge=0.3};
    \pic(m2)at(1.7,-0.15){merge=0.3};
    \coordinate[output](o1)at(2,|-m1-o);
    \coordinate[output](o2)at(o1|-m2-o);
    \draw[-](-0.5,|-b-i)to(b-i);
    \draw[-](b-o1)to(a-i);
    \draw[-](b-o2)to(c-i);
    \draw[-](a-o1)to(m1-i1);
    \draw[-](a-o2)to(m2-i1);
    \draw[-](c-o1)to(m1-i2);
    \draw[-](c-o2)to(m2-i2);
    \draw[-](m1-o)to(o1);
    \draw[-](m2-o)to(o2);
  \end{tikzpicture}  
}

\newcommand{\figGuardAnd}{
  \begin{tikzpicture}[baseline=-0.1cm,color=pluscolor]
    \pic(b)at(0,0){morphism=\(\branch{a}\)};
    \pic(a)at(1,0.3){morphism=\(b\)};
    \pic(m)at(1.7,-0.15){merge=0.3};
    \coordinate[output](o1)at(2,|-a-o1);
    \coordinate[output](o2)at(o1|-m-o);
    \draw[-](-0.5,|-b-i)to(b-i);
    \draw[-](b-o1)to(a-i);
    \draw[-](b-o2)to(m-i2);
    \draw[-](a-o1)to(o1);
    \draw[-](a-o2)to(m-i1);
    \draw[-](m-o)to(o2);
  \end{tikzpicture}  
}

\newcommand{\figGuardOr}{
  \begin{tikzpicture}[baseline=-0.1cm,color=pluscolor]
    \pic(b)at(0,0){morphism=\(\branch{a}\)};
    \pic(c)at(1,-0.3){morphism=\(b\)};
    \pic(m)at(1.7,0.15){merge=0.3};
    \coordinate[output](o1)at(2,|-m-o);
    \coordinate[output](o2)at(o1|-c-o2);
    \draw[-](-0.5,|-b-i)to(b-i);
    \draw[-](b-o1)to(m-i1);
    \draw[-](b-o2)to(c-i);
    \draw[-](c-o1)to(m-i2);
    \draw[-](c-o2)to(o2);
    \draw[-](m-o)to(o1);
  \end{tikzpicture}  
}

\newcommand{\figPredicateAnd}{
  \begin{tikzpicture}[baseline=-0.1cm, color=tensorcolor]
    \pic(c)at(-0.7,0){copy=0.3};
    \pic(p)at(0,|-c-o1){morphism=\(p\)};
    \pic(q)at(p-c|-c-o2){morphism=\(q\)};
    \coordinate[output](o1)at(0.5,|-p-o);
    \coordinate[output](o2)at(o1|-q-o);
    \draw[-](-1,|-c-i)to(c-i);
    \draw[-](c-o1)to(p-i);
    \draw[-](c-o2)to(q-i);
    \draw[-](p-o)to(o1);
    \draw[-](q-o)to(o2);
  \end{tikzpicture}    
}

\newcommand{\figPredicateFalse}{
  \begin{tikzpicture}[baseline=-0.1cm, color=pluscolor]
    \pic(m)at(0,0){bot};
    \pic(z)at(0.4,0){bot};
    \draw[-](-0.4,|-m-i)to(m-i);
    \draw[-](z-o)to(0.8,|-z-o);
  \end{tikzpicture}
}
\usepackage{tikz}
\usepackage{tikz-cd}
\tikzcdset{ampersand replacement=\&} %
\tikzcdset{arrows={color=localblack}}
\usetikzlibrary{babel}

\newcommand{\injectionsTotalDiagramProofFig}{
  \begin{tikzcd}
    {A} \arrow{r}{\discard} \arrow{d}[swap]{\inject} \arrow[phantom]{rd}{\scriptsize{(i)}} \& {1} \arrow{d}{\inject} \arrow{dr}{\id{}} \& {} \\
    {A+B} \arrow{r}[swap]{\discard + \discard} \arrow[bend right]{rr}[swap]{\discard} \& {1+1} \arrow[phantom]{ur}[pos=0.15]{\scriptsize{(ii)}} \arrow[phantom]{d}[pos=0.15]{\scriptsize{(iii)}} \arrow{r}[swap,pos=0.4]{\join} \& {1}\\
    \& {} \&
  \end{tikzcd}
}

\newcommand{\injectionsDeterministicDiagramProofFig}{
  \begin{tikzcd}
    {A} \arrow{rrr}{\cp} \arrow{dd}[swap]{\inject} \arrow[phantom]{rd}{\scriptsize{(i)}} \& \& \& {AA} \arrow{dd}{\inject \tensor \inject} \arrow{dll}[swap]{\inject} \arrow[phantom]{ddl}[pos=0.3]{\scriptsize{(ii)}}  \\
    \& {AA + BB} \arrow{r}[swap]{\inject + \inject} \arrow[phantom]{dr}{\scriptsize{(iii)}} \& {AA + AB + BA + BB} \arrow{dr}{(\lidistr + \lidistr) \dcomp \ridistr} \& \\
    {A+B} \arrow{rrr}[swap]{\cp} \arrow{ru}{\cp + \cp} \& \&{} \& {(A+B)(A+B)}
  \end{tikzcd}
}

\newcommand{\discardCompatibleCoproductsDiagramProofFig}{
  \begin{tikzcd}
    {A} \arrow{r}{\inject} \arrow{dr}[swap]{\discard} \& {A+B} \arrow{d}[pos=0.4]{\discard} \& {B} \arrow{l}[swap]{\inject} \arrow{dl}{\discard}\\
    {} \& {1} \& {}
  \end{tikzcd}
}

\newcommand{\copyCompatibleCoproductsDiagramProofFig}{
  \begin{tikzcd}
    {A} \arrow{r}{\inject} \arrow{d}[swap]{\cp} \& {A+B} \arrow{d}{\cp} \& {B} \arrow{l}[swap]{\inject} \arrow{d}{\cp}\\
    {AA} \arrow{r}[swap]{\inject \tensor \inject} \& {(A+B)(A+B)} \& {BB} \arrow{l}{\inject \tensor \inject}
  \end{tikzcd}
}

\newcommand{\leftDistributorDiagram}{
  \begin{tikzcd}
   {AB} \arrow{r}{\inject} \arrow{dr}[swap]{\id \tensor \inject} \& {AB + AC} \arrow{d}{\lidistr} \& {AC} \arrow{l}[swap]{\inject} \arrow{dl}{\id \tensor \inject}\\
   \& {A(B+C)} \&
  \end{tikzcd}
}

\newcommand{\rightDistributorDiagram}{
  \begin{tikzcd}
    {AC} \arrow{r}{\inject} \arrow{dr}[swap]{\inject \tensor \id} \& {AC + BC} \arrow{d}{\ridistr} \& {BC} \arrow{l}[swap]{\inject} \arrow{dl}{\inject \tensor \id}\\
    \& {(A+B)C} \&
  \end{tikzcd}
}

\AtEndPreamble{
\usepackage{hyperref}
\usepackage{cleveref} %
\crefname{remark}{Remark}{Remarks}
\Crefname{remark}{Remark}{Remarks}
}

\allowdisplaybreaks{} %
\DeclareUnicodeCharacter{2014}{\,---\,}
\makeatletter
\newcommand{\superimpose}[3][\mathord]{#1{\mathpalette\superimpose@{{#2}{#3}}}}
\newcommand{\superimpose@}[2]{\superimpose@@{#1}#2}
\newcommand{\superimpose@@}[3]{%
  \ooalign{%
    \hfil$\m@th#1#2$\hfil\cr
    \hfil$\m@th#1#3$\hfil\cr
  }%
}
\makeatother
\definecolor{med0}{HTML}{1C1B1B} %
\definecolor{med1}{HTML}{261D1D} %
\definecolor{med2}{HTML}{362B2B} %
\definecolor{med3}{HTML}{5E5757} %
\definecolor{med4}{HTML}{FFE983} %
\definecolor{med5}{HTML}{FFF4C2} %
\definecolor{med6}{HTML}{FDF6E3} %
\definecolor{med7}{HTML}{DB7842} %
\definecolor{med8}{HTML}{B32E39} %
\definecolor{med9}{HTML}{821529} %
\definecolor{medA}{HTML}{602E51} %
\definecolor{medB}{HTML}{FFC929} %
\definecolor{medC}{HTML}{60C37E} %
\definecolor{medD}{HTML}{89D7D0} %
\definecolor{medE}{HTML}{3E75DA} %
\definecolor{medF}{HTML}{D0ADE1} %

\colorlet{localblack}{black}
\colorlet{localwhite}{white}
\colorlet{localcolor}{medB}
\colorlet{localgray}{med3}
\colorlet{localblue}{medE}
\colorlet{localred}{med8}
\colorlet{addcolor}{white}
\colorlet{pluscolor}{med8}
\colorlet{tensorcolor}{medE}
\colorlet{copycolor}{tensorcolor}

\let\section\savedsection
\makeatletter
\let\ACM@origbaselinestretch\baselinestretch
\makeatother

\begin{document}
\title[Deriving Program Logics from Distributive Monoidal Categories]{Deriving Program Logics\\ from Distributive Monoidal Categories}
\author{Filippo Bonchi}
\affiliation{\institution{Università di Pisa}\country{Italy}}
\author{Elena Di Lavore}
\author{Mario Rom\'an}
\affiliation{\institution{Tallinn University of Technology}\country{Estonia}}
\author{Sam Staton}
\affiliation{\institution{University of Oxford}\country{UK}}

\begin{abstract}
  We derive multiple program logics—including correctness, incorrectness, and relational Hoare logic—from the axioms of imperative categories: uniformly traced distributive copy-discard categories.
  Rules of program logics follow from the axioms of imperative categories.
  The algebra of guarded commands derived by the categorical structure generalises guarded Kleene algebras with tests.
\end{abstract}

\maketitle

\section{Introduction}

\paragraph{\textbf{Program logics}}

Program logics provide deductive systems for reasoning about program behaviour
under assumptions on their inputs and outputs. Their judgements are typically
written as triples $\triple{p}{c}{q}$,
where \(c\) is a command, \(p\) a precondition, and \(q\) a postcondition. The
meaning of such triples depends on the behavioural property under
consideration. In Hoare logic, for example, a triple is valid whenever every
execution of \(c\) starting from a state satisfying \(p\) terminates in a state
satisfying \(q\). The familiar rule for while loops illustrates this
interpretation:
\begin{equation}\label{eq:while-rule-correctness}
  \inferrule{\triple{p \pand b}{c}{p}}
            {\triple{p}{\while{b}{c}}{\gnot b \pand p}}.
\end{equation}

Program correctness is only one possible interpretation of triples. Over the
years, a wide range of program logics has been developed to reason about
incorrectness, quantitative properties, probabilistic behaviour, and many
other semantic aspects of programs.

A program logic is determined by three ingredients: a semantics for commands,
an interpretation of triples, and a collection of inference rules that is sound
for the chosen semantics. Command semantics may be
partial~\cite{hoare1969axiomatic,benton2004relational},
relational~\cite{winskel1993formal,o2019incorrectness}, or
probabilistic~\cite{kaminski2018thesis,barthe2012probabilistic,zilberstein2023outcome}.
Similarly, triples may express
correctness~\cite{hoare1969axiomatic},
incorrectness~\cite{deVries2011reverse,o2019incorrectness}, or
quantitative properties of
execution~\cite{zilberstein2023outcome,avanzini2025quantitative}. Although
these proof systems exhibit striking similarities, they have traditionally been
developed on a case-by-case basis for each combination of program semantics and
behavioural interpretation.

\paragraph{\textbf{Algebras of programs}}

Soon after the introduction of Floyd and Hoare logics, it was recognised that the calculus of
relations~\cite{pratt1976semantical} provides a natural algebraic semantics for imperative
programs and their logics. Relational composition models sequential
composition, union models non-deterministic choice, and reflexive-transitive
closure models iteration. This observation culminated in Kozen's
\emph{Kleene Algebra with Tests} (KAT)~\cite{kozen1997kleene}, which extends Kleene algebra
with Boolean tests to provide an algebraic treatment of conditionals and loops.

A limitation of KAT is that assignments are treated as atomic commands.
Consequently, simple equivalences arising from data dependencies cannot be derived from the algebraic axioms alone.
For instance, the two programs below are behaviourally equivalent, yet their equivalence requires additional ad hoc axioms beyond the core algebra~\cite{angus2001kleene}.
\begin{equation}\label{eq:assignment-interchange}
  (\assign{x}{z}) \ccomp (\assign{y}{z}) \sequiv (\assign{y}{z}) \ccomp (\assign{x}{z})
\end{equation}

The development of new programming paradigms has led to numerous extensions of
KAT~\cite{hoare2009concurrent,anderson2014netkat,gomes2016modal,foster2016probabilistic,smolka2019guarded,rozowski2023probabilistic}. For example, probabilistic Kleene
algebra~\cite{mciver2005towards,mciver2006probabilistic} provides algebraic foundations for probabilistic computation,
while BiKAT~\cite{antonopoulos2023algebra,gomes2025bigkat} supports reasoning about pairs of programs and forms an
algebraic basis for relational Hoare
logic~\cite{benton2004relational}. Nevertheless, properties of the form \(\forall\exists\), such as
\emph{possibilistic non-interference}~\cite{clarkson2010hyperproperties}, cannot naturally be expressed in
BiKAT. This limitation has motivated proposals such as
TriKAT~\cite{antonopoulos2023algebra}, extending the algebraic framework from reasoning about pairs of
programs to reasoning about triples.

\paragraph{\textbf{Imperative categories}}
We propose the algebraic structure of \emph{imperative categories} as a common foundation for  program logics.
\kl{Imperative categories} are traced coproduct categories—long recognised as
models of control flow~\cite{arbib1980partially}—equipped with a second monoidal structure, \(\tensor\), distributing over coproducts, \(+\). Operationally, coproducts model
control flow, while the tensor corresponds to \emph{program products} in the
sense of Barthe \emph{et al.}~\cite{barthe2011relational,barthe2017coupling}. This additional structure naturally
supports relational reasoning about arbitrary numbers of programs, extending
binary alignments to products of any finite arity. At the same time, it
captures data flow, allowing program equivalences such as~\eqref{eq:assignment-interchange}
to be derived directly within the algebraic framework.

Moreover, the tensor endows imperative categories with the structure of a \kl{copy-discard category}~\cite{corradini1999algebraic,ChoJacobs2019}, a generalisation of cartesian categories
that captures copying and discarding in the presence of computational effects.
\kl{Copy-discard categories} encompass many standard semantic models, including sets and partial functions, sets and relations, sets and stochastic maps, and standard Borel spaces with measurable stochastic kernels.
This structure provides an elegant treatment of variable sharing and free variables across these diverse settings.

A final distinguishing feature of imperative categories is that their trace
satisfies \emph{posetal uniformity}, a recent strengthening of the classical
notion of trace uniformity~\cite{stefanescu1986algebraic,stefanescu1987flowchart}. Posetal uniformity provides precisely the
abstraction needed to reason uniformly about loop invariants, making imperative
categories a natural semantic foundation for a broad spectrum of program
logics.

\paragraph{\textbf{Interpreting triples.}}
The interpretation of program triples rests on comparing programs: the validity
of a Hoare triple \(\triple{p}{c}{q}\) will be defined as an inequality,
\(\assert{p} \ccomp c \pleq c \ccomp \assert{q}\). In the category of relations, where morphisms
are ordered by inclusion, we recover the validity of a partial correctness
triple: it compares the subset \(c \ccomp \assert{q}\) of possible final states with the subset
\(\assert{p} \ccomp c\) of possible outputs of \(c\) on inputs that belong to \(p\).  In
general, we require a poset enrichment on \kl{imperative categories}, leading to
\kl{posetal imperative categories}: poset-enriched categories with traced coproducts and a second monoidal copy-discard structure, interacting by distributivity. 

The most important axiom for this posetal structure is \kl{posetal uniformity},
which justifies reasoning by \emph{loop invariants}. Intuitively, it says that if a property
$p$ is invariant under execution of a command \(c\), then it remains invariant under a loop guarded by $b$. More generally, $\assert{p} \ccomp c \pleq d \ccomp \assert{p}$ implies $\assert{p} \ccomp \while{b}{c} \pleq \while{b}{d} \ccomp \assert{p}$.
With this interpretation, the categorical structure derives the validity of the triple in~\eqref{eq:while-rule-correctness} (\Cref{ex:while-invariants}) %

\subsection{Contributions and synopsis}%
\label{sec:contributions}
\Cref{sec:imperative-categories} introduces \kl{imperative categories}: a semantic universe for imperative programs with monoidal effects.
The appropriate syntax is a generalisation of the guarded command language that allows arbitrary monoidal effects; \kl{imperative categories} give it denotational semantics (\Cref{sec:monoidal-gcl}).
The categorical structure derives an algebra of guarded commands, which generalises guarded Kleene algebra with tests (\Cref{prop:gkat-from-distributive}).
\Cref{sec:poset-enrichment} introduces the poset structure of commands and uses it to derive weakest precondition and strongest postcondition calculi (\Cref{th:tightest-pre-post}).
\Cref{sec:program-logics,sec:relational-program-logics} derive the rules of multiple program logics and relational program logics from the categorical structure (\Cref{thm:hoarelogic,thm:incorrectnesslogic,thm:outcomelikelogic,thm:relationalcorrectness,thm:relationalincorrectness}).

\subsection{Related work}

\paragraph{Categorical program semantics}

Categorical program semantics has a long
tradition~\cite{lambek88:introduction,oles83:category,winskel1993formal}. In
particular, \kl{distributive categories} are since long used to model both
control flow and data flow of
programs~\cite{cockett93,carboni93:distributiveextensive,walters92}. More
specifically, \kl{distributive monoidal categories} with \kl{copy-discard}
structure have naturally appeared in non-deterministic, partial, and stochastic
semantics~\cite{bonchiGS23,liell2025compositional,nester2025elgot}. The approach is compatible
with the long tradition of using monads for
computations~\cite{moggi91,wadler1998marriage,benton1999monads,benton2000monads}.

Arbib and Manes employ traced cocartesian categories to express the control flow
of programs~\cite{arbib1980partially}, generalising Elgot's techniques for the
interpretation of iteration and choice in partial functions
\cite{elgot1975monadic}; apart from their work, categorical semantics
for iteration has been studied
extensively~\cite{bloom1993iteration,simpson2000complete}. Of particular
relevance to our work is the metalanguage for guarded iteration by Goncharov,
Rauch and Schr{\"o}der~\cite{goncharov2021metalanguage}; and the recent
denotational semantics of static single assingment of Ghalayini and Krishnaswami
\cite{ghalayini2024denotational}. When reasoning about the semantics of loops,
we employ Hasuo's generic trace
theory~\cite{hasuo2006generic,hasuo2006generictrace}, which builds on Fiore's
work on coinduction~\cite{fiore1993coinduction,fiore1996coinduction}.
\kl{Uniform traces} need not to exist in cocartesian categories. In our
examples, we ensure the existence of \kl{uniform traces} by relying on partially
additive monads~\cite{jacobs2010coalgebraic}, which ensure a form of
iteration in the Kleisli category less restrictive than additive
monads~\cite{coumans2013scalars} or Kleene monads~\cite{goncharov2010thesis}.
Fixpoint operators are an earlier axiomatisation of recursion~\cite{bloom1993iteration,stefanescu1986algebraic,stefanescu1987flowchart} that builds on the work of Elgot~\cite{elgot1975monadic,bloom1980solutions}, and Scott~\cite{scott97lattice}.
Partially additive categories~\cite{arbib1980partially} support a notion of iteration that generalises iteration theories and Scott semantics; Kleisli categories of partially additive monads are partially additive~\cite{jacobs2010coalgebraic}.
Kleene monads~\cite{goncharov2010thesis} instead ensure that their Kleisli category is additive and supports a Kleene star operator.

\paragraph{Categorical logic and algebras of programs}

The guarded command syntax for programs distinguishes between guards and commands.
We interpret this distinction in the categorical setting following the ideas from effectus theory~\cite{jacobs2015new}, where the logic on guards derives from categorical structure.
The structure of hom-sets in imperative categories generalises that of Kleene algebras with tests~\cite{kozen1997kleene} and their probabilistic variation~\cite{mciver2006probabilistic}, and of guarded Kleene algebras with tests~\cite{smolka2019guarded,gomes2025bigkat} and their probabilistic~\cite{rozowski2023probabilistic} and approximate variants~\cite{gomes2025kat}.
Guards in imperative categories do not in general form a boolean algebra as they are not necessarily deterministic.

\subsubsection*{Program logics}
Since the work of Floyd~\cite{floyd1993assigning} and Hoare~\cite{hoare1969axiomatic,clint1972program} on correctness assertions about programs, much work on program logics has extended the scope of the original logic.
Separation logic~\cite{reynolds2002separation,ohearn2001local} considers programs that access globally shared data, incorrectness logic considers assertions about faults of programs~\cite{deVries2011reverse,o2019incorrectness}, and outcome logics~\cite{zilberstein2023outcome,zilberstein2024outcome,zilberstein2025demonic} provide a synthesis of correctness and incorrectness reasoning.
Verification of probabilistic programs is an active research area that takes another view on program logics studying weakest precondition and strongest postcondition calculi~\cite{kaminski2017weakest,kaminski2018thesis,zhang2022quantitative}.

Relational program logics extend the reasoning of program logics to pairs of programs considering binary relations between their inputs instead of predicates on the inputs of one program alone.
As in the predicate version, relational program logics can focus on correctness assertions about deterministic programs~\cite{benton2004relational}, or be extended to probabilistic semantics~\cite{barthe2009formal,avanzini2025quantitative} and approximate reasoning~\cite{olmedo2014thesis,barthe2012probabilistic,sato2016approximate,aguirre2021preexpectation}.

Categorical approaches to program logics are not new. Manes and Arbib describe
the control flow of Hoare logic with traced cocartesian
categories~\cite{manes2012algebraic}. Outcome logic considers a class of
semantic universes given by Kleisli categories of monads with some extra
structure~\cite{gomes2016modal,zilberstein2023outcome}. Program triples can also be seen as
fibrations over a category of
programs~\cite{mellies2015functors,mellies2016bifibrational} or as functors to
monotone relations~\cite{arthan2009general}. More recently, the structure of
distributive categories as been shown to derive the rules of Hoare logic,
restricted to the relational semantics~\cite{2025tracedtapes}.

\section{Imperative categories}%
\label{sec:imperative-categories}

This section introduces \kl{imperative categories}.
These combine two ideas from the program semantics literature: a copy-discard monoidal structure for the semantics of data flow—how variables are shared across programs—and a traced coproduct structure for the semantics of control flow—if-else statements and while loops.

\subsection{Copy-discard categories}

Monoidal categories allow parallel compositions of morphisms: two programs that run in parallel on distinct variables take semantics as a monoidal product of the corresponding morphisms.
The additional copy-discard structure allow programs to run in parallel while sharing some of the variables.

\begin{example}%
  \label{ex:assignment-interchange}
  The program \(\assign{x_{1}}{e_{1}(x_{1}, x_{3})} \ccomp \assign{x_{2}}{e_{2}(x_{2},x_{3})}\) assigns to the variable \(x_{1}\) the value of the expression \(e_{1}(x_{1}, x_{3})\) and then assigns to \(x_{2}\) the value of \(e_{2}(x_{2},x_{3})\).
  The two assignments share a free variable, \(x_{3}\), but modify distinct variables, the first modifies \(x_{1}\) while the second \(x_{2}\).
  Because of this, we should be able to interchange the order of the two assignments without altering the result.
  \[\assign{x_{1}}{e_{1}(x_{1}, x_{3})} \ccomp \assign{x_{2}}{e_{2}(x_{2},x_{3})} \sequiv \assign{x_{2}}{e_{2}(x_{2},x_{3})} \ccomp \assign{x_{1}}{e_{1}(x_{1}, x_{3})}\]
\end{example}

Programs that share variables are naturally interpreted in \kl{copy-discard categories}.
The algebra of \kl{copy-discard categories} proves valid interchange equations like the one in \Cref{ex:assignment-interchange}.

\begin{definition}
  A \intro{copy-discard category} is a \kl{symmetric monoidal category}, \((\cat{C}, \tensor, 1)\), where every object has a commutative comonoid structure (\Cref{fig:ax-commutative-comonoid}) that is compatible with the monoidal product.
  Explicitly, for every object \(A\), there are a \intro{copy} morphism, \(\cp_{A} \colon A \to A \tensor A\), and a \intro{discard} morphism, \(\discard_{A} \colon A \to 1\), that are associative, \(\cp_{A} \dcomp (\id_{A} \tensor \cp_{A}) = \cp_{A} \dcomp (\cp_{A} \tensor \id_{A})\), unital, \(\cp_{A} \dcomp (\id_{A} \tensor \discard_{A}) = \id_{A} = \cp_{A} \dcomp (\discard_{A} \tensor \id_{A})\), commutative, \(\cp_{A} \dcomp \swap_{A,A} = \cp_{A}\), and compatible with the monoidal product, \(\cp_{A \tensor B} = (\cp_{A} \tensor \cp_{B}) \dcomp (\id_{A} \tensor \swap_{A,B} \tensor \id_{B})\), \(\discard_{A \tensor B} = \discard_{A} \tensor \discard_{B}\), \(\cp_{1} = \id_{1}\), \(\discard_{1} = \id_{1}\).
\end{definition}
\begin{toappendix}
  \begin{figure}[h!]
    \begin{align*}
      \figCoassociativityFirst &= \figCoassociativitySecond & \figCounitality & = \figIdentity & \figCocommutativity &= \figComonoid
    \end{align*}
    \caption{Axioms of a commutative comonoid.}%
    \label{fig:ax-commutative-comonoid}
  \end{figure}
\end{toappendix}

When convenient, we use string diagrams to specify morphisms in monoidal categories.
\kl{Imperative categories} have two monoidal structures, which we will distinguish with two colours:
a \textcolor{tensorcolor}{monoidal product (\(\tensor\))} and a \textcolor{pluscolor}{coproduct (\(+\))}\footnote{Bimonoidal categories do have a diagrammatic language~\cite{comfort2020sheet,bonchiGS23}, but we will not need to look at both monoidal products at the same time, so we use string diagrams in each of them separately for simplicity.}.

\begin{example}%
  \label{ex:assignment-interchange-semantics}
  The programs in \Cref{ex:assignment-interchange} can be given semantics in a \kl{copy-discard category}.
  If we fix an objext, \(X\), and two morphisms, \(\semn{e_1}, \semn{e_2} \colon X \tensor X \to X\), that interpret the two expressions, we define the semantic of the assignment statements.
  \begin{align*}
    &\semn{\assign{x_1}{e_1(x_1,x_3)} \ccomp \assign{x_2}{e_2(x_2,x_3)}} &&\semn{\assign{x_2}{e_2(x_2,x_3)} \ccomp \assign{x_1}{e_1(x_1,x_3)}} \\
    &\defn \figAssignmentLeft &&\defn \figAssignmentRight
  \end{align*}
  By the interchange axiom of monoidal categories and associativity of copy morphisms, the two programs have the same semantics.
\end{example}

We work with monoidal rather than premonoidal categories: stack variables and other premonoidal effects are handled by a state-passing style transformation (\Cref{rem:sps-transformation}).

\begin{toappendix}
\begin{definition}
  A \intro{monad} on a category \(\cat{C}\) is a triple \((T, \eta, \klextend{(-)})\) of a functor \(T \colon \cat{C} \to \cat{C}\), a family of morphisms \(\eta_{A} \colon A \to T(A)\) indexed by objects \(A\) of \(\cat{C}\), and an \intro[Kleisli extension]{}operation on hom-sets \(\klextend{(-)} \colon \cat{C}(A,TB) \to \cat{C}(TA,TB)\) satisfying
  \emph{(i)} \(\klextend{\eta_{A}} = \id_{TA}\),
  \emph{(ii)} \(\eta_{A} \dcomp \klextend{f} = f\), and
  \emph{(iii)} \(\klextend{f} \dcomp \klextend{g} = \klextend{(f \dcomp \klextend{g})}\).
\end{definition}

The Kleisli category of a \kl{monad} \(T \colon \cat{C} \to \cat{C}\) commonly serves as semantics for computations in \(\cat{C}\) with \(T\)-effects~\cite{moggi91}.

\begin{definition}
  For a \kl{monad} \(T\) on a category \(\cat{C}\), its \intro{Kleisli category}, \(\kleisli{T}\), has the same objects as \(\cat{C}\) and the morphisms \(A \to B\) are the morphisms \(A \to T(B)\) in \(\cat{C}\).
  Identities are given by the monad unit, \(\eta_{A}\), and the composition is defined with \kl{Kleisli extensions}, \(f \dcomp \klextend{g}\).
\end{definition}

When the base category has a monoidal structure, we may ask that the monad interacts well with it to ensure that the monoidal structure lifts to the \kl{Kleisli category}.

\begin{definition}
  A \kl{monad}, \((T, \eta, \klextend{(-)})\), on a \kl{symmetric monoidal category}, \((\cat{C}, \monoidal, I, \mswap)\) is \intro[strong monad]{strong} if there is a natural transformation \(\lstr_{A,B} \colon A \monoidal T(B) \to T(A \monoidal B)\), the \intro{left strength}, that is compatible with the monoidal structure and with the monad structure:
  \emph{(i)} \(\lambda_{TA} \dcomp \lstr_{I,A} = T(\lambda_{A})\),
  \emph{(ii)} \(\lstr_{A \monoidal B, C} \dcomp T(\alpha_{A,B,C}) = \alpha_{A,B,TC} \dcomp (\id_{A} \monoidal \lstr_{B,C}) \dcomp \lstr_{A,B \monoidal C}\),
  \emph{(iii)} \((\id_{A} \monoidal \eta_{B}) \dcomp \lstr_{A,B} = \eta_{A \monoidal B}\), and
  \emph{(iv)} \((\id_{A} \monoidal \mu_{B}) \dcomp \lstr_{A,B} = \lstr_{A,TB} \dcomp T(\lstr_{A,B}) \dcomp \mu_{A \monoidal B}\),
  where \(\alpha\), \(\lambda\) and \(\rho\) denote the associator, and left and right unitors of the monoidal category \(\cat{C}\), and \(\mu\) denotes the monad multiplication, \(\mu_{A} = \klextend{\id_{TA}}\).

  A strong monad is \intro[commutative monad]{commutative} if the two morphism of type \(TA \monoidal TB \to T(A \monoidal B)\) obtained by composing strengths and symmetries coincide: \(\lstr_{TA,B} \dcomp T(\rstr_{A,B}) \dcomp \mu_{A \monoidal B} = \rstr_{A,TB} \dcomp T(\lstr_{A,B}) \dcomp \mu_{A \monoidal B}\), where \(\rstr_{A,B} = \mswap_{TA,B} \dcomp \lstr_{B,A} \dcomp T(\mswap_{B,A})\) is the \intro{right strength} obtained by composing the left strength \(\lstr\) with the symmetry \(\mswap\).
\end{definition}
\end{toappendix}

A class of examples of \kl{copy-discard categories} come from \kl{Kleisli categories}.

\begin{lemma}%
  \label{lem:klelisli-copy-discard}
  The \kl{Kleisli category} of a \kl{commutative monad} on a \kl{cartesian category} is a \kl{copy-discard category}.
  The copy-discard structure is inherited from the base category via the monad unit, \(\cp_{A} \dcomp \eta_{A \times A}\) and \(\discard_{A} \dcomp \eta_{1}\).
\end{lemma}

\begin{example}%
  \label{ex:substoch}
  A concrete instance of \Cref{lem:klelisli-copy-discard} is \(\subStoch\), the \kl{Kleisli category} of the \intro{countably-supported subdistribution monad}, \(\subdistr \colon \Set \to \Set\)~\cite{jacobs2010coalgebraic,bowler2025probabilistic}.
  The monad maps a set \(A\) to the set of functions \(\sigma \colon A \to [0,1]\) whose \intro{support}, \(\supp(\sigma) \defn \{a \in A \mid \sigma(a) > 0\}\), is countable and whose total \intro{probability mass} is at most \(1\), \(\mass(\sigma) \defn \sum_{a \in A} \sigma(a) \leq 1\).
  The unit of the monad is the Dirac's delta, \(\eta_{A}(a)(a') \defn [a=a']\), and the Kleisli extension sums over the possible inputs, \(\klextend{f}(\sigma)(b) \defn \sum_{a \in A} \sigma(a) \cdot f(a)(b)\).
  The monoidal structure is the inclusion \(i \colon \subdistr(A) \times \subdistr(B) \to \subdistr(A \times B)\) defined by \(i(\sigma, \tau)(a,b) \defn \sigma(a) \cdot \tau(b)\).
  The value \(f(a)(b)\) indicates the probability that the channel \(f\) returns \(b\) on input \(a\); in other words, it is the probability of \(b\) given \(a\), which suggests the \intro[given]{}notation \(f(b \given a) \defn f(a)(b)\).
  More examples are provided in \Cref{sec:examples-imperative}.
\end{example}

\kl{Copy-discard categories} are strictly more general than cartesian categories.
These are characterised by naturality of the comonoid structure.

\begin{theorem}[\cite{fox1976coalgebras}]%
  \label{th:fox}
  A \kl{copy-discard category}, \(\cat{C}\), is cartesian if and only if the \kl{copy} and \kl{discard} morphisms are the components of two natural transformations, \(\cp \colon \id_{\cat{C}} \to (- \tensor -)\) and \(\discard \colon \id_{\cat{C}} \to 1\).
  Explicitly, for all \(f \colon A \to B\), \(f \dcomp \discard_{B} = \discard_{A}\) and \(f \dcomp \cp_{B} = \cp_{A} \dcomp (f \tensor f)\) (also in \Cref{fig:total-deterministic}).
\end{theorem}

\kl{Copy-discard categories} let us copy and discard variables, but not programs.
This allows for effects like probability: throwing a coin and copying the result is different from throwing two identical coins.
Still, some programs may commute with \kl{discard} or \kl{copy} morphisms.

\begin{definition}
  A morphism \(f \colon A \to B\) in a \kl{copy-discard category} is \intro{total} if it commutes with the \kl{discard} morphism, \(f \dcomp \discard_{B} = \discard_{A}\), and it is \intro{deterministic} if it commutes with the \kl{copy} morphism, \(f \dcomp \cp_{B} = \cp_{A} \dcomp (f \tensor f)\) (\Cref{fig:total-deterministic}).
\end{definition}
\begin{toappendix}
  \begin{figure}[h!]
    \begin{align*}
      \figTotalLeft & = \figDiscard & \figDeterministicLeft &= \figDeterministicRight
    \end{align*}
    \caption{Total (left) and deterministic (right) morphisms.}%
    \label{fig:total-deterministic}
  \end{figure}
\end{toappendix}

The monoidal product on \kl{Kleisli categories} that is inherited from the base \kl{cartesian category} may cease to be cartesian: morphisms with \(T\)-effects may not be \kl{deterministic} or \kl{total}.
The example of \kl{subdistributions} justifies the names \kl{total} and \kl{deterministic}.

\begin{example}
  \kl{Deterministic} morphisms in \(\subStoch\) are partial functions, i.e.\ morphisms \(f \colon A \to B\) where \(f(b \given a) = 0\) or \(f(b \given a) = 1\) for all \(a \in A\) and \(b \in B\); \kl{total} morphisms are total stochastic channels, i.e.\ morphisms \(f \colon A \to B\) where the total \kl{probability mass} of \(f(a)\) is \(1\), for all \(a \in A\).
\end{example}

\begin{remark}[Predicates]%
  \label{rem:predicates}
  Morphisms \(p \colon A \to 1\) in \(\subStoch\) assign to each element \(a \in A\) a number in the real interval \([0,1]\), representing the probability that \(a\) satisfies the property encoded by \(p\).
  In other words, morphisms of type \(A \to 1\) are ``fuzzy predicates''.
  Deterministic morphisms of type \(A \to 1\) assign either \(0\) or \(1\) to each element of \(A\); these correspond to subsets of \(A\) and define boolean predicates.  
  A similar interpretation is true for morphisms of this type in other \kl{copy-discard categories}, so we will call \emph{predicates} the morphisms to the monoidal unit, \(1\).

  The copy-discard structure gives semantics to \emph{assert} statements: we interpret a syntactic predicates, \(p\), as predicates in a \kl{copy-discard category}, \(\semn{p} \colon X \to 1\), and assertions by copying the input.
  \[\semn{\assert{p}} = \cp_X \dcomp (\semn{p} \tensor \id_X) = \figAssert\]
\end{remark}

\begin{remark}[Notation]
  The copy-discard structure, \(\cp_{A}, \discard_{A}\), determines \intro[unbiased copy]{} \(n\)-fold copy morphisms \(\ncp_{A}^{n} \colon A \to A^{n}\), for \(n \in \naturals\), where \(A^{n}\) denotes the \(n\)-fold monoidal product of \(A\) with itself;
  thanks to \kl[copy-discard category]{associativity and unitality}, all the possible compositions of copy and discard morphisms of type \(A \to A^{n}\) are equal, so there is no ambiguity in using the notation \(\ncp_{A}^{n}\).
  With this notation, \(\cp_{A} = \ncp_{A}^{2}\) and \(\discard_{A} = \ncp_{A}^{0}\).
  The copy-discard structure also determines \intro{projections}, \(\proj_{i_{1}, \dots, i_{k}} \colon A^{n} \to A^{k}\), that keep the components in positions \(i_{1}, \dots, i_{k}\) and discard the rest.
  Similarly, we denote with \(\proj_{-i_{1}, \dots, -i_{k}} \colon A^{n} \to A^{n-k}\) the morphism that discards the components in positions \(i_{1}, \dots, i_{k}\) and keeps the rest.
  These morphisms do not necessarily form natural transformations because the monoidal unit is not necessarily terminal.
  The \intro{symmetries}, \(\swap_{A,B} \colon A \tensor B \to B \tensor A\), of a symmetric monoidal category generate arbitrary \intro[permutations]{} permutations.
  For a permutation \(\rho\) on \(n\) elements, we denote with \(\sigma_{A,\rho}\) the corresponding permutation of the objects \(A^{n}\).
\end{remark}

\subsection{Distributive copy-discard categories}

The control flow of programs requires another monoidal structure; programs should be able to split execution threads according to boolean variables and later merge them again.
We require the additional monoidal product to be a coproduct: the dual statement of \Cref{th:fox} ensures that merging execution threads is a natural operation (\Cref{fig:ax-natural-commutative-monoid}).
\begin{toappendix}
\begin{figure}[h!]
  \begin{gather*}
    \figAssociativityFirst = \figAssociativitySecond \qquad\quad \figUnitality = \figIdentity[pluscolor] \qquad\quad \figCommutativity = \figMonoid \\[4pt]
    \figCodeterministicLeft = \figCodeterministicRight \qquad\quad \figCototalLeft = \figZero
  \end{gather*}
  \caption{Axioms of natural commutative monoids; naturality ensures the universal property of coproducts.}%
  \label{fig:ax-natural-commutative-monoid}
\end{figure}
\end{toappendix}

\begin{example}%
  \label{ex:if-else-properties}
  Command composition should distribute on the right of conditional statements.
  \[(\cchoice{f}{b}{g}) \ccomp h \sequiv \cchoice{(f \ccomp h)}{b}{(g \ccomp h)}\]
  Distributivity on the left, instead, requires an additional condition: if the free variables of the condition \(b\) are disjoint from the modified variables of the program \(h\), then the program \(h\) should distribute on the left of the conditional statement.
  \[h \ccomp \cchoice{f}{b}{g} \sequiv \cchoice{(h \ccomp f)}{b}{(h \ccomp g)}\]
\end{example}

The two monoidal structures need to interact appropriately to satisfy the properties in \Cref{ex:if-else-properties}: the data-flow monoidal product, \((\tensor)\), distributes over the control-flow coproduct, \((+)\).

\begin{toappendix}
The correct notion of coproducts for \kl{copy-discard categories} requires that the coproduct injections are \kl{total} and \kl{deterministic}.
This ensures that the copy-discard structure is compatible with coproducts (\Cref{lem:copy-discard-coproducts}).

\begin{definition}
  A \kl{copy-discard category} has \intro{copy-discard coproducts} if it has coproducts and the coproduct injections are \kl{total} and \kl{deterministic}.
  We will denote unbiased finite coproducts with \((\coprod)\), binary coproducts with \((+)\) and the initial object with \(0\).
\end{definition}

\begin{definition}
  A \intro{distributive monoidal category} is a \kl{monoidal category} with finite coproducts such that the canonical morphisms \(\lidistr_{A;B_{1}, \dots, B_{n}} \colon \coprod_{i \leq n} (A \tensor B_{i}) \to A \tensor (\coprod_{i \leq n} B_{i})\) and \(\ridistr_{A_{1}, \dots, A_{n};B} \colon \coprod_{i \leq n} (A_{i} \tensor B) \to (\coprod_{i \leq n} A_{i}) \tensor B\) are isomorphisms.
\end{definition}

With the above definitions, we may restate the definition of \kl{copy-discard category} (\Cref{def:distributive-copy-discard}).

\begin{definition}
  A \kl{distributive copy-discard category} is a \kl{copy-discard category} with \kl{copy-discard coproducts} that is \kl[distributive monoidal category]{distributive monoidal}.
\end{definition}
\end{toappendix}

\begin{definition}%
  \label{def:distributive-copy-discard}
  A \intro{distributive copy-discard category} is a \kl{copy-discard category} with finite coproducts where the \intro[distributors]{}canonical morphisms
  \[\lidistr_{A;B_{1}, \dots, B_{n}} \colon \textstyle{\coprod}_{i \leq n} (A \tensor B_{i}) \to A \tensor (\textstyle{\coprod}_{i \leq n} B_{i}) \quad \text{and} \quad \ridistr_{A_{1}, \dots, A_{n};B} \colon \textstyle{\coprod}_{i \leq n} (A_{i} \tensor B) \to (\textstyle{\coprod}_{i \leq n} A_{i}) \tensor B\]
  are isomorphisms and the coproduct injections are \kl{total} and \kl{deterministic}.
\end{definition}

\begin{toappendix}
\begin{remark}%
  \label{rem:binary-nullary-distributors}
  The structure of a \kl{distributive copy-discard category} determines binary and nullary distributors.
  \begin{align*}
    \ldistr_{A;B_{1}, B_{2}} &\colon A \tensor (B_{1} + B_{1}) \to (A \tensor B_{1}) + (A \tensor B_{2}) & \ldistr_{A;0} &\colon A \tensor 0 \to 0 \\
    \rdistr_{A_{1}, A_{2}; B} &\colon (A_{1} + A_{2}) \tensor B \to (A_{1} \tensor B) + (A_{2} \tensor B) & \rdistr_{0;B} &\colon 0 \tensor B \to 0
  \end{align*}
\end{remark}
\end{toappendix}

\begin{remark}[Notation]
  A category with finite coproducts has natural \intro{injections}, \(\inject_{1} \colon A \to A+B\) and \(\inject_{2} \colon B \to A+B\), natural \intro[initmap]{}unique morphisms from the initial object, \(\initmap_{A} \colon 0 \to A\), natural \intro{copairings}, \(\copairing{f}{g} \colon A+B \to C\), for \(f \colon A \to C\) and \(g \colon B \to C\), natural \intro{join} morphisms, \(\join_{A} \defn \copairing{\id_{A}}{\id_{A}} \colon A + A \to A\), and \intro[cswap]{} symmetries, \(\cswap_{A,B} \colon A+B \to B+A\).
\end{remark}

\begin{example}%
  \label{ex:if-else-semantics}
  The programs in \Cref{ex:if-else-properties} can be given semantics in a \kl{distributive copy-discard category}.
  We fix three morphisms, \(\semn{f}, \semn{g}, \semn{h} \colon X \to X\)—endomorphisms for simplicity, and a morphism \(\semn{b} \colon X \to 1+1\).
  The morphism \(\semn{b}\) determines whether the left or right branch of the conditional statement will be executed; for this we call \emph{guards} the morphisms to \(1+1\).
  Thanks to the distributive and copy-discard structures, a guard \(g \colon X \to 1+1\) determines a morphism of type \intro[branch]{}\(\branch{g} \colon X \to X+X\) that copies the input to the two branches, so that execution can resume after checking the guard condition: \(\branch{g} \defn \cp_X \dcomp (g \tensor \id) \dcomp \rdistr_{1,1;X}\).
  With this, we give semantics to conditional statements, \(\semn{\cchoice{f}{b}{g}} = \branch{\semn{b}} \dcomp \copairing{\semn{f}}{\semn{g}}\), and prove the equalities in \Cref{ex:if-else-properties}.
  The first equality follows by naturality of the copairing maps of coproducts.
  \[\figIfElseLeft = \figIfElseRight\]
  The second equality relies on monoidality and naturality of the distributors.
\end{example}

We have asked that the coproduct injections are \kl{total} and \kl{deterministic}.
This is the right notion of \kl[copy-discard coproduct]{coproduct for copy-discard categories} as it ensures that the copy-discard structure is compatible with coproducts (cf.~\cite{liell2025compositional,bonchiGS23}) and that the structure morphisms of coproducts are \kl{total} and \kl{deterministic} (see \Cref{lem:copy-discard-coproducts} and \Cref{lem:join-zero-deterministic-total}).
When convenient, we will omit the monoidal product from expressions on object and write \(AB + C\) in place of \((A \tensor B) + C\).

\begin{toappendix}
\begin{lemma}%
  \label{lemma:inverse-distributor}%
  The following holds in any \kl{distributive monoidal category}.
  \[\inject_{AA} \dcomp (\inject_{AA} + \inject_{BB}) \dcomp (\lidistr_{A;A,B} + \lidistr_{B;A,B}) \dcomp \ridistr_{A,B;A+B} = \inject_{A} \tensor \inject_{A}\]
\end{lemma}
\begin{proof}
  The distributors are the canonical coproduct maps below.
  \[\leftDistributorDiagram{} \qquad \rightDistributorDiagram\]
  We rewrite the left-hand side using (\ref{eq:distr-proof-one},~\ref{eq:distr-proof-four}) that the distributors are the canonical ones, (\ref{eq:distr-proof-two},~\ref{eq:distr-proof-five}) the properties of coproducts, and~\eqref{eq:distr-proof-three} naturality of injections.
  \begin{align}
    & \inject_{AA} \dcomp (\inject_{AA} + \inject_{BB}) \dcomp (\lidistr_{A;A,B} + \lidistr_{B;A,B}) \dcomp \ridistr_{A,B;A+B} \nonumber\\
    &= \inject_{AA} \dcomp ((\inject_{AA} \dcomp \copairing{\id_{A} \tensor \inject_{A}}{\id_{A} \tensor \inject_{B}}) + (\inject_{BB} \dcomp \copairing{\id_{B} \tensor \inject_{A}}{\id_{B} \tensor \inject_{B}})) \dcomp \ridistr_{A,B;A+B} \label{eq:distr-proof-one}\\
    &= \inject_{AA} \dcomp ((\id_{A} \tensor \inject_{A}) + (\id_{B} \tensor \inject_{B})) \dcomp \ridistr_{A,B;A+B} \label{eq:distr-proof-two}\\
    &= (\id_{A} \tensor \inject_{A}) \dcomp \inject_{A(A+B)} \dcomp \ridistr_{A,B;A+B} \label{eq:distr-proof-three}\\
    &= (\id_{A} \tensor \inject_{A}) \dcomp \inject_{A(A+B)} \dcomp \copairing{\inject_{A} \tensor \id_{A+B}}{\inject_{B} \tensor \id_{A+B}} \label{eq:distr-proof-four}\\
    &= (\id_{A} \tensor \inject_{A}) \dcomp (\inject_{A} \tensor \id_{A+B}) \label{eq:distr-proof-five}\\
    &= \inject_{A} \tensor \inject_{A}. \nonumber
\end{align}
  This concludes the proof.
\end{proof}

\begin{proposition}%
  \label{lem:copy-discard-coproducts}
  A \kl{copy-discard category} that is also \kl[distributive monoidal category]{distributive monoidal} is a \kl{distributive copy-discard category} if and only if the copy and discard morphisms are compatible with coproducts,
  \(\cp_{A+B} = (\cp_{A} + \initmap_{A \tensor B} + \initmap_{B \tensor A} + \cp_{B}) \dcomp (\lidistr_{A;A,B} + \lidistr_{B;A,B}) \dcomp \ridistr_{A,B;A+B}\) and
  \(\discard_{A+B} = (\discard_{A} + \discard_{B}) \dcomp \join_{1}\).
\end{proposition}
\begin{proof}
  Suppose that the copy and discard morphisms are compatible with coproducts.
  We show that \(\inject_{A} \dcomp \discard_{A+B} = \discard_{A}\), i.e.\ that the outer diagram below commutes.
  \[\injectionsTotalDiagramProofFig{}\]
  The diagram (\emph{i}) commutes by naturality of the injection \(\inject_{A}\); the diagram (\emph{ii}) commutes by unitality of the structure morphism of the coproduct \(\join_{I}\); the diagram (\emph{iii}) commutes by hypothesis.
  Similarly, we show that \(\inject_{A} \dcomp \cp_{A+B} = \cp_{A} \dcomp (\inject_{A} \tensor \inject_{A})\), i.e.\ that the outer diagram below commutes.
  We omit the symbol \(\tensor\) for the monoidal product to ease readability.
  \[\injectionsDeterministicDiagramProofFig{}\]
  The diagram (\emph{i}) commutes by naturality of the injection \(\inject_{A}\); the diagram (\emph{ii}) commutes by \Cref{lemma:inverse-distributor}; the diagram (\emph{iii}) commutes by hypothesis.

  Conversely, suppose that the coproduct injections are total and deterministic.
  Then, the two diagrams below commute.
  \[\discardCompatibleCoproductsDiagramProofFig{} \qquad \copyCompatibleCoproductsDiagramProofFig{}\]
  By the universal property of coproducts, we must have \(\discard_{A+B} = \copairing{\discard_{A}}{\discard_{B}} = (\discard_{A} + \discard_{B}) \dcomp \join_{1}\) and equation~\eqref{eq:copy-compatible-proof-eq-one} below.
  Equations (\ref{eq:copy-compatible-proof-eq-two},~\ref{eq:copy-compatible-proof-eq-three}) follow from properties of coproducts, while (\ref{eq:copy-compatible-proof-eq-four},~\ref{eq:copy-compatible-proof-eq-five}) follow from the canonicity of distributors.
  \begin{align}
    &\cp_{A+B} \nonumber\\
    & = \copairing{\cp_{A} \dcomp (\inject_{A} \tensor \inject_{A})}{\cp_{B} \dcomp (\inject_{B} \tensor \inject_{B})} \label{eq:copy-compatible-proof-eq-one}\\
    & = (\cp_{A} + \cp_{B}) \dcomp \copairing{\inject_{A} \tensor \inject_{A}}{\inject_{B} \tensor \inject_{B}} \label{eq:copy-compatible-proof-eq-two}\\
    & = (\cp_{A} + \cp_{B}) \dcomp ((\id_{A} \tensor \inject_{A}) + (\id_{B} \tensor \inject_{B})) \dcomp \copairing{\inject_{A} \tensor \id_{A+B}}{\inject_{B} \tensor \id_{A+B}} \label{eq:copy-compatible-proof-eq-three}\\
    & = (\cp_{A} + \cp_{B}) \dcomp ((\id_{A} \tensor \inject_{A}) + (\id_{B} \tensor \inject_{B})) \dcomp \ridistr_{A,B;A+B} \label{eq:copy-compatible-proof-eq-four}\\
    & = (\cp_{A} + \cp_{B}) \dcomp ((\inject_{AA} \dcomp \lidistr_{A;A,B}) + (\inject_{BB} \dcomp \lidistr_{B;A,B})) \dcomp \ridistr_{A,B;A+B} \label{eq:copy-compatible-proof-eq-five}\\
    & = (\cp_{A} + \cp_{B}) \dcomp (\inject_{AA} + \inject_{BB}) \dcomp (\lidistr_{A;A,B} + \lidistr_{B;A,B}) \dcomp \ridistr_{A,B;A+B} \nonumber\\
    & = (\cp_{A} + \initmap_{AB} + \initmap_{BA} + \cp_{B}) \dcomp (\lidistr_{A;A,B} + \lidistr_{B;A,B}) \dcomp \ridistr_{A,B;A+B} \nonumber
\end{align}
\end{proof}

\begin{lemma}%
  \label{lem:join-zero-deterministic-total}
  In a \kl{distributive copy-discard category}, the structure morphisms of coproducts, \(\join_{A}\) and \(\initmap_{A}\), are \kl{total} and \kl{deterministic}.
\end{lemma}
\begin{proof}
  By initiality of \(0\), we obtain that \(\cp_{0} \dcomp (\initmap_{A} \tensor \initmap_{A}) = \initmap_{A} \dcomp \cp_{A}\) and that \(\discard_{0} = \initmap_{A} \dcomp \discard_{A}\).
  By the hypothesis on the discard maps, \(\discard\), and by naturality of \(\join\), we obtain that the morphisms \(\join_{A}\) are total: \(\discard_{A+A} = (\discard_{A} + \discard_{A}) \dcomp \join_{1} = \join_{A} \dcomp \discard_{A}\).
  Determinism of the morphisms \(\join_{A}\) follows by~\eqref{eq:join-deterministic-proof-one} the hypothesis on the copy maps, \(\cp\); (\ref{eq:join-deterministic-proof-two},~\ref{eq:join-deterministic-proof-three}) the canonicity of the distributors; (\ref{eq:join-deterministic-proof-four},~\ref{eq:join-deterministic-proof-six}) naturality of \(\join\); and~\eqref{eq:join-deterministic-proof-five} the properties of coproducts.
  \begin{align}
    & \cp_{A+A} \dcomp (\join_{A} \tensor \join_{A}) \nonumber\\
    & = (\cp_{A} + \initmap_{AA} + \initmap_{AA} + \cp_{A}) \dcomp (\lidistr_{A;A,A} + \lidistr_{A;A,A}) \dcomp \ridistr_{A,A;A+A} \dcomp (\join_{A} \tensor \join_{A}) \nonumber\\
    & = (\cp_{A} + \cp_{A}) \dcomp ((\inject_{AA} \dcomp \lidistr_{A;A,A}) + (\inject_{AA} \dcomp \lidistr_{A;A,A})) \dcomp \ridistr_{A,A;A+A} \dcomp (\join_{A} \tensor \join_{A}) \label{eq:join-deterministic-proof-one}\\
    & = (\cp_{A} + \cp_{A}) \dcomp ((\id_{A} \tensor \inject_{A}) + (\id_{A} \tensor \inject_{A})) \dcomp \ridistr_{A,A;A+A} \dcomp (\join_{A} \tensor \join_{A}) \label{eq:join-deterministic-proof-two}\\
    & = (\cp_{A} + \cp_{A}) \dcomp ((\id_{A} \tensor \inject_{A}) + (\id_{A} \tensor \inject_{A})) \label{eq:join-deterministic-proof-three}\\
    & \qquad\dcomp ((\inject_{A} \tensor \id_{A+A}) + (\inject_{A} \tensor \id_{A+A})) \dcomp \join_{(A+A)(A+A)} \dcomp (\join_{A} \tensor \join_{A}) \\
    & = (\cp_{A} + \cp_{A}) \dcomp ((\inject_{A} \tensor \inject_{A}) + (\inject_{A} \tensor \inject_{A})) \dcomp ((\join_{A} \tensor \join_{A}) + (\join_{A} \tensor \join_{A})) \dcomp \join_{AA} \label{eq:join-deterministic-proof-four}\\
    & = (\cp_{A} + \cp_{A}) \dcomp \join_{AA} \label{eq:join-deterministic-proof-five}\\
    & = \join_{A} \dcomp \cp_{A} \label{eq:join-deterministic-proof-six}
  \end{align}
\end{proof}
\end{toappendix}

All \kl{monads} are \kl[commutative monad]{commutative} with respect to coproducts.
We say that a \kl{monad} on a \kl{distributive copy-discard category} is \kl{commutative} when it is so with respect to the monoidal product \((\tensor)\).
Their \kl{Kleisli categories} are also \kl{distributive copy-discard categories}.

\begin{propositionrep}%
  \label{prop:kleisli-distributive-cd}%
  The \kl{Kleisli category} of a \kl{commutative monad}, \(T \colon \cat{C} \to \cat{C}\), on a \kl{distributive copy-discard category}, \(\cat{C}\), is also a \kl{distributive copy-discard category}.
\end{propositionrep}
\begin{proof}
  This is a known fact.
\end{proof}

\begin{example}
  As a consequence of \Cref{prop:kleisli-distributive-cd}, the \kl{Kleisli category} of the \kl{countably-supported subdistribution monad} is a \kl{distributive copy-discard category}.
\end{example}

\subsection{Traced coproduct categories}

Traced monoidal structures have been related to iteration since their definition~\cite{joyal1996traced}.
When the monoidal product is a coproduct, trace operators are equivalent to fixpoint operators~\cite{hasegawa97recursion,simpson2000complete}, an earlier axiomatisation of recursion~\cite{bloom1993iteration,stefanescu1986algebraic,stefanescu1987flowchart}.

\begin{example}%
  \label{ex:loop-invariants}
  When a loop body, \(f\), preserves a predicate, \(p\), then the whole loop should also preserve it: if
  \(\assert{p} \ccomp f \sequiv f \ccomp \assert{p}\), then \(\assert{p} \ccomp \while{b}{f} = \while{b}{f} \ccomp \assert{p}\).
\end{example}

The equation in \Cref{ex:loop-invariants} is a particular case of the uniformity axiom for traces.

\begin{toappendix}
\begin{definition}
  A \kl{symmetric monoidal category}, \((\cat{C}, \monoidal, \unit, \mswap)\), is \intro[uniformly-traced monoidal]{uniformly traced} if there is an operator, \(\trace_{S} \colon \cat{C}(S \monoidal A, S \monoidal B) \to \cat{C}(A,B)\) for all objects \(S\), \(A\) and \(B\) of \(\cat{C}\), that satisfies the axioms below (also in \Cref{fig:ax-uniform-trace}).
  \begin{fleqn}[2\parindent]
  \begin{align}
    & \trace_{S}((\id_{S} \monoidal u) \dcomp f \dcomp (\id_{S} \monoidal v)) = u \dcomp \trace_{S}(f) \dcomp v & \tag{Tightening}\label{ax:tightening-mon}\\
    & \trace_{S}(f \monoidal g) = \trace_{S}(f) \monoidal g & \tag{Strength}\label{ax:strength-mon}\\
    & \trace_{T} \trace_{S} f = \trace_{S \monoidal T}f & \tag{Joining}\label{ax:joining-mon}\\
    & \trace_{\unit} f = f & \tag{Vanishing}\label{ax:vanishing-mon}\\
    & \trace_{A} \mswap_{A,A} = \id_{A} & \tag{Yanking}\label{ax:yaking-mon}\\
    & \text{if } f \dcomp (u \monoidal \id_{B}) = (u \monoidal \id_{A}) \dcomp g \text{, then } \trace_{S} f = \trace_{T} g & \tag{Uniformity}\label{ax:uniformity-mon}
  \end{align}
  \end{fleqn}
\end{definition}
\end{toappendix}

\begin{definition}%
  \label{def:uniform-trace}
  A category, \(\cat{C}\), with coproducts is \intro[uniformly-traced coproduct category]{} \intro{uniformly traced} if there is an operator, \(\trace_{S} \colon \cat{C}(S+A, S+B) \to \cat{C}(A,B)\) for all objects \(S\), \(A\) and \(B\) of \(\cat{C}\), that satisfies the axioms below.
  \begin{fleqn}[2\parindent]
  \begin{align}
    & \trace_{S}((\id_{S} + u) \dcomp f \dcomp (\id_{S} + v)) = u \dcomp \trace_{S}(f) \dcomp v & \tag{Tightening}\label{ax:tightening}\\
    & \trace_{S}(f + g) = \trace_{S}(f) + g & \tag{Strength}\label{ax:strength}\\
    & \trace_{T} \trace_{S} f = \trace_{S+T}f & \tag{Joining}\label{ax:joining}\\
    & \trace_{0} f = f & \tag{Vanishing}\label{ax:vanishing}\\
    & \trace_{A} \cswap_{A,A} = \id_{A} & \tag{Yanking}\label{ax:yaking}\\
    & \text{if } f \dcomp (u + \id_{B}) = (u + \id_{A}) \dcomp g \text{, then } \trace_{S} f = \trace_{T} g & \tag{Uniformity}\label{ax:uniformity}
  \end{align}
  \end{fleqn}
  In other words, a \kl{uniformly-traced coproduct category} is a \kl{uniformly-traced monoidal category} where the monoidal product is a coproduct.
\end{definition}
\begin{toappendix}
  \begin{figure}[h!]
    \begin{gather*}
      \figTraceUniformLeftIf = \figTraceUniformRightIf \quad \Rightarrow \quad \figTrace{f} = \figTrace{g}
    \end{gather*}
    \caption{Uniformity axiom for traces.}%
    \label{fig:ax-uniform-trace}
  \end{figure}
\end{toappendix}
The uniformity axiom appears in the literature to capture a notion of ``simulation equivalence''~\cite{stefanescu1986algebraic,stefanescu1987flowchart,hasegawa02uniformity}, it ensures that the free construction of trace preserves coproducts~\cite{2025tracedtapes}, and allows reasoning with loop invariants.

\begin{example}%
  \label{ex:while-invariants}
  Traces give semantics to \emph{while} loops; uniformity allows reasoning about them.
  Given a morphism, \(\semn{f} \colon X \to X\), a guard, \(\semn{b} \colon X \to 1+1\), and a predicate \(\semn{p} \colon X \to 1\), we define the semantics of the loop with the trace.
  \[\semn{\while{b}{f}} = \trace_X(\join_X \dcomp \branch{\semn{b}} \dcomp (\semn{f} + \id_X)) = \figSemanticsWhile{\branch{\semn{b}}}{\semn{f}}\]
  By the copy-discard structure, \emph{assert} statements commute with guards, \(\semn{\assert{p}} \dcomp \branch{\semn{b}} = \branch{\semn{b}} \dcomp (\semn{\assert{p}} + \semn{\assert{p}})\).
  From this equality, the assumption and naturality of the monoids, we obtain that \(\copairing{\semn{\assert{p}}}{\semn{\assert{p}}} \dcomp \branch{\semn{b}} \dcomp (\semn{f} + \id) = \join \dcomp \branch{\semn{b}} \dcomp ((\semn{f} \dcomp \semn{\assert{p}}) + \semn{\assert{p}})\).
  By uniformity, we obtain that \(\semn{\assert{p} \ccomp \while{b}{f}} = \semn{\while{b}{f} \ccomp \assert{p}}\).
\end{example}

\begin{toappendix}
\begin{remark}%
  \label{rem:sliding}
  The better known \intro{sliding axiom} follows from\ \ref{ax:uniformity}: \(((u + \id_{A}) \dcomp h) \dcomp (u + \id_{B}) = (u + \id_{A}) \dcomp (h \dcomp (u + \id_{B})) \) implies \(\trace_{S}((u + \id_{A}) \dcomp h) = \trace_{T}(h \dcomp (u + \id_{B}))\).
\end{remark}
\end{toappendix}

\begin{toappendix}
\begin{remark}[Feedback and trace]
  Feedback~\cite{katis1997bicategories,katis02feedback}, or guarded trace~\cite{milius2017guard,goncharov2021metalanguage}, is a generalisation of trace that drops the\ \ref{ax:yaking} axiom: the feedback loop is not instantaneous but creates a delay, \(\partial_{A} = \trace_{A}(\cswap_{A,A})\).
\end{remark}
\end{toappendix}

Some conditions ensure that the Kleisli category of a \kl{monad} has a \kl{uniform coproduct trace}~\cite{jacobs2010coalgebraic}.

\begin{definition}[{\cite[Definition~4.2]{jacobs2010coalgebraic}}]
  A monad \(T\) on a category \(\cat{C}\) with countable coproducts and products is \intro{partially additive} if its \kl{Kleisli category} is poset-enriched with a zero object and the morphisms \(\beta_{\underline{X}} \colon T(\coprod_{n} X_{n}) \to \prod_{n} T(X_{n})\), defined by pairing the canonical maps \(\coprod_{n} X_{n} \to T(X_{i})\), are monic and form a cartesian natural transformation.
\end{definition}

Recall that a $\ncat{Dcpo}_{\bot}$-enriched category is a category where each homset has countable directed joins and a bottom element that are both preserved by composition.

\begin{theorem}[\protect{\cite[Theorem 5.2]{jacobs2010coalgebraic}}]%
  \label{thm:Jacobs}%
  Consider a category, $\cat{C}$, with countable coproducts and a partially additive \kl{monad}, $T$, on it such that its \kl{Kleisli category}, $\kleisli{T}$, is $\ncat{Dcpo}_{\bot}$-enriched and has monotone cotuplings.
  Then, this \kl{Kleisli category} has a \kl{uniform coproduct-trace},
  $(\kleisli{T}, +, 0, \trace)$. %
\end{theorem}

\begin{example}[\protect{\cite[Example~5.3]{jacobs2010coalgebraic}}]
  The \kl{countably-supported subdistribution monad} satisfies the assumptions of \Cref{thm:Jacobs}.
  As a consequence, its Kleisli category has a \kl{uniform coproduct-trace}.
  For a morphism \(f \colon S+A \to S+B\), its trace is given by \(\trace_{S}(f)(b \given a) = \sum_{n \in \naturals} \sum_{s_{1}, \dots , s_{n} \in S}(f(s_{1} \given a) \cdot f(s_{2} \given s_{1}) \cdots f(s_{n} \given s_{n-1}) \cdot f(b \given s_{n}))\).
\end{example}

In a traced coproduct category, there are \intro[finmap]{}morphisms \(\finmap_{A} \colon A \to 0\) that we may interpret as \(\while{\mathsf{true}}{\noop}\) (\Cref{eq:loop-forever}).
The uniformity axiom makes these morphisms unique, ensuring that the initial object \(0\) is also terminal.
\begin{equation}%
  \label{eq:loop-forever}
  \finmap_{A} \defn \trace(\copairing{\id_{A}}{\id_{A}}) = \figLoopForever
\end{equation}

\begin{lemmarep}
  In a \kl{uniformly traced cocartesian category}, the initial object is also terminal.
\end{lemmarep}
\begin{proof}
  Define morphisms \(\finmap_{A} \colon A \to 0\) as \(\finmap_{A} \defn \trace_{A}(\copairing{\id_{A}}{\id_{A}})\).
  These assemble into a natural transformation: by the universal property of coproducts, we have
  \begin{gather*}
    (f + f) \dcomp \copairing{\id_{B}}{\id_{B}} = \copairing{f}{f} = \copairing{\id_{A}}{\id_{A}} \dcomp f\\[2pt]
    \figUniformityZeroProofIfLeft = \figUniformityZeroProofIfRight ;
  \end{gather*}
  by uniformity, we obtain
  \begin{gather*}
    f \dcomp \finmap_{B} \defn f \dcomp \trace_{B}(\copairing{\id_{B}}{\id_{B}}) = \trace_{A}(\copairing{\id_{A}}{\id_{A}}) = \finmap_{A}\\[2pt]
    \figUniformityZeroProofThenLeft = \figUniformityZeroProofThenRight.
  \end{gather*}
  Therefore, for every object \(A\), there is a unique morphism \(\finmap_{A} \colon A \to 0\); this makes the object \(0\) terminal.
\end{proof}

\begin{remark}[Notation]
  The morphisms to the terminal object, \(\finmap_{A} \colon A \to 0\), determine \intro[coproduct projections]{}natural projections \(\cproj_{1} \colon A + B \to A\) and \(\cproj_{2} \colon A + B \to B\);
  the zero object \(0\) determines natural \intro{zero morphisms} \(\zeromap_{A,B} \defn \finmap_{A} \dcomp \initmap_{B} \colon A \to B\).
\end{remark}

\begin{toappendix}
\begin{lemma}%
  \label{lemma:unrolling}
  For a morphism \(f \colon S \to S + A\) in a traced-coproduct category,
  \[\trace_{S}(\join \dcomp f) = f \dcomp \copairing{\trace_{S}(\join \dcomp f)}{\id_{S}}.\]
\end{lemma}
\begin{proof}
  This follows by naturality of the copairing morphisms, \(\join\), and by the \kl{sliding axiom}, which is a consequence of~\ref{ax:uniformity} (see \Cref{rem:sliding}).
\end{proof}
\end{toappendix}

\subsection{Imperative categories}

\kl{Imperative categories} require all the structure seen so far, so that they can give semantics to the data flow and the control flow of imperative programs.

\begin{definition}
  An \intro{imperative category} is a \kl{distributive copy-discard category} with a \kl{uniform coproduct-trace} that is compatible with the monoidal product:
  \[\trace_{S}(f) \tensor \id_{C} = \trace_{S \tensor C} (\ridistr_{S,A; C} \dcomp (f \tensor \id_{C}) \dcomp \rdistr_{S,B;C}).\]
\end{definition}

\Cref{thm:Jacobs} and \Cref{prop:kleisli-distributive-cd} give us a source of examples of \kl{imperative categories}.

\begin{corollary}%
  \label{cor:kleisli-imperative}%
  The \kl{Kleisli category} of a \kl[commutative monad]{\(\tensor\)-commutative} \kl{partially additive monad} on a \kl{distributive category} satisfying the assumptions of \Cref{thm:Jacobs} is an \kl{imperative category}.
\end{corollary}

\begin{example}%
  \label{ex:predicates-from-guards}
  Given a guard \(b \colon A \to 1+1\) in an \kl{imperative category}, we can obtain two predicates, corresponding to \(b\) and its negation, by composing with the \kl{coproduct projections}: \(\pred{b} = b \dcomp \cproj_1\) and \(\pred{(\gnot b)} = b \dcomp \cproj_2\).
\end{example}

\subsection{Examples}%
\label{sec:examples-imperative}

This section presents more examples of \kl{imperative categories}.
The examples cover deterministic semantics, nondeterministic semantics, and probabilistic semantics, both discrete (\Cref{ex:substoch}) and measurable.

\begin{example}[Deterministic semantics]\AP%
  Consider the category \intro[sets]{\(\Set\)} of sets and functions.
  The \intro{maybe} monad on \(\Set\) acts on objects as \(\maybe(X) = X+1\); its unit is the inclusion \(\eta_{X} \colon X \to X+1\); and the Kleisli extension of a function \(f \colon X \to Y+1\) is \(\klextend{f}(x) = f(x)\) for \(x \in X\), and \(\klextend{f}(\ast) = \ast\), where \(\ast\) denotes the element of \(1\).
  Morphisms in its \kl{Kleisli category}, \intro[category of partial functions]{\(\Par\)}, specify partial functions.
\end{example}

\begin{example}[Relational semantics]\AP%
  Consider the \intro{powerset} monad on \(\Set\).
  Its action on objects is \(\powerset(X) = \{E \subseteq X\}\); its unit \(\eta_{X}(x) = \{x\}\) maps each element \(x \in X\) to the singleton \(\{x\}\); and the Kleisli extension of a function \(f \colon X \to \powerset(Y)\) is \(\klextend{f}(E) = \{f(x) \in Y \mid x \in E\}\).
  Morphisms in its \kl{Kleisli category}, \intro[category of relations]{\(\Rel\)}, are relations.
\end{example}

\begin{example}[Measurable probabilistic semantics]\AP%
  Consider the category \intro[standard Borel spaces]{\(\StdBorel\)} of standard Borel spaces and measurable functions between them.
  A subdistribution on a standard Borel space \((X,\mathcal{A}_{X})\) is a measurable function \(\sigma \colon (X,\mathcal{A}_{X}) \to ([0,1], \mathcal{B})\) whose total probability mass \(\sigma(X)\) is at most \(1\), where \(\mathcal{B}\) is the Borel \(\sigma\)-algebra on the interval \([0,1]\).
  The \intro[measurable subdistribution]{subdistribution} monad on \(\StdBorel\)~\cite{giry82:categorical,panangaden1999} maps a standard Borel space \(X\) to the standard Borel space \(\subgiry(X)\) of subdistributions on it with the \(\sigma\)-algebra generated by the set of evaluation maps \(\mathsf{ev}_{U} \colon \subgiry(X) \to [0,1]\) for all the measurable subsets \(U\) of \(X\).
  Morphisms in its \kl{Kleisli category}, \(\BorelsubStoch\), are subprobability kernels.
\end{example}

\begin{example}[Nonexamples]
  The categories \(\Set\) of sets and functions, \(\Stoch\) of sets and total probabilistic channels, and \(\BorelStoch\) of standard Borel spaces and total probability kernels are \kl{distributive copy-discard categories} but not \kl{imperative categories} as they are known to lack a \kl{coproduct trace}. 
\end{example}

We apply \Cref{thm:Jacobs} and \Cref{prop:kleisli-distributive-cd} to obtain that the \kl{Kleisli categories} of these monads are \kl{imperative categories}.

\begin{proposition}%
  [{\cite[Example~4.4]{jacobs2010coalgebraic} and~\cite[Section 7]{jacobs2016effectuses}}]%
  \label{prop:partially-additive-monads}
  The \kl{maybe} monad, \kl{powerset} monad, and \kl[discrete subdistributions]{subdistributions} monad %
  on the \kl{distributive category} of sets and functions, \(\Set\), are partially additive.
  The \kl[measurable subdistributions]{subdistributions} monad on the \kl{distributive category} \(\StdBorel\) is a partially additive monad.
\end{proposition}

\begin{corollaryrep}%
  \label{cor:examples-kleisli-imperative}
  The \kl{Kleisli categories} of the \kl{maybe} monad, \kl{powerset} monad, and \kl[discrete subdistributions]{subdistributions} monad %
  on the \kl{distributive category} of sets and functions, \(\Set\), and the \kl{Kleisli category} of the \kl[measurable subdistributions]{subdistributions} monad on the \kl{distributive category} of standard Borel spaces and measurable functions, \(\StdBorel\), are \kl{imperative categories}.
\end{corollaryrep}
\begin{proof}
  We need to check the assumptions of \Cref{thm:Jacobs}.
  By \Cref{prop:partially-additive-monads}, all these monads are \kl{partially additive}.
  For the monads on \(\Set\), the remaining assumptions of \Cref{thm:Jacobs} are already checked in~\cite{jacobs2010coalgebraic}.
  We are left to check the last two conditions of \Cref{thm:Jacobs} for the monad \(\subgiry\) on \(\StdBorel\).
  The countable coproduct of standard Borel spaces is again standard Borel, so \(\StdBorel\) has countable coproducts.
  The \kl{Kleisli category} of \(\subgiry\) is poset-enriched with the pointwise order and it has a bottom element, the zero subdistribution.
  Moreover, hom-sets are DCPOs because the supremum of an increasing sequence of measurable functions is defined pointwise and bounded increasing sequences of real numbers have a supremum.
  Finally, cotuplings are monotone because they are so pointwise.

  Finally, all these monads are known to be \(\tensor\)-\kl{commutative}.
  By \Cref{cor:kleisli-imperative}, the \kl{Kleisli categories} of all these monads are \kl{imperative categories}.
\end{proof}

\begin{remark}[SPS-transformation]%
  \label{rem:sps-transformation}
  Beyond these \kl{Kleisli} examples, premonoidal and graded models can be treated by a state-passing style transformation.
  \emph{Pre}monoidal categories~\cite{powerR97:premonoidalnotions} are weaker than monoidal categories, in that the order of side effects matters.
  Mutable stack variables are the archetypal premonoidal effect, but they need no premonoidal structure in this paper: the semantics of \Cref{sec:monoidal-gcl} passes the store around explicitly, as usual in program logics.
  Every premonoidal category can be faithfully embedded into a monoidal category, by adding an extra ``global state'' variable that is modified by all the effectful programs~\cite{jeffrey1997:premonoidal,effectful22}.
  Similarly, every graded monoidal category can be faithfully embedded into a monoidal category by adding an extra variable for every grade~\cite{earnshaw_et_al:LIPIcs.CSL.2026.28}.
  This allows us to reason about interchange of programs even when they produce global effects or depend on names.

  For example, the Kleisli category of the state monad, \(\mathsf{State}_R(A) = (R \times A)^R\), on the category \(\Set\) of sets and functions, \(\kleisli{\mathsf{State}_R}\), can be embedded in a category, \(\Set_R\), where the objects are sets together with an additional ``syntactic'' object, \(R\): a morphism \(A \to B\) in \(\kleisli{\mathsf{State}_R}\) becomes a morphism \(A \times R \to B \times R\).
  The morphisms in \(\Set_R\) whose domain and codomain contains at most one occurrence of \(R\) are valid morphisms in \(\kleisli{\mathsf{State}_R}\).

  Another example comes from the graded \kl{copy-discard category} \(\ncat{Imp}\) of stochastic channels with named nondeterministic choice~\cite{liell2025compositional}.
  This can be embedded in a \kl{copy-discard category} \(\ncat{Imp}^{\tensor}\) that has all the objects of \(\ncat{Imp}\) with an extra object, \(R\), that represent the named nondeterministic choices: the morphism \(\mathsf{knight}_a \colon 1 \to A\) in \(\ncat{Imp}\), representing nondeterministic choice in \(A\), becomes a morphism \(\mathsf{knight} \colon R \to A\) in \(\ncat{Imp}^{\tensor}\).
\end{remark}
\section{A monoidal guarded command language}%
\label{sec:monoidal-gcl}
The guarded command language~\cite{dijkstra1975guarded} is a simple syntax for imperative programs; it has appeared with slight variants as \textsf{While} language and \textsf{Imp} language~\cite{winskel1993formal}; it has been extended to probabilistic programs by adding probabilistic choice~\cite{morgan1996probabilistic,morgan1999pgcl,kozen1979semantics}. %
This section defines the syntax of a monoidal version of the guarded command language and gives it semantics in \kl{imperative categories}.
The focus of this work is on the semantics, so we keep the syntax simple.
We assume that programs have access to a fixed number of variables, \(N \in \naturals\), and that these are all of the same type, interpreted as some object \(X\) of an \kl{imperative category} \(\cat{C}\).
Of course, one could extend the syntax of the monoidal guarded command language with proper types and their derivations.
The store is then an explicit object, \(X^{N}\), that is transformed by commands: an assignment, \(\assign{x_{i}}{e}\), rebinds the \(i\)-th component of the store to a freshly computed value.

\subsection{The monoidal guarded command language and its semantics}
The usual guarded command language has three kinds of terms: expressions, predicates and commands.
In this context, the predicates serve three roles: testing properties of programs, generating values that satisfy the predicate, and branching the control flow of programs.
When restricting to relational semantics, conflating these three roles is appropriate because the morphisms \(A \to 1\), morphisms \(1 \to A\) and total morphisms \(A \to 1+1\) all encode the same information: subsets of \(A\).
However, this is not true in general.
The monoidal guarded command language, then, needs three different kinds of terms to distinguish between the three roles of boolean predicates, so it needs two additional kinds of terms: states for generating values and guards for branching the control flow.

Terms are interpreted as morphisms in an imperative category \(\cat{C}\).
The semantics of an expression is a total and deterministic morphism of type \(X^N \to X\), where \(X^N\) denotes the \(N\)-fold \(\tensor\)-product of \(X\) with itself; the semantics of a guard is a morphism of type \(X^N \to 1 + 1\); the semantics of a predicate is a morphism of type \(X^N \to 1\); the semantics of a state is a morphism of type \(1 \to X^N\); the semantics of a command is a morphism of type \(X^N \to X^N\).

The syntax for each kind of terms is defined inductively starting from a signature.

\begin{definition}
  A \intro{signature} is a pair \(\sig = (\gens, \arity)\) of a set \(\gens\) and a function \(\arity \colon \gens \to \naturals\).
  A \intro{monoidal signature} is a triple \(\sig = (\gens, \arity, \carity)\) of a set \(\gens\) and two functions \(\arity, \carity \colon \gens \to \naturals\).
  The elements of the set \(\gens\) are called \intro{generators}, and the functions \(\arity\) and \(\carity\) assign an \intro{arity} and a \intro{coarity} to them.
\end{definition}

We now define the syntax for each kind of terms and the corresponding semantics.
We fix four \kl{signatures}, \(\esig\), \(\gsig\), \(\psig\) and \(\ssig\), for expressions, guards, predicates and states, respectively, and a \kl{monoidal signature}, \(\csig\), for commands.
For each \(N \in \naturals\), we define expressions, guards, predicates, states and commands on \(N\) variables; we assume that \(N\) is large enough to interpret any generator, that is, \(N \geq \arity(g)\) for each \(g \in \esig + \gsig +\psig +\ssig + \csig\) and \(N \geq \carity(g)\) for each \(g \in \csig\).

The syntax that we propose is mostly standard.
The only subtleties arise when generalising guards, predicates and states appropriately.

\paragraph{Expressions.}
Expression terms are generated by variables and application of a generator to other expression terms.
The semantics of expressions are morphisms of type \(X^{N} \to X\), defined inductively below, starting form an assignment that chooses a deterministic and total morphism \(\gsemn{e_0} \colon X^{\arity(e_0)} \to X\) in \(\cat{C}\) for each generator \(e_0 \in \egens\).

\begin{fleqn}[\parindent]
  \begin{alignat*}{3}
  & \text{Terms} && && \text{Semantics} \\
  e \Coloneqq &\ x_i && i \leq N && \proj_{i}\\
  \mid &\ e_0(e_1, \dots, e_{\arity(e_0)}) &\quad & e_0 \in \egens &\qquad\qquad\qquad & \ncp_{X^N}^{\arity(e_0)} \dcomp (\semn{e_1} \tensor \cdots \tensor \semn{e_{\arity(e_0)}}) \dcomp \gsemn{e_0}
\end{alignat*}
\end{fleqn}

The semantics of a variable is the projection on the appropriate coordinate; the semantics of the application \(e_0(e_1, \dots, e_{\arity(e_0)})\) copies the input variables \(\arity(e_0)\) times, computes the interpretations of the subterms \(e_1, \dots, e_{\arity(e_0)}\), and applies the interpretation of \(e_{0}\) to their result.
By induction, one can see that the semantics of expressions are always total and deterministic morphisms.

\begin{example}%
  \label{ex:signature-expressions}
  We consider the signature for expressions to contain constants for the integers and binary operations for the arithmetic operations.
  \[\esig = \{n \colon 0 \st n \in \integers\} \union \{(+), (-), (\times) \colon 2\}\]
  In each of the semantics examples in \Cref{sec:examples-imperative}, we may choose the integers as base type, \(X = \integers\), and interpret the expression generators as expected.
\end{example}

\paragraph{Guards.}
The role of guards is to split the control flow in two branches according to some condition on the input variables.
Their semantics are morphisms of type \(X^{N} \to 1 + 1\) which, by the construction in \Cref{ex:if-else-semantics}, determine morphisms \(X^N \to X^N + X^N\).

\begin{fleqn}[\parindent]
\begin{alignat*}{3}
  & \text{Terms} && && \text{Semantics} \\
  b \Coloneqq &\ \gleft && && \discard_{X^N} \dcomp \inject_{1} \\
  \mid &\ \gright && && \discard_{X^N} \dcomp \inject_{2} \\
  \mid &\ b_0(e_1, \dots, e_{\arity(b_0)}) &\quad & b_0 \in \ggens &\qquad\qquad\qquad & \ncp_{X^N}^{\arity(b_0)} \dcomp (\semn{e_1} \tensor \cdots \tensor \semn{e_{\arity(b_0)}}) \dcomp \gsemn{b_0} \\
  \mid &\ b_1 \gand b_2 && && \cp_{X^N} \dcomp (\semn{b_1} \tensor \semn{b_2}) \dcomp \andmorph\\
  \mid &\ \gnot b && && \semn{b} \dcomp \cswap_{1,1}\\
  \mid &\ \gchoice{b_{1}}{b}{b_{2}} && && \branch{\semn{b}} \dcomp \copairing{\semn{b_{1}}}{\semn{b_{2}}} \\
  \mid &\ \gmute{x_{i_{1}}, \dots, x_{i_{k}}}{b}{t} && && (\proj_{-i_1, \dots, -i_k} \tensor \semn{t}) \dcomp \swap_{(\dots, i_1, \dots, i_k)} \dcomp \semn{b}
\end{alignat*}
\end{fleqn}

For each generator \(b_0 \in \ggens\), its semantics is a morphism \(\gsemn{b_0} \colon X^{\arity(b_0)} \to 1+1\) in \(\cat{C}\).
The semantics of the constant guards, \(\gleft\) and \(\gright\), are the ``constant left'' and ``constant right'' morphisms that always choose the left and right component of the coproduct \(1+1\), respectively; we may interpret the left component as \emph{true} and the right component as \emph{false}.
The semantics of guard application is analogous to that of expression application.
The semantics of conjunction relies on a morphism, \(\andmorph \colon (1+1) \tensor (1+1) \to 1+1\), representing the boolean \emph{and} gate, \(\andmorph \defn \rdistr_{1,1; 1+1} \dcomp (\id_{1} + \copairing{\copairing{\id_{1}}{\id_{1}}}{\id_{1}})\).
The semantics of negation simply swaps \emph{true} and \emph{false} by composing with the symmetry morphism, \(\cswap_{1,1} \colon 1 + 1 \to 1 + 1\).
Guarded choice of guards, \(\gchoice{b_{1}}{b}{b_{2}}\) chooses with the guard \(b\) which guard to use for branching.
Intuitively, the guard \(\gmute{x_{i_{1}}, \dots, x_{i_{k}}}{b}{t}\) binds the variables \(x_{i_{1}}, \dots, x_{i_{k}}\) in \(b\) according to the state \(t\); \Cref{ex:signature-states-par-rel,ex:signature-states-stoch} will make this intuition precise.
For now, we discuss the interpretation of this guard construct in \(\Rel\).
In the category \(\Rel\), there are morphisms \(\nondetvar \colon 1 \to X\) that are the opposite relation of the \emph{true} predicate, \(\nondetvar(\ast) = X\).
For a guard \(b\), the interpretation of \(\gmute{x_{i}}{b}{\nondetvar}\) is the existential quantification of the variable \(x_{i}\) in the guard \(b\),
\(\semn{\pmute{x_{i}}{p}{\nondetvar}}(x_{1}, \dots, x_{N}) = \{j \in 1+1 \st \exists y \in X\ . \ j \in \semn{b}( (x_{1}, \dots, x_{i-1}, y, x_{i+1}, \dots x_{n}))\}\).

\begin{example}%
  \label{ex:signature-guards-par-rel}
  We define a signature for guards, meant to be interpreted in \(\Par\), containing equality and inequality guards.
  If we extend the semantics to relations, we may add nondeterministic choice to the signature.
  \begin{align*}
    \gsig^{\Par} &\defn \{(=), (\leq) : 2\} & \gsig^{\Rel} & \defn \{(=), (\leq) : 2\} \union \{\nondetchoice \colon 0\}
  \end{align*}
  The interpretations of equality and inequality are as expected.
  The interpretation of \((\nondetchoice)\) gives nondeterministic choice, where \(\ast_{1}\) and \(\ast_{2}\) indicate the two elements of \(1+1\).
  \begin{align*}
    \gsemn{\nondetchoice} &\colon 1 \to 1+1 & \gsemn{\nondetchoice}(\ast) & \defn \{\ast_{1}, \ast_{2}\}
  \end{align*}
\end{example}

\begin{example}%
  \label{ex:signature-guards-stoch}
  For probabilistic semantics, we add biased coinflips to the signature.
  \[\gsig^{\subStoch} \defn \{(=), (\leq) : 2\} \union \{\coin{p} \colon 0 \st p \in (0,1)\}\]
  Their interpretations are Bernoulli distributions with parameter \(p\).
  Note that we exclude \(p=1\) and \(p=0\) because they would give guards that are already present, the constant guards \(\gleft\) and \(\gright\).
  \begin{align*}
    \gsemn{\coin{p}} &\colon 1 \to 1+1 & \gsemn{\coin{p}}(\ast) & \defn p \cdot \ket{\ast_{1}} + (1-p) \cdot \ket{\ast_{2}}
  \end{align*}
\end{example}

\paragraph{Predicates.}
The role of predicates is to test properties of programs.
Their semantics are morphisms of type \(X^N \to 1\).
Their syntax is different from that of guards because different categorical structure is available for combining them; for example, predicates cannot, in general, be negated while guards can.

\begin{fleqn}[\parindent]
\begin{alignat*}{3}
  & \text{Terms} && && \text{Semantics} \\
  p \Coloneqq &\ \ptrue && && \discard_{X^N} \\
  \mid &\ \pfalse && && \zeromap_{X^N,1} \\
  \mid &\ p_0(e_1, \dots, e_{\arity(p_0)}) &\quad & p_0 \in \pgens &\qquad\qquad\qquad & \ncp_{X^N}^{\arity(p_0)} \dcomp (\semn{e_1} \tensor \cdots \tensor \semn{e_{\arity(p_0)}}) \dcomp \gsemn{p_0} \\
  \mid &\ p_1 \pand p_2 && && \cp_{X^N} \dcomp (\semn{p_1} \tensor \semn{p_2}) \\
  \mid &\ \pchoice{p_1}{b}{p_2} && && \branch{\semn{b}} \dcomp \copairing{\semn{p_1}}{\semn{p_2}} \\
  \mid &\ \pred{b} && && \semn{b} \dcomp \cproj_{1} \\
  \mid &\ \psub{p}{x_i}{e} && i \leq N && \cp_{X^N} \dcomp (\semn{e} \tensor \proj_{-i}) \dcomp \swap_{(i, 1, \dots, i-1, i+1, \dots, N)} \dcomp \semn{p} \\
  \mid &\ \pmute{x_{i_{1}}, \dots, x_{i_{k}}}{p}{t} && && (\proj_{-i_1, \dots, -i_k} \tensor \semn{t}) \dcomp \swap_{(\dots, i_1, \dots, i_k)} \dcomp \semn{p}
\end{alignat*}
\end{fleqn}

The semantics of predicates is defined inductively starting from an assignment of a morphism \(\gsemn{p_0} \colon X^{\arity(p_0)} \to 1\) for every generator \(p_0 \in \pgens\).
We discuss the predicate constructs and their semantics.
The two constant predicates, \(\ptrue\) and \(\pfalse\), are interpreted as the discard morphism, \(\semn{\ptrue} = \discard_{X^{N}}\), and the zero morphism, \(\semn{\pfalse} = \zeromap_{X^{N},1}\), respectively.
The application of a generator to expression terms is interpreted analogously to expression and guard application.
The conjunction of predicates tests both predicates on the same input variables.
The guarded disjunction of predicates uses a guard to choose which predicate to test (see \Cref{ex:if-else-semantics} for the definition of \(\branch{\semn{b}}\)).
Substituting a variable with an expression in a predicate amounts to computing the expression and then testing the predicate on the result. 
For a guard, \(b\), we may consider its corresponding predicate, \(\pred{b}\), obtained by forcing the \emph{false} branch to fail (as in \Cref{ex:predicates-from-guards}).
The semantics of the last construct will be better discussed once we introduce states as well.
The permutation \(\swap_{(\dots, i_1, \dots, i_k)} \colon X^N \to X^N\) in \((\cat{C}, \tensor, 1)\) is needed to keep the order of the variables consistent: it inserts the last \(k\) variables between the first \(N-k\) in positions \(i_1, \dots, i_k\).
The interpretation of the predicate \(\pmute{x_{i_{1}}, \dots, x_{i_{k}}}{p}{t}\) is similar to the analogous construct for guards: it binds the variables \(x_{i_{1}}, \dots, x_{i_{k}}\) in \(p\) according to the state \(t\).
We discuss the interpretation of this predicate construct in \(\Rel\).
For a predicate \(p\), the interpretation of \(\pmute{x_{i}}{p}{\nondetvar}\) is the existential quantification of the variable \(x_{i}\) in the predicate \(p\),
\(\semn{\pmute{x_{i}}{p}{\nondetvar}} = \{(x_{1}, \dots, x_{i-1}, y, x_{i+1}, \dots x_{n}) \in X^{N} \st \exists x_{i}\ .\ (x_{1}, \dots, x_{N}) \in \semn{p}\}\).

\begin{example}
  For predicates, we will not consider any additional generator and fix \(\pgens = \emptyset\) for all our examples.
  The interpretation of predicate combinators is worth describing explicitly in common semantics examples.
  In the category of partial functions and in that of relations, morphisms of type \(X^{N} \to 1\) are in bijection with subsets of \(X^{N}\), so predicate terms are interpreted as predicates in the usual sense.
  Every guard, \(b\), determines a predicate, \(\pred{b}\), by diverging in the second branch, \(\semn{\pred{b}} = \{x \in X^{N} \st \semn{b}(x) = \ast_{1}\}\).
  Conjunction of predicates corresponds to intersection of the corresponding subsets, \(\semn{p_{1} \pand p_{2}} = \semn{p_{1}} \intersection \semn{p_{2}}\).
  Guarded choice of predicates corresponds to their `guarded union'.
  \[\semn{\pchoice{p_{1}}{b}{p_{2}}} = \{x \in X^{N} \st x \in \semn{p_{1}} \land x \in \semn{\pred{b}}\} \union \{x \in X^{N} \st x \in \semn{p_{2}} \land x \in \semn{\pred{(\gnot b)}}\}\]
  For the particular case of the guard for nondeterministic choice, guarded union particularises to union, \(\semn{\pchoice{p_{1}}{\nondetchoice}{p_{2}}} = \semn{p_{1}} \union \semn{p_{2}}\), because the predicates associated to nondeterministic choice and its negation are both the `true' predicate, \(\semn{\pred{\nondetchoice}} = \semn{\pred{(\gnot \nondetchoice)}} = \semn{\ptrue} = X\).
\end{example}

\begin{example}
  In the category \(\subStoch\), morphisms of type \(X^{N} \to 1\) assign to each element of \(X^{N}\) a number in the real interval \([0,1]\), which we may interpret as expectations: suppose we are given a joint subdistribution, \(s \colon 1 \to X^{N}\), and an expectation \(p \colon X^{N} \to 1\), then the composition \(s \dcomp p \colon 1 \to 1\) returns a real number in \([0,1]\), \(s \dcomp p = \sum_{x \in X^{N}} s(x) \cdot p(x)\), the expected value of the discrete random variable \(p\) on the discrete probability space \((X^{N},s)\).
  Predicates in \(\subStoch\), then, correspond to expectations.
  The interpretation of predicate conjunction is pointwise multiplication, \(\semn{p_{1} \pand p_{2}}(x) = \semn{p_{1}}(x) \cdot \semn{p_{2}}(x)\).
  The interpretation of choice guarded by coinflips is convex combinations, \(\semn{\pchoice{p_{1}}{\coin{r}}{p_{2}}}(x) = r \cdot \semn{p_{1}}(x) + (1-r) \cdot \semn{p_{2}}(x)\).
\end{example}

\paragraph{States.}
The role of states is to generate, or sample, values according to some conditions; their semantics are morphisms of type \(1 \to X^{N}\).

\begin{fleqn}[\parindent]
\begin{alignat*}{3}
  & \text{Terms} && && \text{Semantics} \\
  s \Coloneqq &\ \sfalse && && \zeromap_{1,X^{N}} \\
  \mid &\ [u_1, \dots, u_{N}]s_0 &\quad & s_0 \in \sgens &\qquad\qquad\qquad & \gsemn{s_0} \dcomp \ncp_{X^{\arity(s_0)}}^{N} \dcomp (\semn{u_1} \tensor \cdots \tensor \semn{u_{N}}) \\
  \mid &\ \smute{x_{i_1}, \dots, x_{i_k}}{s}{t} && && \semn{s} \dcomp (\proj_{-i_1, \dots, -i_k} \tensor \semn{t}) \dcomp \swap_{(\dots, i_1, \dots, i_k)} \\
  \mid &\ \schoice{s_1}{b}{s_2} && && \semn{b} \dcomp \copairing{\semn{s_1}}{\semn{s_2}} \\
  \mid &\ \restrict{s}{p} && && \semn{s} \dcomp \cp_{X^N} \dcomp (\id_{X^N} \tensor \semn{p}) \\
  \mid &\ \ssub{s}{x_i}{e} && i \leq N && \semn{s} \dcomp \cp_{X^N} \dcomp (\semn{e} \tensor \proj_{-i}) \dcomp \swap_{(i, 1, \dots, i-1, i+1, \dots, N)}
\end{alignat*}
\end{fleqn}

The syntax of states is, in some parts, dual to that of predicates.
The asymmetry is due to the asymmetry of the \kl{copy-discard structure} in most of the semantics that we wish to consider.
There is a constant state \(\sfalse\), but not a constant state \(\top\); there is guarded disjunction of states, but no conjunction of states; state application and variable substitution look similar syntactically, but their semantics is dual to the corresponding constructs for predicates.
Finally, when sampling from a state \(s\), we may resample some of its variables with another state \(t\), \(\smute{x_{i_1}, \dots, x_{i_k}}{s}{t}\), or we may restrict to samples that satisfy a predicate \(p\), \(\restrict{s}{p}\).

The \kl{arity} function of states specifies their coarity and the variables appearing in a state term represent output variables.
The semantics of the constant state is the zero morphism, \(\semn{\sfalse} = \zeromap_{1,X^{N}}\).
The semantics of state application samples \(\arity(s_{0})\) values from the state \(s_{0}\) and, then, applies the expressions \(u_{1}, \dots, u_{N}\) to these values to compute the outputs.
The semantics of guarded disjunction of two states uses a guard to choose which state to sample from.
The semantics of variable substitution samples values from a state and then computes the expression with the sampled values.
The semantics of the state \(\smute{x_{i_1}, \dots, x_{i_k}}{s}{t}\) samples from the state \(s\) first, then discards the values \(x_{i_{1}}, \dots, x_{i_k}\) and resamples them from the state \(t\) of arity \(k\).
The semantics of the state \(\restrict{s}{p}\) samples from the state \(s\) and tests these values with \(p\) before returning them.

\begin{example}%
  \label{ex:signature-states-par-rel}
  Morphisms of type \(1 \to X^{N}\) in partial functions are either elements of \(X^{N}\) or the zero morphism; in relations, they are in bijection with subsets of \(X^{N}\) and, as a consequence, with morphisms \(X^{N} \to 1\).
  When considering semantics in partial functions, we consider a generator for each integer, \(\ssig^{\Par} \defn \{n \colon 1 \st n \in \integers\}\).
  When considering semantics in relations, we add a generator for nondeterministic assignment, \(\ssig^{\Rel} \defn \{n \colon 1 \st n \in \integers\} \union \{\nondetvar \colon 1\}\), whose interpretation is the whole set \(X\): \(\gsemn{\nondetvar} \colon 1 \to X\) and \(\gsemn{\nondetvar}(\ast) = X\).

  The interpretation of \(\restrict{s}{p}\) in \(\Rel\) is just the intersection of the corresponding predicates because morphisms of type \(X^{N} \to 1\) are in bijection with those of type \(1 \to X^{N}\).
  For the same reason, the interpretation of \(\smute{x_{i_1}, \dots, x_{i_k}}{s}{t}\) corresponds to existential quantification of \(s\) on the variables \(x_{i_1}, \dots, x_{i_k}\), which are asked to satisfy \(t\) instead; explicitly, in the case of \(k=1\) this is \(\semn{\smute{x_{i}}{s}{t}} = \{(x_{1}, \dots, x_{i-1}, y, x_{i+1}, \dots x_{n}) \in X^{N} \st \exists x_{i} \ (x_{1}, \dots, x_{N}) \in \semn{s} \land y \in \semn{t}\}\).
\end{example}

\begin{example}%
  \label{ex:signature-states-stoch}
  Morphisms of type \(1 \to X^{N}\) in \(\subStoch\) are joint subdistributions of \(N\) variables.
  As generators for states, we take constants for the integers and for uniform distributions, \(\ssig^{\subStoch} \defn \{n \colon 1 \st n \in \integers\} \union \{\unif{n} \colon 1 \st n \in \integers \land n \geq 1\}\); the interpretation of the constants \(n\) is the Dirac's delta on \(n\), \(\gsemn{n} \defn 1 \cdot \ket{n}\); the interpretation of \(\unif{n}\) is the uniform distribution on \(n\) elements.
  \begin{align*}
    \gsemn{\unif{n}} & \colon 1 \to \integers & \gsemn{\unif{n}} & \defn \textstyle{\sum_{i=0}^{n-1}} \frac{1}{n} \cdot \ket{i}
  \end{align*}
  The semantics of state combinators in \(\subStoch\) is more interesting because morphisms of type \(1 \to X^{N}\) and morphisms of type \(X^{N} \to 1\) are not in bijection.
  The interpretation of choice guarded by coinflips is convex combination, \(\semn{\schoice{s_{1}}{\coin{r}}{s_{2}}}(x) = r \cdot \semn{s_{1}}(x) + (1-r) \cdot \semn{s_{2}}(x)\).
  The interpretation of \([u_1, \dots, u_{N}]s_0\) applies the expressions \(\semn{u_{1}}, \dots, \semn{u_{N}}\) to the result of sampling from \(\gsemn{s_{0}}\).
  In particular, the expressions can be just variables, \(u_{i} = x_{j(i)}\); in this case, the values sampled from \(\gsemn{s_{0}}\) are copied, permuted and discarded according to the function \(j\).
  The interpretation of the operation \(\smute{x_{i_{1}}, \dots, x_{i_{k}}}{s}{t}\) marginalises the variables \(x_{i_{1}}, \dots, x_{i_{k}}\) in \(s\) and resamples them from \(t\);
  in the case where \(t\) is the generator for a uniform distribution, we obtain \(\semn{\smute{x_{i}}{s}{\unif{n}}}(x_{1}, \dots, x_{N}) = \sum_{y \in X} s(x_{1}, \dots, x_{i-1},y,x_{i+1}, \dots, x_{N}) \cdot \unif{n}(x_{i})\).
  By applying this construct repeatedly to uniform distributions, we obtain \(N\) independent uniformly distributed samples.
  \[\semn{\smute{x_{N}}{\cdots (\smute{x_{2}}{[x_{1}, \dots, x_{1}] \unif{n_{1}}}{\unif{n_{2}}}) \cdot \cdot }{\unif{n_{N}}}} = \semn{\unif{n_{1}}} \tensor \cdots \tensor \semn{\unif{n_{N}}}\]
  The interpretation of testing a state with a predicate reweights the subdistribution, \(\semn{\restrict{s}{p}}(x) = \semn{s}(x) \cdot \semn{p}(x)\); suppose for simplicity that \(N=1\) and take \(s\) to be a uniform distribution, then \(\semn{\restrict{\unif{n}}{p}} = \sum_{i=0}^{n-1} \frac{\semn{p}(i)}{n} \cdot \ket{i}\) is not a uniform distribution anymore.
  Finally, the interpretation of predicate operation \(\pmute{x_{i_{1}}, \dots, x_{i_{k}}}{t}{p}\) takes the expected value of the variables \(x_{i_{1}}, \dots, x_{i_{k}}\) only; explicitly, for \(t\) the uniform distribution on \(n\),
  \[\semn{\pmute{x_{i}}{\unif{n}}{p}}(x_{1}, \dots, x_{n}) = \sum_{y \in X} \semn{p}(x_{1}, \dots, x_{i-1},y,x_{i+1}, \dots, x_{N}) \cdot \semn{\unif{n}}(y).\]
\end{example}

\paragraph{Commands.}
Finally, we define the syntax and semantics of commands.
The semantics of commands are endomorphisms of type \(X^{N} \to X^{N}\); for each command generator, \(c_{0} \in \cgens\), its semantics is chosen to be a morphism \(\gsemn{c_{0}} \colon X^{\arity(c_{0})} \to X^{\carity(c_{0})}\).

\begin{figure}[h!]
\begin{fleqn}[\parindent]
\begin{alignat*}{3}
  & \text{Terms} && && \text{Semantics} \\
  c \Coloneqq &\ \noop && && \id_{X^N} \\
  \mid &\ \abort && && \zeromap_{X^N, X^{N}} \\
  \mid &\ [u_{1}, \cdots, u_{N}] c_{0}(e_{1}, \cdots, e_{\arity(c_{0})}) && c_{0} \in \cgens && \cp_{X^{N}} \dcomp (\id_{X^{N}} \tensor (\ncp_{X^{N}}^{\arity(c_{0})} \dcomp (\semn{e_{1}} \tensor \cdots \tensor \semn{e_{\arity(c_{0})}})  \\
              & && && \dcomp \gsemn{c_{0}})) \dcomp \ncp_{X^{N+ \carity(c_{0})}}^{N} \dcomp (\semn{u_{1}} \tensor \cdots \tensor \semn{u_{N}})\\
  \mid &\ c_1 \ccomp c_2 && && \semn{c_1} \dcomp \semn{c_2} \\
  \mid &\ \cchoice{c_1}{b}{c_2} && && \branch{\semn{b}} \dcomp \copairing{\semn{c_1}}{\semn{c_2}} \\
  \mid &\ \while{b}{c} && && \trace_{X^N}(\copairing{\id_{X^N}}{\id_{X^N}} \dcomp \branch{\semn{b}} \dcomp (\semn{c} + \id_{X^N})) \\
  \mid &\ \assert{p} && && \cp_{X^N} \dcomp (\id_{X^N} \tensor \semn{p}) \\
  \mid &\ \assign{x_i}{e} && i \leq N && \cp_{X^N} \dcomp (\semn{e} \tensor \proj_{-i}) \dcomp \swap_{(i, 1, \dots, i-1, i+1, \dots, N)} \\
  \mid &\ \samp{(x_{i_1}, \dots, x_{i_k})}{s} &\quad & i_j \leq N &\quad & (\proj_{-i_1, \dots, -i_k} \tensor \semn{s}) \dcomp \swap_{(\dots, i_1, \dots, i_k)}
\end{alignat*}
\end{fleqn}
\caption{Syntax of commands and its semantics.}%
\label{def:commands}
\end{figure}

Let us briefly discuss the syntax and semantics of commands.
The two constants, \(\noop\) and \(\abort\), are interpreted as the identity, \(\semn{\noop} = \id_{x^{N}}\), and the zero morphism, \(\semn{\abort} = \zeromap_{X^{N}, X^{N}}\), respectively.
Expressions can be applied to both inputs and outputs of commands; the semantics of \([u_{1}, \cdots, u_{N}] c_{0}(e_{1}, \cdots, e_{\arity(c_{0})})\) computes the expressions \(e_i\) first, then uses these values as inputs to \(c_0\) and finally returns the values of the expressions \(u_j\) applied to both the outputs of \(c_0\) and the inputs of the expressions \(e_i\).
A command composition executes the two commands sequentially.
An if-else statements uses a guard to decide which command to execute.
The semantics of while loops relies on the trace operator.
Predicate assertions test predicates on the input variables before returning them.
The semantics of variable assignment returns the values \(x_{1}, \dots, x_{i-1}, e(x_{1}, \dots, x_{N}), x_{i+1}, \dots, x_{N}\).
The semantics of sampling discard the variables \(x_{i_{1}}, \dots, x_{i_{k}}\) to be sampled and resamples their values from the state \(s\).

\begin{example}%
  \label{ex:local-command}
  With the signature \(\ssig^{\Rel}\) defined in \Cref{ex:signature-states-par-rel}, we can derive an operation on commands, \(\mathsf{local}\ x. c \defn (\assign{y}{x}) \ccomp (\samp{x}{\nondetvar}) \ccomp c \ccomp (\assign{x}{y})\), that makes the variable \(x\) local in the command \(c\) (with the auxiliary variable \(y\) that is not free in \(c\)).
\end{example}

\begin{example}
  With the signature \(\gsig^{\subStoch}\) defined in \Cref{ex:signature-guards-stoch}, we can express conditional statements where the guard can be both a boolean expression or a probabilistic choice.
\end{example}

\subsection{The algebra of guarded commands}

This section studies the algebras of guards, predicates, states and commands.
The properties that we prove are essential for the results in \Cref{sec:program-logics}.
We relate the algebra of commands with guarded Kleene algebras with tests (GKAT)~\cite{smolka2019guarded}: commands satisfy all the axioms of GKAT that do not involve iteration.
The next \Cref{sec:poset-enrichment} introduces inequalities and derives the remaining GKAT axioms for iteration.

\paragraph{Guards}
We study the algebra of guards in an \kl{imperative category} and single out which axioms hold for total and for deterministic guards.
In particular, \kl{total} and \kl{deterministic} guards form a boolean algebra.
In general, we do not restrict guards to be \kl{deterministic} or \kl{total}, so their algebra is more general: we choose as fundamental operations a guarded disjunction and negation; conjunction is derived.

\begin{definition}
  A \intro{guard algebra} is a quadruple, \((B, (\gchoice{-}{(-)}{-}), \gleft, \gnot)\), of a set, \(B\), a ternary operation, \((\gchoice{-}{(-)}{-}) \colon B \times B \times B \to B\), an element, \(\gleft \in B\), and an involution, \(\gnot(-) \colon B \to B\), satisfying the equations below, for all \(a,b,c,d,e,f \in B\).
  \begin{align*}
    a & = \gchoice{a}{\gleft}{b} & \gnot (\gchoice{a}{b}{c}) & = \gchoice{\gnot a}{b}{\gnot c} \\
    a & = \gchoice{\gleft}{a}{\gnot \gleft} & \gchoice{(\gchoice{a}{b}{c})}{d}{(\gchoice{a}{e}{c})} & = \gchoice{a}{(\gchoice{b}{d}{e})}{c} \\
    \gchoice{c}{b}{a} & = \gchoice{a}{\gnot b}{c} & \gchoice{(\gchoice{a}{b}{c})}{d}{(\gchoice{e}{b}{f})} & = \gchoice{(\gchoice{a}{d}{e})}{b}{(\gchoice{c}{d}{f})}
  \end{align*}
  A \intro{total guard algebra} is a \kl{guard algebra} that additionally satisfies \(\gchoice{\gleft}{b}{\gleft} = \gleft\) for all \(b \in B\).
  A \intro{deterministic guard algebra} is a \kl{guard algebra} that additionally satisfies \(\gchoice{a}{b}{(\gchoice{c}{b}{d})} = \gchoice{a}{b}{d}\).
\end{definition}

Conjunction of guards is derived from guarded disjunction, but it satisfies all the boolean algebra axioms only for deterministic and total guards.

\begin{definition}%
  \label{def:conjunction-disjunction-guards}
  In a \kl{guard algebra}, there are derived conjunction and disjunction operations, \(a \gand b \defn \gchoice{b}{a}{\gnot \gleft}\) and \(b_{1} \gor b_{2} \defn \gnot (\gnot b_{1} \gand \gnot b_{2})\), and constants \emph{true} and \emph{false}, \(\gleft\) and \(\gright \defn \gnot \gleft\).
\end{definition}

\begin{propositionrep}
  Conjunction in \kl{guard algebra} (\Cref{def:conjunction-disjunction-guards}) always forms a monoid, \((B, \gand, \gleft)\).
  \begin{align*}
    (b_{1} \gand b_{2}) \gand b_{3} & \gequiv b_{1} \gand (b_{2} \gand b_{3}) & b \gand \gleft &\gequiv b & \gleft \gand b &\gequiv b
  \end{align*}
  If the \kl{guard algebra} is \kl[total guard algebra]{total}, then the monoid is also commutative and has an absorbing element.
  \begin{align*}
    b_{1} \gand b_{2} & \gequiv b_{2} \gand b_{1} & b \gand \gnot \gleft & \gequiv \gnot \gleft %
  \end{align*}
  The disjunction in a \kl{guard algebra} is also a monoid, \((B, \gor, \gright)\).
  If the \kl{guard algebra} is \kl[deterministic guard algebra]{deterministic}, then the two monoids satisfy distributivity and idempotency.
  \begin{align*}
    b_{1} \gand (b_{2} \gor b_{3}) &\gequiv (b_{1} \gand b_{2}) \gor (b_{1} \gand b_{3}) & b \gand b & \gequiv b \\
    b_{1} \gor (b_{2} \gand b_{3}) &\gequiv (b_{1} \gor b_{2}) \gand (b_{1} \gor b_{3}) & b \gor b & \gequiv b
  \end{align*}
  If the \kl{guard algebra} is \kl[deterministic guard algebra]{deterministic} and \kl[total guard algebra]{total}, then the involution satisfies complements.
  \begin{align*}
    b \gand \gnot b & \gequiv \gright & b \gor \gnot b & \gequiv \gleft
  \end{align*}
  In particular, a \kl[total guard algebra]{total} and \kl{deterministic guard algebra} determines a boolean algebra.
\end{propositionrep}
\begin{proof}
  Associativity and unitality.
  \begin{align*}
    & (a \gand b) \gand c && b \gand \gleft && \gleft \gand b \\
    & = \gchoice{c}{(\gchoice{b}{a}{\gnot \gleft})}{\gnot \gleft} && = \gchoice{\gleft}{b}{\gnot \gleft} && = \gchoice{b}{\gleft}{\gnot \gleft} \\
    & = \gchoice{(\gchoice{c}{b}{\gnot \gleft})}{a}{(\gchoice{c}{\gnot \gleft}{\gnot \gleft})} && = b && = b\\
    & = \gchoice{(\gchoice{c}{b}{\gnot \gleft})}{a}{\gnot \gleft} && && \\
    & = a \gand (b \gand c) && &&
  \end{align*}
  Commutativity (using totality).
  \begin{align*}
    & a \gand b \\
    & = \gchoice{b}{a}{\gnot \gleft} \\
    & = \gchoice{(\gchoice{\gleft}{b}{\gnot \gleft})}{a}{(\gchoice{\gnot \gleft}{b}{\gnot \gleft})} \\
    & = \gchoice{(\gchoice{\gleft}{a}{\gnot \gleft})}{b}{(\gchoice{\gnot \gleft}{a}{\gnot \gleft})} \\
    & = \gchoice{a}{b}{\gnot \gleft} \\
    & = b \gand a
  \end{align*}
  Absorption (using totality).
  \[b \gand \gnot \gleft = \gchoice{\gnot \gleft}{b}{\gnot \gleft} = \gnot \gleft\]
  Distributivity (using determinism).
  \begin{align*}
    & (a \gand b) \gor (a \gand c) \\
    & = \gchoice{\gleft}{\gchoice{b}{a}{\gnot \gleft}}{(\gchoice{c}{a}{\gnot \gleft})} \\
    & = \gchoice{(\gchoice{\gleft}{b}{(\gchoice{c}{a}{\gnot \gleft})})}{a}{(\gchoice{\gleft}{\gnot \gleft}{(\gchoice{c}{a}{\gnot \gleft})})} \\
    & = \gchoice{(\gchoice{\gleft}{b}{(\gchoice{c}{a}{\gnot \gleft})})}{a}{(\gchoice{c}{a}{\gnot \gleft})} \\
    & = \gchoice{(\gchoice{\gleft}{b}{c})}{a}{\gnot \gleft} \\
    & = a \gand (b \gor c)
  \end{align*}
  Idempotency (using determinism).
  \begin{align*}
    & b \gand b \\
    & = \gchoice{b}{b}{\gnot \gleft} \\
    & = \gchoice{(\gchoice{\gleft}{b}{\gnot \gleft})}{b}{\gnot \gleft} \\
    & = \gchoice{\gleft}{b}{\gnot \gleft} \\
    & = b
  \end{align*}
  Complements (using determinism and totality).
  \begin{align*}
    & b \gand \gnot b \\
    & = \gchoice{\gnot b}{b}{\gnot \gleft} \\
    & = \gnot (\gchoice{b}{b}{\gleft}) \\
    & = \gnot (\gchoice{(\gchoice{\gleft}{b}{\gnot \gleft})}{b}{\gleft}) \\
    & = \gnot (\gchoice{\gleft}{b}{\gleft}) \\
    & = \gnot \gleft
  \end{align*}
\end{proof}

Guard algebras describe the behaviour of guards, i.e.\ morphisms of type \(A \to 1+1\), in an \kl{imperative category}.

\begin{propositionrep}%
  \label{lemma:logic-guards}
  The guards in an \kl{imperative category}, i.e.\ the morphisms of type \(A \to 1+1\) for some object \(A\), form a \kl{guard algebra}; the \kl{total} guards form a \kl{total guard algebra}; the \kl{deterministic} guards form a \kl{deterministic guard algebra}.
  In particular, the \kl{total} and \kl{deterministic} guards form a \kl{boolean algebra}.
  \Cref{fig:guards-guard-algebra} shows how the operations in a \kl{guard algebra} derive from the structure of \kl{imperative category}.
\end{propositionrep}
\begin{proof}
  This proof is easy to see with string diagrams.
  The rules follow from naturality of the zero morphisms, the symmetries and the copairing morphisms.
\end{proof}

\begin{figure}[h!]
  \begin{align*}
    \gchoice{a}{b}{c} &\defn \figGuardChoice & a \gand b &\defn \figGuardAnd & a \gor b &\defn \figGuardOr
  \end{align*}
  \caption{Definition of guarded disjunction, conjunction and disjunction of guards in an \kl{imperative category}.}%
  \label{fig:guards-guard-algebra}
\end{figure}

In the following, we will see that guards act on predicates, states and commands, determining the structure of a \emph{guarded monoid}.

\begin{definition}%
  \label{def:guarded-monoid}
  A \intro{guarded monoid} is a monoid with an absorbing element, \((C, \cdot, 1, 0)\), with a guard algebra, \((B, (\gchoice{-}{(-)}{-}), \gleft, \gnot)\), and a \(B\)-indexed binary operation on \(C\), \((\mchoice{-}{b}{-}) \colon C \times C \to C\) satisfying the axioms below, for all \(a,b,c \in B\) and all \(f,g,h,l \in C\).
  \begin{align*}
    \mchoice{f}{\gleft}{g} & = f & (\mchoice{f}{b}{g}) \cdot h & = \mchoice{(f \cdot h)}{b}{(g \cdot h)} \\
    \mchoice{f}{\gnot b}{g} & = \mchoice{g}{b}{f} & \mchoice{(\mchoice{f}{b}{g})}{c}{(\mchoice{h}{b}{l})} & = \mchoice{(\mchoice{f}{c}{h})}{b}{(\mchoice{g}{c}{l})} \\
    \mchoice{f}{\gchoice{a}{b}{c}}{g} & = \mchoice{(\mchoice{f}{a}{g})}{b}{(\mchoice{f}{c}{g})} & &
  \end{align*}
  A \intro{totally guarded monoid} is a \kl{guarded monoid} whose \kl{guard algebra} is \kl[total guard algebra]{total} and that additionally satisfies \(\mchoice{1}{b}{1} = 1\).
  A \intro{deterministically guarded monoid} is a \kl{guarded monoid} whose \kl{guard algebra} is \kl[deterministic guard algebra]{deterministic} and that additionally satisfies \(\mchoice{f}{b}{(\mchoice{g}{b}{h})} = \mchoice{f}{b}{h}\).
\end{definition}

In \kl{guarded monoids}, every guard determines an element of the monoid, via the guarded disjunction: \(b^{\circ} \defn \mchoice{1}{b}{0}\).
This operation allows us to compare \kl{guarded monoids} with guarded Kleene algebras~\cite{smolka2019guarded}.

\begin{propositionrep}%
  \label{prop:morph-guard-to-monoid}
  In a \kl{guarded monoid}, the operation \((-)^{\circ} \colon (B, \gand, \gleft) \to (C, \cdot, 1)\), defined as \(b^{\circ} \defn \mchoice{1}{b}{0}\), is a monoid morphism such that \((\gnot \gleft)^{\circ} = 0\).
  If the monoid is \kl[totally guarded monoid]{totally} and \kl[deterministically guarded monoid]{deterministically guarded}, then \(\mchoice{(b^{\circ} \cdot f)}{b}{g} = \mchoice{f}{b}{g}\) and the guarded choice satisfies a form of associativity, \(\mchoice{(\mchoice{f}{b}{g})}{c}{h} = \mchoice{f}{b \gand c}{(\mchoice{g}{c}{h})}\).
\end{propositionrep}
\begin{proof}
  Multiplicativity.
  \begin{align*}
    & (b \gand c)^{\circ} \\
    & = \mchoice{1}{b \gand c}{0} \\
    & = \mchoice{1}{\gchoice{c}{b}{\gnot \gleft}}{0} \\
    & = \mchoice{(\mchoice{1}{c}{0}) }{b}{(\mchoice{1}{\gnot \gleft}{0})}\\
    & = \mchoice{c^{\circ}}{b}{0} \\
    & = \mchoice{1 \cdot c^{\circ}}{b}{0 \cdot c^{\circ}} \\
    & = (\mchoice{1}{b}{0}) \cdot c^{\circ} \\
    & = b^{\circ} \cdot c^{\circ}
  \end{align*}
  Unit and zero.
  \begin{align*}
    & \gleft^{\circ} && (\gnot \gleft)^{\circ} \\
    & = \mchoice{1}{\gleft}{0} && = \mchoice{1}{\gnot \gleft}{0} \\
    & = 1 && = 0
  \end{align*}
  If the guards are deterministic,
  \begin{align*}
    & \mchoice{(b^{\circ} \cdot f)}{b}{g} \\
    & = \mchoice{((\mchoice{1}{b}{0}) \cdot f)}{b}{g} \\
    & = \mchoice{(\mchoice{f}{b}{0})}{b}{g} \\
    & = \mchoice{f}{b}{g}.
  \end{align*}
  If the guards are total and deterministic,
  \begin{align*}
    & \mchoice{f}{b \gand c}{(\mchoice{g}{c}{h})} \\
    & = \mchoice{(\mchoice{f}{c}{(\mchoice{g}{c}{h})} )}{b}{(\mchoice{g}{c}{h})} \\
    & = \mchoice{(\mchoice{f}{c}{h} )}{b}{(\mchoice{g}{c}{h})} \\
    & = \mchoice{(\mchoice{f}{b}{g} )}{c}{(\mchoice{h}{b}{h})} \\
    & = \mchoice{(\mchoice{f}{b}{g} )}{c}{((\mchoice{1}{b}{1}) \cdot h)} \\
    & = \mchoice{(\mchoice{f}{b}{g} )}{c}{h}
  \end{align*}
\end{proof}

\begin{corollary}%
  \label{cor:morph-guard-to-monoid}
  In a guarded monoid, \(b^{\circ} \cdot f = \mchoice{f}{b}{0}\) and \((\gnot b)^{\circ} \cdot f = \mchoice{0}{b}{f}\).
\end{corollary}

\begin{corollaryrep}
  A \kl{guarded monoid} satisfies the sequence axioms (S1-S5), the skew commutativity axiom (U2) and the right distributivity axiom (U5) of a guarded Kleene algebra with tests~\cite{smolka2019guarded}; a \kl{totally guarded monoid} additionally satisfies the idempotence axiom (U1); a \kl{deterministically guarded monoid} additionally satisfies the guardedness axiom (U4); a \kl[totally guarded monoid]{totally} and \kl{deterministically guarded monoid} satisfies skew associativity axiom (U3).
\end{corollaryrep}
\begin{proof}
  Axioms (S1-S5, U2, U5) follow from the \Cref{def:guarded-monoid}.
  Totality imposes (U1).
  \Cref{prop:morph-guard-to-monoid} gives (U3, U4).
\end{proof}

\paragraph{Predicates}
Morphisms of type \(X \to 1\) in a monoidal category, and in an \kl{imperative category} in particular, are its \emph{predicates}.
Their algebra is more restricted than the algebra of guards because, in general, predicates cannot be negated.
However, they do form a \kl{guarded commutative monoid}.

\begin{propositionrep}%
  \label{prop:predicates-guarded-monoid}
  The predicates in an \kl{imperative category} form a \intro{guarded commutative monoid}, i.e.\ a \kl{guarded monoid} whose underlying monoid is commutative (\Cref{fig:predicate-algebra}); the multiplication of the underlying monoid is predicate conjunction, \((\pand)\), the unit is the \emph{true} predicate, \(\ptrue\), and the absorbing element is the \emph{false} predicate, \(\pfalse\).
  For deterministic predicates, conjunction is idempotent, \(p \pand p = p\).
  \begin{align*}
    p \pand q &\defn \figPredicateAnd & \ptrue &\defn \figDiscard & \pfalse &\defn \figPredicateFalse 
  \end{align*}
\end{propositionrep}
\begin{proof}
  This proof is easily seen in string diagrams.
\end{proof}

\begin{toappendix}

\begin{figure}[h!]
  \begin{align*}
    (p \pand q) \pand r &\sequiv p \pand (q \pand r)
    & p \pand q &\sequiv q \pand p
    & p \pand \ptrue &\sequiv p \\
    (\pchoice{p}{b}{q}) \pand r & \sequiv \pchoice{(p \pand r)}{b}{(q \pand r)}
    & \pchoice{p}{\gnot b}{q} & \sequiv \pchoice{q}{b}{p}
    & p \pand \pfalse &\sequiv \pfalse \\
    \pchoice{(\pchoice{p}{a}{q})}{b}{(\pchoice{p}{c}{q})} & \sequiv \pchoice{p}{\gchoice{a}{b}{c}}{q}
    & \pchoice{p}{\gleft}{q} & \sequiv p
    & \pred{\gleft} &\sequiv \ptrue \\
    \pchoice{(\pchoice{p}{c}{r})}{b}{(\pchoice{q}{c}{s})} & \sequiv \pchoice{(\pchoice{p}{b}{q})}{c}{(\pchoice{r}{b}{s})}
    & \pred{(b_{1} \gand b_{2})} &\sequiv \pred{b_{1}} \pand \pred{b_{2}}
    & \pred{\gright} &\sequiv \pfalse
  \end{align*}
  \caption{Predicates form a \kl{guarded commutative monoid} (\Cref{prop:predicates-guarded-monoid}). The operation \(\pred{(-)}\) is a morphism of monoids with annihilators (\Cref{cor:morph-guard-to-predicates}).}%
  \label{fig:predicate-algebra}
\end{figure}
\end{toappendix}
From \Cref{prop:morph-guard-to-monoid} and \Cref{cor:morph-guard-to-monoid}, we obtain an operation from guards to predicates.

\begin{corollary}%
  \label{cor:morph-guard-to-predicates}
  There is a monoid morphism, \(\pred{(-)}\), from guards to predicates.
  This is defined as \(\pred{b} \defn \pchoice{\ptrue}{b}{\pfalse}\).
  It satisfies the axioms in \Cref{fig:predicate-algebra} and, additionally, \(\pred{b} \pand p = \pchoice{p}{b}{\pfalse}\).
\end{corollary}

\paragraph{Commands}
We study the algebra of commands, i.e.\ endomorphisms, in an \kl{imperative category}.
All the properties that follow are necessary for the results in \Cref{sec:program-logics}.
Commands in an \kl{imperative category} form a \kl{guarded monoid}; thanks to this, we can compare the algebra of commands with guarded Kleene algebras.

\begin{propositionrep}%
  \label{prop:commands-guarded-monoid}
  The commands in an \kl{imperative category} form a \kl{guarded monoid} (\Cref{fig:commands-guarded-monoid}); the multiplication of the underlying monoid is composition, \((\ccomp)\), the unit is the identity, \(\noop\), the absorbing element is the zero morphism, \(\abort\), and the guarded disjunction is defined in \Cref{ex:if-else-semantics}. %
  Commands with \kl{total} guards form a \kl{totally guarded monoid} (\Cref{fig:commands-tot-guarded-monoid}); commands with \kl{deterministic} guards form a \kl{deterministically guarded monoid} (\Cref{fig:commands-det-guarded-monoid}).
\end{propositionrep}
\begin{proof}
  This proof is easily seen in string diagrams.
\end{proof}

\begin{toappendix}
\begin{figure}[h!]
  \begin{align*}
    f \ccomp (g \ccomp h) &\sequiv (f \ccomp g) \ccomp h & f \ccomp \noop & \sequiv f \sequiv \noop \ccomp f & f \ccomp \abort &\sequiv \abort \sequiv \abort \ccomp f
  \end{align*}
  \begin{align*}
    \cchoice{f}{\gleft}{g} & \sequiv f
    & \cchoice{f}{\gnot b}{g} & \sequiv \cchoice{g}{b}{f}
  \end{align*}
  \begin{align*}
    (\cchoice{f}{b}{g}) \ccomp h & \sequiv \cchoice{(f \ccomp h)}{b}{(g \ccomp h)} \\
    \cchoice{f}{(\gchoice{a}{b}{c})}{g} & \sequiv \cchoice{(\cchoice{f}{a}{g})}{b}{(\cchoice{f}{c}{g})} \\
    \cchoice{(\cchoice{f}{b}{g})}{c}{(\cchoice{h}{b}{l})} & \sequiv \cchoice{(\cchoice{f}{c}{h})}{b}{(\cchoice{g}{c}{l})}
  \end{align*}
  \caption{Commands form a \kl{guarded monoid} (\Cref{prop:commands-guarded-monoid}).}%
  \label{fig:commands-guarded-monoid}
\end{figure}

\begin{figure}[h!]
  \begin{align*}
    \cchoice{(\cchoice{f}{b}{g})}{c}{h} &\sequiv \cchoice{f}{(b \gand c)}{(\cchoice{g}{c}{h})}
  \end{align*}
  \caption{Commands with deterministic guards form a \kl{deterministically guarded monoid} (\Cref{prop:commands-guarded-monoid}).}%
  \label{fig:commands-det-guarded-monoid}
\end{figure}

\begin{figure}[h!]
  \begin{align*}
    \cchoice{\noop}{b}{\noop} &\sequiv \noop
  \end{align*}
  \caption{Commands with total guards form a \kl{totally guarded monoid} (\Cref{prop:commands-guarded-monoid}).}%
  \label{fig:commands-tot-guarded-monoid}
\end{figure}
\end{toappendix}

\begin{lemmarep}%
  \label{cor:while-unrolling}%
  \label{lemma:deterministic-while}
  Commands in an \kl{imperative category} satisfy the unrolling axiom (W1) of guarded Kleene algebras with tests~\cite{smolka2019guarded}: \(\while{b}{f} = \cchoice{(f \ccomp \while{b}{f})}{b}{\noop}\).
  If the guard is deterministic, we obtain \(\while{b}{c} \sequiv (\while{b}{(\assert{\pred{b}} \ccomp c)}) \sequiv  (\while{b}{c}) \ccomp \assert{\pred{(\gnot b)}}\).
\end{lemmarep}
\begin{proof}
  This follows from the properties of trace (\Cref{lemma:unrolling}) and the definition of the \(\mathsf{while}\) construct.
\end{proof}

\begin{corollary}
  Commands in an \kl{imperative category} with \kl{deterministic} and \kl{total} guards satisfy the sequence axioms (S1-S5), the guarded union axioms (U1-U5), and the unrolling axiom (W1) of guarded Kleene algebras with tests~\cite{smolka2019guarded}.
\end{corollary}

From \Cref{prop:morph-guard-to-monoid} applied to the guarded monoid of commands (\Cref{prop:commands-guarded-monoid}), we obtain an operation from guards to commands.

\begin{corollaryrep}%
  \label{cor:assert-guards}%
  \label{cor:duplicate-deterministic-guards}
  There is a monoid morphism, \(\assert{\pred{(-)}}\), from guards to commands defined as \(\assert{\pred{b}} \defn \cchoice{\noop}{b}{\abort}\).
  In particular, it satisfies the equations below.
  \begin{align*}
    \cchoice{f}{b}{\abort} &= \assert{\pred{b}} \ccomp f & \cchoice{\abort}{b}{f} &= \assert{\pred{(\gnot b)}} \ccomp f
  \end{align*}
  Moreover, deterministic guards can be duplicated,
  \(\cchoice{f}{b}{g} = \cchoice{(\assert{\pred{b}} \ccomp f)}{b}{g}\).
\end{corollaryrep}
\begin{proof}
  This follows from the distributive structure.
\end{proof}

The next result relates the guarded monoid of predicates (\Cref{prop:predicates-guarded-monoid}) with the guarded monoid of commands (\Cref{prop:commands-guarded-monoid}).

\begin{lemmarep}%
  \label{lem:assert-monoid-morph}
  In an \kl{imperative category}, predicate assertion is a guarded monoid morphism.
  \begin{align*}
    \assert{(p \pand q)} & = \assert{p} \ccomp \assert{q} & \assert{\ptrue} & = \noop \\
    \assert{(\pchoice{p}{b}{q})} &= \cchoice{(\assert{p})}{b}{(\assert{q})} & \assert{\pfalse} &= \abort
  \end{align*}
\end{lemmarep}
\begin{proof}
  This follows from string diagrammatic reasoning.
\end{proof}

\begin{toappendix}
\begin{proposition}%
  \label{cor:guards-predicate-interchange}
  Consider a guard, \(b\), a predicate, \(p\), and two commands, \(c_{1}\) and \(c_{2}\), and their semantics in an \kl{imperative category}.
  Then,
  \[\assert{p} \ccomp (\cchoice{c_{1}}{b}{c_{2}}) \sequiv \cchoice{(\assert{p} \ccomp c_{1})}{b}{(\assert{p} \ccomp c_{2})}.\]
\end{proposition}
\begin{proof}
  This follows from \(\assert{p} \dcomp \cp \dcomp (b \tensor \id) = \cp \dcomp (b \tensor \assert{p})\) and from \(\branch{b} \defn \cp \dcomp (b \tensor \id) \dcomp \rdistr_{1,1;X}\).
\end{proof}

\begin{proposition}%
  \label{prop:ifelse-deterministic}
  Consider a deterministic guard, \(b\), and two commands, \(c_{1}\) and \(c_{2}\), and their semantics in an \kl{imperative category}.
  Then, \(\branch{b} \sequiv \branch{b} \ccomp (\assert{\pred{b}} + \assert{\pred{(\gnot b)}})\), and, in particular, \(\cchoice{c_{1}}{b}{c_{2}} \sequiv \cchoice{(\assert{\pred{b}} \ccomp c_{1})}{b}{(\assert{\pred{(\gnot b)}} \ccomp c_{2})}\).
\end{proposition}
\begin{proof}
  This follows from the distributive structure.
\end{proof}
\end{toappendix}

\begin{toappendix}
\begin{lemma}%
  \label{lem:deterministic-assignment}
  The semantics of variable assignment in an \kl{imperative category}, \(\semn{\assign{x_i}{e}}\), is deterministic.
\end{lemma}
\begin{proof}
  Easy to see by induction on the grammar for expressions.
\end{proof}

\begin{proposition}%
  \label{prop:predicate-state-actions}%
  \label{lem:states-restrict-assert}%
  \label{cor:assert-subst}
  The monoid of commands acts on states and predicates.
  The action also preserves the absorbing element.
  Also, \((\schoice{s_1}{b}{s_2}) \ccomp c \sequiv \schoice{(s_1 \ccomp c)}{b}{(s_2 \ccomp c)}\) and \(\pchoice{(c_1 \ccomp q)}{b}{(c_2 \ccomp q)} \sequiv (\cchoice{c_1}{b}{c_2}) \ccomp q\).
  The actions, moreover, satisfy the equations below.
  \begin{align*}
    \psub{p}{x_i}{e} &\sequiv (\assign{x_i}{e}) \ccomp p
    & \pmute{x_{i_1}, \dots, x_{i_k}}{p}{t} &\sequiv (\samp{x_{i_1}, \dots, x_{i_k}}{t}) \ccomp p \\
    \ssub{s}{x_i}{e} &\sequiv s \ccomp (\assign{x_i}{e})
    & \smute{x_{i_1}, \dots, x_{i_k}}{s}{t} &\sequiv s \ccomp (\samp{x_{i_1}, \dots, x_{i_k}}{t})\\
    \assert{p} \ccomp q &\sequiv p \pand q
    & \restrict{s}{p} &\sequiv s \ccomp \assert{p}
  \end{align*}
  In particular, we obtain that \(\assert{(\psub{p}{x_i}{e})} \ccomp (\assign{x_i}{e}) \sequiv (\assign{x_i}{e}) \ccomp \assert{p}\).
\end{proposition}
\begin{proof}
  The equalities follow by definition of the predicate and state constructs.
\end{proof}
\end{toappendix}

\subsection{Free and modified variables}

Thanks to the monoidal structure and to the interpretation of variable assignment via SPS-transformation, the semantics of the monoidal guarded command language can reason about free and modified variables; these are essential for deriving rules about variables and mutation in \Cref{sec:program-logics,sec:relational-program-logics}.
We define free and modified variables by induction.
Variables are modified by commands only.

\begin{definition}
  The set \(\modv{c}\) of \intro[modified variables]{variables modified by a command} \(c\) is defined by induction on the grammar of the monoidal guarded command language.
\begin{fleqn}[\parindent]
\begin{alignat*}{2}
  & \text{Command } c && \text{Modified variables } \modv{c} \\
  & \noop && \emptyset \\
  & \abort && \{x_{1}, \dots, x_{N}\} \\
  & [u_{1}, \cdots, u_{N}] c_{0}(e_{1}, \cdots, e_{\arity(c_{0})}) &\qquad\qquad& \{x_{i} \st u_{i} \neq x_{i}\} \\
  & c_{1} \ccomp c_{2} && \modv{c_{1}} \union \modv{c_{2}} \\
  & \cchoice{c_1}{b}{c_2} && \modv{c_{1}} \union \modv{c_{2}} \\
  & \while{b}{c} && \modv{c} \\
  & \assert{p} && \emptyset \\
  & \assign{x_i}{e} && \{x_{i}\} \\
  & \samp{(x_{i_1}, \dots, x_{i_k})}{s} && \{x_{i_1}, \dots, x_{i_k}\}
\end{alignat*}
\end{fleqn}
\end{definition}

\begin{toappendix}
We present the definition of free variables in terms of input variables.

\begin{definition}%
  \label{def:free-variables}
  The set of free variables of a command is the union of the sets of modified variables and input variables (\Cref{def:input-variables}), \(\freev{c} = \inv{c} \union \modv{c}\); the set of free variables of guards and predicates coincide with that of input variables, \(\freev{b} = \inv{b}\) and \(\freev{p} = \inv{p}\).
\end{definition}

We now define input variables for expressions, guards, predicates, states and commands.

\begin{definition}[Input variables]%
  \label{def:input-variables}
  We define the set of input variables for expressions.
\begin{fleqn}[\parindent]
  \begin{alignat*}{2}
  & \text{Expression } e && \text{Input variables } \inv{e} \\
  & x_i && \{x_{i}\}\\
  & e_0(e_1, \dots, e_{\arity(e_0)}) &\qquad\qquad & \textstyle{\Union_{i=1}^{\arity(e_0)}} \inv{e_{i}}
\end{alignat*}
\end{fleqn}
We define the set of input variables for guards.
\begin{fleqn}[\parindent]
\begin{alignat*}{2}
  & \text{Guard } b && \text{Input variables } \inv{b} \\
  & \gleft && \emptyset \\
  & \gright && \emptyset \\
  & b_0(e_1, \dots, e_{\arity(b_0)}) &\qquad\qquad & \textstyle{\Union_{i=1}^{\arity(b_0)}} \inv{e_{i}} \\
  & b_1 \gand b_2 && \inv{b_{1}} \union \inv{b_{2}}\\
  & \gnot b && \inv{b}
\end{alignat*}
\end{fleqn}
We define the set of input variables for predicates.
\begin{fleqn}[\parindent]
\begin{alignat*}{2}
  & \text{Predicate } p && \text{Input variables } \inv{p} \\
  & \ptrue && \emptyset \\
  & \pfalse && \emptyset \\
  & p_0(e_1, \dots, e_{\arity(p_0)}) & \qquad\qquad & \textstyle{\Union_{i=1}^{\arity(p_0)}} \inv{e_{i}} \\
  & p_1 \pand p_2 && \inv{p_{1}} \union \inv{p_{2}} \\
  & \pchoice{p_1}{b}{p_2} && \inv{p_{1}} \union \inv{p_{2}} \union \inv{b} \\
  & \pred{b} && \inv{b} \\
  & \psub{p}{x_i}{e} && (\inv{p} \setminus \{x_{i}\}) \union \inv{e} \\
  & \pmute{x_{i_{1}}, \dots, x_{i_{k}}}{p}{t} && \inv{p} \setminus \{x_{i_{1}}, \dots, x_{i_{k}}\}
\end{alignat*}
\end{fleqn}
For states, \(\inv{s} = \emptyset\) always.
We define the set of input variables for commands.
\begin{fleqn}[\parindent]
\begin{alignat*}{2}
  & \text{Command } c && \text{Input variables } \inv{c} \\
  & \noop && \emptyset \\
  & \abort && \emptyset \\
  & [u_{1}, \cdots, u_{N}] c_{0}(e_{1}, \cdots, e_{\arity(c_{0})}) &\qquad\qquad& \textstyle{\Union_{i=1}^{\arity(c_{0})}} \inv{e_{i}} \union (\textstyle{\Union_{j=1}^{N}} \inv{u_{j}} \setminus \{x_{N+1}, \dots, x_{N+\carity(c_{0})}\}) \\
  & c_{1} \ccomp c_{2} && \inv{c_{1}} \union (\inv{c_{2}} \setminus \modv{c_{1}}) \\
  & \cchoice{c_1}{b}{c_2} && \inv{c_{1}} \union \inv{c_{2}} \union \inv{b} \\
  & \while{b}{c} && \inv{c} \union \inv{b} \\
  & \assert{p} && \inv{p} \\
  & \assign{x_i}{e} && \inv{e} \\
  & \samp{(x_{i_1}, \dots, x_{i_k})}{s} && \emptyset
\end{alignat*}
\end{fleqn}
\end{definition}
\end{toappendix}

Similarly, we can define free variables by induction as usual (\Cref{def:free-variables}).
With the sets of free and modified variables, we can prove valid interchange axioms.

\begin{propositionrep}%
  \label{prop:commands-interchange}
  The semantics of commands in an \kl{imperative category} satisfy the interchange axioms below.
\begin{gather*}
  \inferrule[]{ \freev{c_{1}} \intersection \modv{c_{2}} = \emptyset \\ \freev{c_{2}} \intersection \modv{c_{1}} = \emptyset }{ c_{1} \ccomp c_{2} = c_{2} \ccomp c_{1} }
  \\
  \inferrule[]
      { \freev{b} \intersection \modv{d} = \emptyset }
      { d \ccomp (\cchoice{c_{1}}{b}{c_{2}}) = \cchoice{(d \ccomp c_{1})}{b}{(d \ccomp c_{2})} }
  \\
  \inferrule[]
  { \freev{c_{1}} \intersection \modv{c_{2}} = \emptyset \\ \freev{c_{2}} \intersection \modv{c_{1}} = \emptyset \\ \freev{b} \intersection \modv{c_{2}} = \emptyset }{ (\while{b}{c_{1}}) \ccomp c_{2} = c_{2} \ccomp \while{b}{c_{1}} }
\end{gather*}
\end{propositionrep}
\begin{proof}
  This is a consequence of the distributive structure.
\end{proof}

As a consequence, we obtain interchanges between assertions and commands, and between sampling and commands.
\begin{corollary}%
  \label{cor:assert-sample-interchange}
  The semantics of commands in an \kl{imperative category} satisfies the interchange axioms below.
  \begin{gather*}
    \inferrule[assert-interchange]
    { \freev{p} \intersection \modv{c} = \emptyset }{ \assert{p} \ccomp c = c \ccomp \assert{p} }
    \qquad\qquad
    \inferrule[sample-interchange]
    { x_{i_{1}}, \dots, x_{i_{k}} \notin \freev{c} }{ (\samp{(x_{i_1}, \dots, x_{i_k})}{s}) \ccomp c = c \ccomp (\samp{(x_{i_1}, \dots, x_{i_k})}{s}) }
  \end{gather*}
\end{corollary}

\begin{toappendix}
\begin{definition}
  A guard, \(b\), in the monoidal guarded command language is \intro[constant guard]{constant} if it has no free variables, \(\freev{b} = \emptyset\).
\end{definition}

\begin{corollary}%
  \label{cor:constant-guards-interchange}
  Consider a constant guard, \(b\), and three commands, \(c\), \(c_{1}\) and \(c_{2}\), and their semantics in an \kl{imperative category}.
  Then, \(c \ccomp (\cchoice{c_{1}}{b}{c_{2}}) \sequiv \cchoice{(c \ccomp c_{1})}{b}{(c \ccomp c_{2})}\) and
  \(c \ccomp \branch{b} \sequiv \branch{b} \ccomp (c+c)\).
\end{corollary}
\begin{proof}
  The proof uses that \(c \dcomp \cp \dcomp (b \tensor \id) = \cp \dcomp (b \tensor c)\) because \(\freev{b} = \emptyset\), and that \(\branch{b} \defn \cp \dcomp (b \tensor \id) \dcomp \rdistr_{1,1;X}\).
\end{proof}
\end{toappendix}
\section{Program refinement via poset enrichment}%
\label{sec:poset-enrichment}
This section introduces poset-enriched imperative categories to reason about program refinement and express statements of program logics.

\begin{example}
  If commands are interpreted in the \kl{imperative category} \(\Rel\) of sets and relations, the validity of a triple in Floyd-Hoare logic~\cite{hoare1969axiomatic} can be expressed in terms of relation containment: a triple \(\triple{p}{c}{q}\) is valid whenever \(\semn{\assert{p} \ccomp c} \subseteq \semn{c \ccomp \assert{q}}\).

  If commands are interpreted in the \kl{imperative category} \(\subStoch\) of sets and probabilistic channels, the validity of a correctness triple~\cite{kozen1985probabilistic} can be expressed in terms of pointwise inequality of channels: a triple \(\triple{p}{c}{q}\) is valid whenever \(\semn{p}(x) \leq \semn{c \ccomp q}(x)\) for all program states \(x\).
\end{example}

We define the categorical structure needed to express validity of program triples.
The poset enrichment needs to be compatible with both monoidal structures of an \kl{imperative category}; additionally, we require the posetal version of the uniformity axiom (\Cref{fig:ax-posetal-uniformity}).

\begin{toappendix}
We rewrite the definition of \kl{imperative posetal bicategory} in terms of poset enrichment of its two monoidal structures.
We first define poset enrichment for a generic monoidal structure and for a generic uniformly-traced monoidal structure.

\begin{definition}
  A \intro{poset-enriched monoidal category}, \((\cat{C}, \monoidal, \unit, \leq)\), is a \kl{monoidal category}, \((\cat{C}, \monoidal, \unit)\), where the hom-sets are posets, \((\cat{C}(A,B), \leq)\) is a poset for all objects \(A\) and \(B\) of \(\cat{C}\), and compositions and monoidal products preserve the order, if \(f \leq f'\) and \(g \leq g'\) then \(f \dcomp g \leq f' \dcomp g'\) and \(f \monoidal g \leq f' \monoidal g'\).
\end{definition}

\begin{definition}%
  \label{def:poset-enriched-uniformly-traced-monoidal}
  A \intro{poset-enriched uniformly-traced monoidal category}, \((\cat{C}, \monoidal, \unit, \trace, \leq)\), is a \kl{uniformly-traced monoidal category}, \((\cat{C}, \monoidal, \unit, \trace)\), whose underlying monoidal category is a \kl{poset-enriched monoidal category}, \((\cat{C}, \monoidal, \unit, \leq)\), that additionally satisfies \kl{posetal uniformity}: if \((u \monoidal \id) \dcomp f \leq g \dcomp (u \monoidal \id)\), then \(\trace_{S}(f) \leq \trace_{T}(g)\); and if \((u \monoidal \id) \dcomp f \geq g \dcomp (u \monoidal \id)\), then \(\trace_{S}(f) \geq \trace_{T}(g)\) (also in \Cref{fig:ax-posetal-uniformity}).
\end{definition}

We instantiate the two definitions above for the appropriate monoidal structures.

\begin{definition}
  A \intro{poset-enriched copy-discard category} is a \kl{copy-discard category} whose underlying \kl{monoidal category} is a \kl{poset-enriched monoidal category}.
\end{definition}

\begin{definition}
  A \intro{poset-enriched uniformly-traced coproduct category}, \((\cat{C}, +, 0, \trace, \leq)\), is a \kl{uniformly-traced coproduct category}, \((\cat{C}, +, 0, \trace)\), whose underlying cocartesian monoidal category is a \kl{poset-enriched monoidal category}, \((\cat{C}, +, 0, \leq)\), that additionally satisfies \kl{posetal uniformity}: if \((u + \id) \dcomp f \leq g \dcomp (u + \id)\), then \(\trace_{S}(f) \leq \trace_{T}(g)\); and if \((u + \id) \dcomp f \geq g \dcomp (u + \id)\), then \(\trace_{S}(f) \geq \trace_{T}(g)\) (also in \Cref{fig:ax-posetal-uniformity}).
\end{definition}

Finally, an \kl{imperative posetal bicategory} requires both a \kl[poset-enriched copy-discard category]{poset-enriched copy-discard structure} and a \kl[poset-enriched uniformly-traced coproduct category]{poset-enriched uniformly-traced coproduct structure}.

\begin{definition}
  An \kl{imperative posetal bicategory}, \((\cat{C}, \tensor, 1, +, 0, \trace, \pleq)\), is an imperative category, \((\cat{C}, \tensor, 1, +, 0, \trace)\), with a poset enrichment, \((\pleq)\), that makes the underlying copy-discard category a \kl{poset-enriched copy-discard category}, \((\cat{C}, \tensor, 1, \pleq)\), and also makes the underlying traced coproduct category a \kl{poset-enriched uniformly-traced coproduct category}, \((\cat{C}, +, 0, \trace, \pleq)\).
\end{definition}

\begin{figure}[h!]
  \begin{gather*}
    \figTraceUniformLeftIf \pleq \figTraceUniformRightIf \quad \Rightarrow \quad \figTrace{f} \pleq \figTrace{g}\\
    \figTraceUniformLeftIf \pgeq \figTraceUniformRightIf \quad \Rightarrow \quad \figTrace{f} \pgeq \figTrace{g}
  \end{gather*}
  \caption{\intro{Posetal uniformity} axioms.}%
  \label{fig:ax-posetal-uniformity}
\end{figure}
\end{toappendix}

\begin{definition}
  An \intro{imperative posetal bicategory}, \((\cat{C}, \tensor, 1, +, 0, \trace, \pleq)\), is an imperative category, \((\cat{C}, \tensor, 1, +, 0, \trace)\), with a poset enrichment, \((\cat{C}, \pleq)\), that makes both the underlying monoidal categories, \((\cat{C}, \tensor, 1)\) and \((\cat{C}, +, 0)\), \kl{poset-enriched monoidal categories}, and that additionally satisfies \kl{posetal uniformity}: if \(f \dcomp (u + \id) \leq (u + \id) \dcomp g\), then \(\trace_{S}(f) \leq \trace_{T}(g)\); and if \(f \dcomp (u + \id) \geq (u + \id) \dcomp g\), then \(\trace_{S}(f) \geq \trace_{T}(g)\) (also in \Cref{fig:ax-posetal-uniformity}).
\end{definition}

We define semantic inequality of guarded commands with respect to the axioms of \kl{imperative posetal bicategories}.

\begin{definition}
  Two terms, \(t\) and \(u\), of the monoidal guarded command language are \intro{semantically comparable}, \(t \sleq u\), if their semantics can be compared, \(\semn{t} \pleq \semn{u}\), using the axioms of \kl{imperative posetal bicategories}.
  Similarly, the terms \(t\) and \(u\) are \intro{semantically equivalent}, \(t \sequiv u\), if their semantics can be proven equal, \(\semn{t} = \semn{u}\), using the axioms of \kl{imperative categories}.
\end{definition}

\subsection{Examples}%
\label{sec:examples-poset-imperative}

All the examples of \kl{imperative category} presented in \Cref{sec:examples-imperative} are also poset-enriched.
The trace in these examples always gives least fixpoints.

\begin{example}
  The category of sets and partial functions, \(\Par\), is poset-enriched with the extension partial order: for \(f,g \colon A \to B\), \(f \pleq g\) if and only if they coincide on the domain of \(f\), i.e., for all \(a\) where \(f\) is defined, \(f(a) = g(a)\).

  The category of sets and relations, \(\Rel\), is poset-enriched with relation inclusion: for \(f,g \colon A \to B\), \(f \pleq g\) if and only if \(f \subseteq g\) as subsets of \(A \times B\).

  The category of sets and countably-supported subprobability kernels, \(\subStoch\), is poset-enriched with the pointwise order: for \(f,g \colon A \to B\), \(f \pleq g\) if and only if, for all \(a \in A\) and \(b \in B\), \(f(b \given a) \leq g(b \given a)\) as real numbers in the interval \([0,1]\).

  The category of standard Borel spaces and measurable subprobability kernels, \(\BorelsubStoch\), is similarly poset-enriched with the pointwise order: for \(f,g \colon (A, \Sigma_{A}) \to (B, \Sigma_{B})\), \(f \pleq g\) if and only if, for all \(a \in A\) and \(E \in \Sigma_{B}\), \(f(E \given a) \leq g(E \given a)\) as real numbers in the interval \([0,1]\).
\end{example}

These poset enrichments are all \(\tensor\)-monoidal~\cite{2025order} and are easily seen to be \(+\)-monoidal as well.
We are only left to prove posetal uniformity.
Starting from \Cref{thm:Jacobs}, and exploiting a result that generalises forward and backward simulations as lax and oplax coalgebra morphisms~\cite{hasuo2006generic}, we prove that the monoidal trace of the theorem above is not just a \kl{uniform trace} but also a \kl[posetal uniformity]{posetally uniform trace}.

\begin{propositionrep}%
  \label{Jacobsposetaluniform}%
  Under the conditions of \Cref{thm:Jacobs}, the Kleisli category of a monad, $\kleisli{T}$, has a \kl{posetally uniform trace}.
\end{propositionrep}
\begin{proof}
  We first recall the construction of the monoidal trace in \Cref{thm:Jacobs}.
  Hereafter, identities (e.g. $\id_B \colon B \to B$), injections ($\inject_U \colon U \to  U+A$) and coproducts ($+$) are all in $\kleisli{T}$.
  Moreover, we write $\coprod_{n\in \naturals} B$ for the countable coproduct of an object $B$ with itself and $\nabla \colon \coprod_{n\in \naturals} B \to B$ for the countable copairing of $id_B$.

  For each $f\colon U+A \to U+B$ in $\kleisli{T}$, one defines $\hat{f} \colon (U+A) \to (U+A)+B$ as $f \dcomp ( \inject_U + \id_B)$.
  This is a coalgebra for the functor $\id + B \colon \kleisli{T} \to \kleisli{T}$.
  One can show that $\coprod_{n\in \naturals} B$ carries a final coalgebra for such functor and thus one has a unique coalgebra morphism $\omega_{\hat{f}} \colon (U+A) \to \coprod_{n\in \naturals} B$.
  The operation \(\trace \colon \kleisli{T}(U+A, U+B) \to \kleisli{T}(A,B)\) defined as
  \begin{equation}\label{JacobTrace}
    \trace(f) = \inject_A \dcomp \omega_{\hat{f}} \dcomp \nabla
  \end{equation}
  provides a \kl{uniform coproduct trace}~\cite[Theorem~5.2]{jacobs2010coalgebraic}.

  In order to prove \kl{posetal uniformity} we rely on a previous result~\cite[Proposition 5.6]{hasuo2006generic}, stated under the same conditions of \Cref{thm:Jacobs} but restricted to the case $\cat{C}=\cat{Set}$;
  one can carefully check that its proof also works for arbitrary categories $\cat{C}$ with countable coproducts.

  Take $f\colon U+A \to U+B$, $g\colon V+A \to V+B$ and $u \colon U \to V$ in $\kleisli{T}$  and assume that
  \begin{equation}\label{lax}
    f \dcomp (u + \id_{B}) \geq (u + \id_{A}) \dcomp g\text{.}
  \end{equation}
  As for $\hat{f}\colon U \to U+B$, we define the coalgebra $\hat{g}\colon V \to V+B$ and consider the unique coalgebra morphim $\omega_{\hat{g}} \colon V \to  \coprod_{n\in \naturals} B$.
  From~\eqref{lax}, one easily derive that
  \[\hat{f}\dcomp ((u +\id_{A}) \oplus \id_{B}) \geq  (u \oplus \id_{A}) \dcomp \hat{g}\text{,}\]
  namely, $(u +\id_{A})$ is a \emph{lax-coalgebra morphism}~\cite{hasuo2006generic} from $\hat{f}$ to $\hat{g}$.
  Now $(u +\id_{A}) \dcomp \omega_{\hat{g}} \colon U+A \to \coprod_{n\in \naturals} B$ is also a lax-coalgebra morphism.
  By~\cite[Proposition~5.6]{hasuo2006generic}, the unique coalgebra morphism $\omega_{\hat{f}} \colon U+A \to \coprod_{n\in \naturals} B$ is the \emph{greatest} lax coalgebra morphism and thus
  \begin{equation}\label{keyineq}
    \omega_{\hat{f}} \geq (u +\id_{A}) \dcomp \omega_{\hat{g}} \text{.}
  \end{equation}
  We can then conclude with the following derivation.
  \begin{align}
    \trace(f) =
    & \inject_A \dcomp \omega_{\hat{f}} \dcomp \nabla \tag{\ref{JacobTrace}}\\
    & \geq  \inject_A \dcomp (u +\id_{A}) \dcomp \omega_{\hat{g}} \dcomp \nabla \tag{\ref{keyineq}}\\
    & = \inject_A \dcomp \omega_{\hat{g}} \dcomp \nabla \tag{coproduct}\\
    & = \trace(g)\tag{\ref{JacobTrace}}
  \end{align}
  For proving the other implication, one proceeds by reversing the inequalities and use the fact that, by~\cite[Proposition~5.6]{hasuo2006generic}, $\omega_{\hat{f}}$ is the smallest \emph{oplax} coalgebra morphism.
\end{proof}

\begin{corollaryrep}
  The \kl{Kleisli categories} of the \kl{maybe} monad, \kl{powerset} monad, and \kl[discrete subdistributions]{subdistributions} monad %
  on the \kl{distributive category} \(\Set\), and of the \kl[measurable subdistributions]{subdistributions} monad on the \kl{distributive category} \(\StdBorel\) are \kl{posetal imperative categories}.
\end{corollaryrep}
\begin{proof}
  For the monads on \(\Set\), the assumptions of \Cref{thm:Jacobs} are already checked in~\cite{jacobs2010coalgebraic}.
  We now check the conditions for the monad \(\subgiry\) on \(\StdBorel\).
  The countable coproduct of standard Borel spaces is again standard Borel, so \(\StdBorel\) has countable coproducts.
  The \kl{Kleisli category} of \(\subgiry\) is poset-enriched with the pointwise order and it has a bottom element, the zero subdistribution.
  Moreover, hom-sets are DCPOs because the supremum of an increasing sequence of measurable functions is defined pointwise and bounded increasing sequences of real numbers have a supremum.
  Finally, cotuplings are monotone because they are so pointwise.
\end{proof}

\begin{remark}%
  \label{rem:conditionals}
  The poset enrichment in the examples that we consider is defined by \intro{conditionals}, a property of some \kl{copy-discard categories} that asks for a factorization of morphisms of type \(X \to A \tensor B\): for every morphism \(f \colon X \to A \tensor B\) there exist a morphism \(c_{f} \colon A \tensor X \to B\) such that \(f = \cp_{X} \dcomp ((f \dcomp \proj_{A} \dcomp \cp_{A}) \tensor \id_{X}) \dcomp (\id_{A} \tensor c_{f})\).
  In the category \(\subStoch\), conditionals are given by conditional distributions: for a joint distribution \(f \colon X \to A \tensor B\), a conditional is defined by \(c_{f}(b \given a,x) \defn \frac{f(a,b \given x)}{\sum_{b' \in B} f(a,b' \given x)}\) whenever the denominator is nonzero, and \(c_{f}(b \given a,x) \defn 0\) otherwise.
  Conditionals interact with the distributive structure and allow us to interchange guards with commands (\Cref{lem:conditionals-if-else}): for every morphism \(f \colon A \to B\) and every guard \(b \colon B \to 1+1\) there are \(f_{1}, f_{2} \colon A \to B\) such that \(f \dcomp \branch{b} = \branch{(f \dcomp b)} \dcomp (f_{1} + f_{2})\).
\end{remark}

\begin{toappendix}
\begin{lemma}%
  \label{lem:conditionals-if-else}
  Let \(f \colon A \to B\) and \(b \colon B \to 1+1\) be morphisms in an \kl{imperative category} \(\cat{C}\) with \kl{conditionals}.
  Then, there are \(f_{1}, f_{2} \colon A \to B\) such that \(f \dcomp \branch{b} = \branch{(f \dcomp b)} \dcomp (f_{1} + f_{2})\).
\end{lemma}
\begin{proof}
  A conditional \(c \colon (1+1) \tensor A \to B\) of the morphism \(f \dcomp \cp_{B} \dcomp (b \tensor \id)\) determines, by distributivity, two morphisms \(f_{1}, f_{2} \colon A \to B\) satisfying \(f \dcomp \branch{b} = \branch{(f \dcomp b)} \dcomp (f_{1} + f_{2})\).
\end{proof}
\end{toappendix}

The next section explores the consequences of poset-enrichment and posetal uniformity for reasoning about programs.

\subsection{Guarded command language with inequalities}

The first consequence of posetal uniformity is reasoning with loop invariants: if a property is preserved by the loop body, then it should be preserved by the whole loop.

\begin{propositionrep}%
  \label{prop:while-uniformity}
  If \(\assert{p} \ccomp c \sleq d \ccomp \assert{p}\), then \(\assert{p} \ccomp \while{b}{c} \sleq \while{b}{d} \ccomp \assert{p}\).
\end{propositionrep}
\begin{proof}
  This follows from posetal uniformity (\Cref{fig:ax-posetal-uniformity}), from \(\assert{p} \dcomp \cp \dcomp (b \tensor \id) = \cp \dcomp (b \tensor \assert{p})\) and from \(\branch{b} \defn \cp \dcomp (b \tensor \id) \dcomp \rdistr_{1,1;X}\).
\end{proof}

The axioms of guarded Kleene algebras are stronger than the axioms of a posetally uniform trace.
If the trace also gives least, or greatest, fixpoints, then the commands with deterministic and total guards satisfy all the axioms of GKAT but the fixpoint axiom: we obtain a posetal version of it.

\begin{definition}%
  \label{def:trace-fixpoints}
  The trace of an \kl{imperative posetal bicategory}, \(\cat{C}\), gives \intro{least (greatest) fixpoints} if, for all morphisms \(f,g \colon X \to X\) and \(b \colon X \to 1+1\), the morphism \((\while{b}{f}) \dcomp g = \trace (\join \dcomp \branch{b} \dcomp (f+\id)) \dcomp g\) is the least (greatest) fixpoint of the function \(\phi_{b,f,g} \colon \cat{C}(X,X) \to \cat{C}(X,X)\) defined by
  \[\phi_{b,f}(u) \defn \cchoice{(f \dcomp u)}{b}{g} = \branch{b} \dcomp \copairing{f \dcomp u}{g}.\]
\end{definition}

Least, or greatest, fixpoints give that hom-sets have a bottom, or top, element, the zero morphisms.

\begin{propositionrep}%
  \label{prop:zero-bot-happy-trace}
  If the trace of an \kl{imperative posetal bicategory}, \(\cat{C}\), gives \kl{least (greatest) fixpoints}, then the zero morphisms, \(\zeromap_{X,Y}\), are the least (greatest) elements of the hom-sets \(\cat{C}(X,Y)\).
  Moreover, in both cases, \(\while{b}{\noop} = \assert{\pred{(\gnot b)}}\) for all deterministic and total guards \(b\).
\end{propositionrep}
\begin{proof}
  For any \(r \colon X \to Y\), \(\cchoice{r}{\gleft}{\noop} = r\).
  By least (greatest) fixpoints, we obtain that \(\zeromap = \while{\gleft}{\noop} \pleq (\pgeq) r\).

  Suppose the trace gives least fixpoints, then we obtain that \(\while{b}{\noop} \pgeq \while{b}{\abort} = \assert{\pred{(\gnot b)}}\).
  For the other inequality, observe that
  \begin{align*}
    & \cchoice{\assert{(\gnot b)}}{b}{\noop} = \cchoice{\assert{\pred{b}} \ccomp \assert{\pred{(\gnot b)}}}{b}{\noop} \\
    & = \cchoice{\abort}{b}{\noop} = \assert{\pred{(\gnot b)}}
  \end{align*}
  by determinism and totality.
  By least fixpoints, we obtain that \(\while{b}{\noop} \leq \assert{\pred{(\gnot b)}}\).
  The case with greatest fixpoints is dual.
\end{proof}

\begin{toappendix}
\begin{lemma}%
  \label{lem:det-while-uniformity}
  For a guard \(c \colon X \to 1+1\) and a deterministic guard \(b \colon X \to 1+1\) in an \kl{imperative category},
  \[\assert{\pred{b}} \dcomp \trace_{X}(\join \dcomp \branch{c} \dcomp (b + \id)) = \while{c}{\id} \dcomp (\initmap + \assert{\pred{b}}).\]
\end{lemma}
\begin{proof}
  This follows from uniformity.
\end{proof}

\begin{lemmarep}%
  \label{lem:loop-tightening}
  If the trace of an \kl{imperative posetal bicategory}, \(\cat{C}\), gives \kl{least (greatest) fixpoints}, then, for all deterministic and total guards, \(b, c \colon X \to 1+1\), and all morphisms \(f \colon X \to X\),
  \[\while{b}{(\cchoice{f}{c}{\id})} = \while{b}{(\assert{(\pred{c})} \dcomp f)}.\]
\end{lemmarep}
\begin{proof}
  The proof uses \Cref{lem:det-while-uniformity} and \Cref{prop:zero-bot-happy-trace}.
\end{proof}
\end{toappendix}

Finally, we compare the algebra of guarded commands with GKAT.
We give a semantic version of the fixpoint axiom, which prevents us from having a side condition like the one of the fixpoint axiom of GKAT; intuitively, the fixpoint is unique when the loop body is performing at least one action.
Without this side condition, the fixpoint axiom (with equality) is unsound because it would imply \(\noop = \while{\gleft}{\noop} = \abort\) (since \(\noop = \cchoice{(\noop \ccomp \noop)}{\gleft}{\noop}\)).
However, the posetal fixpoint axioms are still sound without side conditions.

\begin{theoremrep}%
  \label{prop:gkat-from-distributive}
  In an \kl{imperative posetal bicategory} where the trace gives \kl{least (or greatest) fixpoints}, the commands with deterministic and total guards satisfy the guarded union axioms (U1-U5), the sequence axioms (S1-S5), the unrolling axiom (W1), the tightening axiom (W2) and a posetal version of the fixpoint axiom (W3): if \(u \pgeq \cchoice{(f \ccomp u)}{b}{g}\), then \(u \pgeq (\while{b}{f}) \ccomp g\) (or, if \(u \pleq \cchoice{(f \ccomp u)}{b}{g}\), then \(u \pleq (\while{b}{f}) \ccomp g\)).
\end{theoremrep}
\begin{proof}
  The proof uses \Cref{lem:loop-tightening} and \Cref{def:trace-fixpoints}.
\end{proof}

\subsection{Tightest-precondition and tightest-postcondition calculi}

With the poset structure on commands, we can reason about preconditions and postconditions.
We are interested in deciding the inequalities below, for a command \(c\), predicates \(p\) and \(q\), and states \(s\) and \(t\).
\begin{align*}
  p & \sleq c \ccomp q & s \ccomp c & \sleq t \\
  p & \sgeq c \ccomp q & s \ccomp c & \sgeq t
\end{align*}
For deciding an inequality \(p \sleq c \ccomp q\), we are interested in determining the weakest precondition of the command \(c\) with respect to the predicate \(q\): the largest predicate \(\mathsf{wpre}(c,q)\) satisfying \(\semn{\mathsf{wpre}(c,q)} \pleq \semn{c \ccomp q}\).
Similarly, for deciding an inequality \(p \sgeq c \ccomp q\), we are interested in determining the strongest precondition of the command \(c\) with respect to the predicate \(q\): the smallest predicate \(\mathsf{spost}(c,q)\) satisfying \(\semn{\mathsf{spost}(c,q)} \pgeq \semn{c \ccomp q}\).

We define an inductive rewriting procedure that computes a predicate \(\tpre{c}{q}\) satisfying \(\tpre{c}{q} \sequiv c \ccomp q\); we call this predicate the \emph{tightest precondition} of the command \(c\) with respect to the predicate \(q\).
Tightest preconditions give weakest preconditions or strongest preconditions depending on which inequality we are interested in.
Dually, we define an inductive rewriting procedure that computes a state \(\tpost{c}{s}\) satisfying \(\tpost{c}{s} \sequiv s \ccomp c\); we call this state the \emph{tightest postcondition} of the command \(c\) with respect to the state \(s\).
Tightest postconditions give weakest postconditions or strongest postconditions depending on which inequality we are interested in.

\begin{figure}[h!]
\begin{fleqn}[\parindent]
\begin{alignat*}{3}
  & \text{Command } c &\qquad\qquad& \tpre{c}{q} &\qquad\qquad& \tpost{c}{s} \\
  &\noop && q && s \\
  &\abort && \pfalse && \sfalse \\
  &c_1 \ccomp c_2 && \tpre{c_{1}}{\tpre{c_{2}}{q}} && \tpost{c_{2}}{\tpost{c_{1}}{s}} \\
  &\cchoice{c_1}{b}{c_2} && \pchoice{\tpre{c_{1}}{q}}{b}{\tpre{c_{2}}{q}} && \schoice{\tpost{c_{1}}{s_{1}}}{s \ccomp b}{\tpost{c_{2}}{s_{2}}} \\
  &\while{b}{c} && \fix{p}{\pchoice{\tpre{c}{p}}{b}{q}} && \fix{s}{\pchoice{\tpost{c}{s_{1}}}{s \ccomp b}{s_{2}}} \\
  &\assert{p} && p \pand q && \restrict{s}{q} \\
  &\assign{x_i}{e} && \psub{q}{x_{i}}{e} && \ssub{s}{x_{i}}{e} \\
  &\samp{(x_{i_1}, \dots, x_{i_k})}{t} && \pmute{x_{i_1}, \dots, x_{i_k}}{q}{t} && \smute{x_{i_1}, \dots, x_{i_k}}{t}{s}
\end{alignat*}
\end{fleqn}
\caption{Rules defining the tightest precondition of a command \(c\) with respect to postcondition \(q\) and the tightest postcondition of a command \(c\) with respect to a precondition \(s\). We denote \(\gmute{x_{1}, \dots, x_{N}}{b}{s}\) with \(s \ccomp b\).}%
\label{fig:tightest-pre-post}
\end{figure}
For simplicity, we have fixed the empty signature for commands, \(\csig = \emptyset\).

\begin{theoremrep}%
  \label{th:tightest-pre-post}
  For a command \(c\), a predicate \(q\) and a state \(s\) in a \kl{imperative posetal bicategory} with \kl{conditionals} (\Cref{rem:conditionals}) and \kl{fixpoints} (\Cref{def:trace-fixpoints}), the tightest precondition calculus and the tightest postcondition calculus in \Cref{fig:tightest-pre-post} are adequate.
  \begin{align*}
    \semn{\tpre{c}{q}} &= \semn{c} \dcomp \semn{q} & \semn{\tpost{c}{s}} &= \semn{s} \dcomp \semn{c}
  \end{align*}
  In particular, we obtain strongest and weakest pre- and postcondition calculi.
  \begin{align*}
    p \leq \semn{\tpre{c}{q}} &\Leftrightarrow p \leq \semn{c} \dcomp \semn{q} & t \leq \semn{\tpost{c}{s}} &\Leftrightarrow t \leq \semn{s} \dcomp \semn{c} \\
    p \geq \semn{\tpre{c}{q}} &\Leftrightarrow p \geq \semn{c} \dcomp \semn{q} & t \geq \semn{\tpost{c}{s}} &\Leftrightarrow t \geq \semn{s} \dcomp \semn{c} \\
  \end{align*}
\end{theoremrep}
\begin{proof}
  This follows from \Cref{prop:predicate-state-actions} and \Cref{lem:conditionals-if-else}.
\end{proof}

\section{Program logics}%
\label{sec:program-logics}
Program triples are tuples containing a precondition \kl{predicate}, a \kl{command} and a postcondition \kl{predicate}.
Program logics are concerned with proving the \emph{validity} of a triple, but what \emph{validity} means depends on the program logic in question and the properties it is concerned with.

For instance, the program triples $\triple{p}{c}{q}$ and $\triple{s}{c}{t}$ may mean any of the inequalities in \Cref{fig:program-triples}, for a command \(c\), predicates \(p\) and \(q\), and states \(s\) and \(t\).

\begin{figure}[h!]
\begin{align*}
  & &&\text{State} && \text{Predicate} && \text{Assertion}\\
  & \text{Correctness} && s \ccomp c \sleq t && p \sleq c \ccomp q && \assert{p} \ccomp c \sleq c \ccomp \assert{q} \\
  & \text{Incorrectness} && s \ccomp c \sgeq t && p \sgeq c \ccomp q && \assert{p} \ccomp c \sgeq c \ccomp \assert{q}
\end{align*}
\caption{Inequalities that define validity of program triples \(\triple{p}{c}{q}\) or \(\triple{s}{c}{t}\).}\label{fig:program-triples}
\end{figure}

This section expresses program logics in the language of imperative categories.
The next section introduces couplings to cover relational program logics in a similar fashion.
This level of generality allows us to instantiate the rules that we prove here in all the examples of \Cref{sec:examples-poset-imperative}.

Each program logic defines validity of triples with one of the inequalities above.
Hoare logic~\cite{hoare1969axiomatic} uses \(\assert{p} \ccomp c \sleq c \ccomp \assert{q}\),
incorrectness logic~\cite{deVries2011reverse,o2019incorrectness} uses \(s \ccomp c \sgeq t\), pre-expectation reasoning~\cite{batz2021relatively} uses \(p \sleq c \ccomp q\).
These are only three of the possibilities outlined above, but nothing prevents us from considering the other ones as well.

The structure of imperative categories allows us to derive rules for any chosen triple shape:
the posetal enrichment is crucial for interpreting validity of triples;
the categorical structure ensures the \(\nooprule\) and \(\comprule\) rules;
the monoidal copy-discard structure gives the \(\assignrule\) and \(\samplerule\) rules;
the distributive coproducts give the rules for choice;
the posetal-uniform trace gives the rules for loops.

\begin{definition}
  A \intro{semantics} for the \kl{monoidal guarded command language} over the \kl{signatures} \((\esig, \gsig, \psig, \ssig, \csig)\) is a tuple, \((\cat{C}, X, N, \gsemn{-})\), of an \kl{imperative posetal bicategory}, \(\cat{C}\), an object in it, \(X \in \cat{C}\), a natural number, \(N \in \naturals\), and a function, \(\gsemn{-}\) as described in \Cref{sec:monoidal-gcl}.
  Explicitly, the function \(\gsemn{-}\) maps each expression generator of arity \(n\), \((e, n) \in \esig\), to a deterministic and total morphism \(\gsemn{e} \colon X^{n} \to X\) in \(\cat{C}\); each guard generator of arity \(n\), \((g, n) \in \gsig\), to a morphism \(\gsemn{g} \colon X^{n} \to 1+1\); each predicate generator of arity \(n\), \((p,n) \in \psig\), to a morphism \(\gsemn{p} \colon X^{n} \to 1\); each state generator of coarity \(n\), \((s,n) \in \ssig\), to a morphism \(\gsemn{s} \colon 1 \to X^{n}\); and each command generator of arity \(m\) and coarity \(n\), \((c,m,n) \in \csig\), to a morphism \(\gsemn{c} \colon X^{m} \to X^{n}\).
\end{definition}

As shown in \Cref{sec:monoidal-gcl}, a \kl{semantics} determines an interpretation, \(\semn{-}\), for each term of the \kl{monoidal guarded command language}.

\subsection{Correctness triples}
This section considers \kl{assertion-correctness triples}.
In the category \(\Rel\) of sets and relations, these are known as \emph{Hoare triples}~\cite{hoare1969axiomatic}.

\begin{definition}%
  \AP%
  An \intro{assertion-correctness triple}, $\triple{p}{c}{q}$, consists of a \kl{command}, $c$, and two \kl{predicates}, \(p\) and \(q\).
  An \kl{assertion-correctness triple} is \intro[valid assertion-correctness triple]{valid} in a \kl{semantics}, \((\cat{C}, X, N, \gsemn{-})\), if $\assert{p} \ccomp c \sleq c \ccomp \assert{q}$.
\end{definition}

We derive the rules of Hoare logic~\cite{hoare1969axiomatic} as presented by Winskel's reference book~\cite{winskel1993formal}.
Additionally, we include rules for nondeterministic choice and iteration that accommodate examples outside of the category of relations.

\begin{toappendix}
\begin{figure}[h!]
  \begin{gather*}
    \inferrule[skip]{ }{ \triple{p}{\noop}{p} }
    \qquad
    \inferrule[comp]
      { \triple{p}{c_1}{q} \\ \triple{q}{c_2}{r} }
      { \triple{p}{c_1 \ccomp c_2}{r} }
    \qquad
    \inferrule[assign]
      {  }
      { \triple{ \psub{p}{x_{i}}{e} }{ \assign{x_{i}}{e} }{ p }}
    \\
    \inferrule[choice]
      { \triple{p}{c_1}{q} \\
        \triple{p}{c_2}{q}
      }
      { \triple{p}{\cchoice{c_1}{b}{c_2}}{q}
      }
    \quad
    \inferrule[loop]
      { \triple{p}{c}{p} }
      { \triple{p}{ \while{b}{c} }{p} }
    \quad
    \inferrule[unroll]
      { \triple{p}{ \cchoice{(c \ccomp \while{b}{c})}{b}{\noop} }{ q } }
      { \triple{p}{ \while{b}{c} }{ q } }
   \\
    \inferrule[ifelse]
      { \triple{ p \pand \pred{b} }{ c_{1} }{ q } \\
        \triple{ p \pand \pred{(\gnot b)} }{ c_{2} }{ q } \\
        \semn{b} \mbox{ deterministic }}
      { \triple{p}{\cchoice{c_{1}}{b}{c_{2}}}{q} }
    \quad
    \inferrule[while]
      { \triple{ p \pand \pred{b} }{c}{p} \\
        \semn{b} \mbox{ deterministic }
      }
      { \triple{p}{\while{b}{c}}{ \pred{(\lnot b)} \pand p} }
  \end{gather*}

  \begin{gather*}
    \inferrule[monotone]
      { \semn{p_1} \pleq \semn{p_2} \\ \triple{ p_{2}}{c}{q_{2}} \\ \semn{q_2} \pleq \semn{q_1} }
      { \triple{p_{1}}{c}{q_{1}} }
    \qquad
    \inferrule[and]
      { \triple{p_1}{c}{q_{1}} \\ \triple{p_2}{c}{q_2} }
      { \triple{p_{1} \pand p_2}{c}{q_{1} \pand q_2} }
    \qquad
    \inferrule[fail]
      { }
      { \triple{p}{\abort}{q} }
    \\
    \inferrule[assert]
      { \semn{q \pand r} \pleq \semn{\pfalse} }
      { \triple{\pchoice{p}{b}{q}}{\assert{r}}{p \pand \pred{b}} }
     \qquad
    \inferrule[constancy]
    { \triple{p}{c}{q} \\ \modv{c} \intersection \freev{r} = \emptyset }
    { \triple{ p \pand r }{c}{ q \pand r } }
    \qquad
    \inferrule[top]
      { }
      { \triple{p}{c}{\ptrue} }
    \qquad
    \inferrule[bot]
      { }
      { \triple{\pfalse}{c}{q} }
  \end{gather*}
  \caption{Derivation rules for \kl{assertion-correctness triples} (\Cref{thm:hoarelogic}).}%
  \label{fig:assertion-correctness-rules}
\end{figure}
\end{toappendix}

\begin{theoremrep}%
  \label{thm:hoarelogic}
  The rules in \Cref{fig:assertion-correctness-rules} derive \kl{valid assertion-correctness triples} in any \kl{imperative posetal bicategory} where \(\zeromap_{A,B} \pleq f\) and \(f \dcomp \discard_{B} \pleq \discard_{A}\) for all morphisms \(f \colon A \to B\).
\end{theoremrep}
\begin{proof}
  \textsc{skip}: use \Cref{prop:commands-guarded-monoid}.
  \[\assert{p} \ccomp \noop \sequiv \assert{p} \sequiv \noop \ccomp \assert{p}\]
  \textsc{comp}: use \Cref{prop:commands-guarded-monoid} (and from now on, we will use associativity implicitly).
  \begin{align*}
    \assert{p} \ccomp (c_1 \ccomp c_2)
    &\sequiv (\assert{p} \ccomp c_{1}) \ccomp c_{2} \sleq (c_1 \ccomp \assert{q}) \ccomp c_2 \\
    &\sequiv c_{1} \ccomp (\assert{q} \ccomp c_{2})
    \sleq c_1 \ccomp (c_2 \ccomp \assert{r}) \sequiv (c_{1} \ccomp c_{2}) \ccomp \assert{r}
  \end{align*}
  \textsc{assign}: use \Cref{cor:assert-subst}.
  \[\assert{(\psub{p}{x_i}{e})} \ccomp (\assign{x_i}{e}) \sequiv (\assign{x_i}{e}) \ccomp \assert{p}\]
  \textsc{choice}: use \Cref{cor:guards-predicate-interchange} and \Cref{prop:commands-guarded-monoid}.
  \begin{align*}
    & \assert{p} \ccomp (\cchoice{c_{1}}{b}{c_{2}}) & &\\
    & \sequiv \cchoice{(\assert{p} \ccomp c_{1})}{b}{(\assert{p} \ccomp c_{2})} && \text{(\Cref{cor:guards-predicate-interchange})}\\
    &\sleq \cchoice{(c_{1} \ccomp \assert{q})}{b}{(c_{2} \ccomp \assert{q})} && \text{(Assumption)}\\
    &\sequiv (\cchoice{c_{1}}{b}{c_{2}}) \ccomp \assert{q} && \text{(\Cref{prop:commands-guarded-monoid})}
  \end{align*}
  \textsc{ifelse}: use \Cref{cor:guards-predicate-interchange}, \Cref{lem:assert-monoid-morph} and \Cref{prop:commands-guarded-monoid}.
  \begin{align*}
    & \assert{p} \ccomp (\cchoice{c_{1}}{b}{c_{2}}) & &\\
    & \sequiv \assert{p} \ccomp (\cchoice{(\assert{\pred{b}} \ccomp c_{1})}{b}{(\assert{\pred{(\gnot b)}} \ccomp c_{2})}) & & \text{(\Cref{cor:duplicate-deterministic-guards})}\\
    & \sequiv \cchoice{(\assert{p} \ccomp \assert{\pred{b}} \ccomp c_{1})}{b}{(\assert{p} \ccomp \assert{\pred{(\gnot b)}} \ccomp c_{2})} && \text{(\Cref{cor:guards-predicate-interchange})}\\
    & \sequiv \cchoice{(\assert{(p \pand \pred{b})} \ccomp c_{1})}{b}{(\assert{(p \pand \pred{(\gnot b)})} \ccomp c_{2})} && \text{(\Cref{lem:assert-monoid-morph})}\\
    &\sleq \cchoice{(c_{1} \ccomp \assert{q})}{b}{(c_{2} \ccomp \assert{q})} && \text{(Assumption)}\\
    &\sequiv (\cchoice{c_{1}}{b}{c_{2}}) \ccomp \assert{q} && \text{(\Cref{prop:commands-guarded-monoid})}
  \end{align*}
  \textsc{loop}: by assumption, \(\assert{p} \ccomp c \sleq c \ccomp \assert{p}\); by \Cref{prop:while-uniformity}, we obtain \(\assert{p} \ccomp \while{b}{c} \sleq \while{b}{c} \ccomp \assert{p}\).\\
  \textsc{while}
  \begin{align*}
    & \assert{p} \ccomp \assert{\pred{b}} \ccomp c && \\
    & \sequiv \assert{(p \pand \pred{b})} \ccomp c && \text{(\Cref{lem:assert-monoid-morph})}\\
    & \sleq c \ccomp \assert{p} && \text{(Assumption)}
  \end{align*}
  Then,
  \begin{align*}
    & \assert{p} \ccomp \while{b}{c} && \\
    & \sequiv \assert{p} \ccomp \while{b}{(\assert{\pred{b}} \ccomp c)} && \text{(\Cref{lemma:deterministic-while})} \\
    & \sleq \while{b}{c} \ccomp \assert{p} && \text{(\Cref{prop:while-uniformity})} \\
    & \sequiv \while{b}{c} \ccomp \assert{\pred{(\gnot b)}} \ccomp \assert{p} && \text{(\Cref{lemma:deterministic-while})} \\
    & \sequiv \while{b}{c} \ccomp \assert{(\pred{(\gnot b)} \pand p)} && \text{(\Cref{lem:assert-monoid-morph})}
  \end{align*}
  \textsc{unroll}
  \begin{align*}
    & \assert{p} \ccomp \while{b}{c} && \\
    & \sequiv \assert{p} \ccomp \cchoice{(c \ccomp \while{b}{c})}{b}{\noop} && \text{(\Cref{cor:while-unrolling})}\\
    & \sleq (\cchoice{(c \ccomp \while{b}{c})}{b}{\noop}) \ccomp \assert{q} && \text{(Assumption)}\\
    & \sequiv \while{b}{c} \ccomp \assert{q} && \text{(\Cref{cor:while-unrolling})}
  \end{align*}
  \textsc{monotone}: use the assumptions.
  \[\assert{p_{1}} \ccomp c \sleq \assert{p_{2}} \ccomp c  \sleq c \ccomp \assert{q_{2}} \sleq c \ccomp \assert{q_{1}}\]
  \textsc{and}: use \Cref{lem:assert-monoid-morph}.
  \begin{align*}
    & \assert{(p_{1} \pand p_{2})} \ccomp c && \\
    & \sequiv \assert{p_{1}} \ccomp \assert{p_{2}} \ccomp c && \text{(\Cref{lem:assert-monoid-morph})} \\
    & \sleq \assert{p_{1}} \ccomp c \ccomp \assert{q_{2}} && \text{(Assumption)} \\
    & \sleq c \ccomp \assert{q_{1}} \ccomp \assert{q_{2}} && \text{(Assumption)} \\
    & \sequiv c \ccomp \assert{(q_{1} \pand q_{2})} && \text{(\Cref{lem:assert-monoid-morph})}
  \end{align*}
  \textsc{assert}
  \begin{align*}
    & \assert{(\pchoice{p}{b}{q})} \ccomp \assert{r} && \\
    & \sequiv \assert{((\pchoice{p}{b}{q}) \pand r)} && \text{(\Cref{lem:assert-monoid-morph})}\\
    & \sequiv \assert{(\pchoice{(p \pand r)}{b}{(q \pand r)})} && \text{(\Cref{prop:predicates-guarded-monoid})}\\
    & \sleq \assert{(\pchoice{(p \pand r)}{b}{\pfalse})} && \text{(Assumption)} \\
    & \sequiv \assert{(\pchoice{(\ptrue \pand r \pand p)}{b}{(\pfalse \pand r \pand p)})} && \text{(\Cref{prop:predicates-guarded-monoid})} \\
    & \sequiv \assert{((\pchoice{\ptrue}{b}{\pfalse}) \pand r \pand p)} && \text{(\Cref{prop:predicates-guarded-monoid})} \\
    & \sequiv \assert{(r \pand p \pand (\pchoice{\ptrue}{b}{\pfalse}))} && \text{(\Cref{prop:predicates-guarded-monoid})} \\
    & \sequiv \assert{(r \pand p \pand \pred{b})} && \text{(\Cref{cor:morph-guard-to-predicates})}\\
    & \sequiv \assert{r} \ccomp \assert{(p \pand \pred{b})} && \text{(\Cref{lem:assert-monoid-morph})}
  \end{align*}
  \textsc{top}
  \begin{align*}
    & \assert{p} \ccomp c && \\
    & \sleq \assert{\ptrue} \ccomp c && \text{(Hypothesis)} \\
    & \sequiv \noop \ccomp c && \text{(\Cref{lem:assert-monoid-morph})} \\
    & \sequiv c \ccomp \noop && \text{(\Cref{prop:commands-guarded-monoid})}\\
    & \sequiv c \ccomp \assert{\ptrue} && \text{(\Cref{lem:assert-monoid-morph})}
  \end{align*}
  \textsc{bot}
  \begin{align*}
    & \assert{\pfalse} \ccomp c && \\
    & \sequiv \abort \ccomp c && \text{(\Cref{lem:assert-monoid-morph})} \\
    & \sequiv c \ccomp \abort && \text{(\Cref{prop:commands-guarded-monoid})}\\
    & \sleq c \ccomp \assert{p} && \text{(Hypothesis)}
  \end{align*}
  \textsc{fail}: use \Cref{prop:commands-guarded-monoid} (\(\abort\) is absorbing).
  \begin{align*}
    \assert{p} \ccomp \abort \sequiv \abort \sequiv \abort \ccomp \assert{q}
  \end{align*}
  \textsc{constancy}: use \Cref{cor:assert-sample-interchange}.
  \begin{align*}
    & \assert{(p \pand r)} \ccomp c && \\
    & \sequiv \assert{p} \ccomp \assert{r} \ccomp c && \text{(\Cref{lem:assert-monoid-morph})}\\
    & \sequiv \assert{p} \ccomp c \ccomp \assert{r} && \text{(\Cref{cor:assert-sample-interchange})}\\
    & \sleq c \ccomp \assert{q} \ccomp \assert{r} && \text{(Assumption)}\\
    & \sequiv c \ccomp \assert{(q \pand r)} && \text{(\Cref{lem:assert-monoid-morph})}
  \end{align*}
\end{proof}

\begin{remark}
  In the imperative category \(\Rel\) of sets and relations, \kl{assertion-correctness triples} are equivalent to state-correctness triples: $\assert{p} \ccomp c \sleq c \ccomp \assert{q}$ if and only if $p\op \ccomp c \sleq q\op$.
  Predicates have, in general, a richer logic compared to states.
  Therefore, we chose the former triple shape.
\end{remark}

\begin{example}[Floyd-Hoare logic]%
  \label{ex:assertion-correctness-par}
  Consider the \kl{posetal imperative bicategory} \(\Par\) of sets and partial functions with the set \(\integers\) of integers as chosen object and fix a number \(N\) of variables.

  Consider the signatures \((\esig, \gsig^{\Par}, \emptyset, \ssig^{\Par}, \emptyset)\), where \(\esig\) contains symbols for the arithmetic operations and constants for the integers (\Cref{ex:signature-expressions}), \(\gsig^{\Par}\) contains symbols for equality and inequality between expressions (\Cref{ex:signature-guards-par-rel}) and \(\ssig^{\Par}\) contains constants for the integers (\Cref{ex:signature-states-par-rel}); the signatures for predicates and commands are empty.
  The guarded command language that we obtain extends the \textsf{Imp} language~\cite{winskel1993formal} with the \(\abort\) command and the \(\pfalse\) predicate.

  We instantiate \kl{assertion-correctness triples} with the signature just defined.
  The program logic that we obtain has the rules of Floyd-Hoare logic~\cite{hoare1969axiomatic} with two additional ones—\textsc{bot} and \textsc{fail}—for the extra \(\abort\) command and \(\pfalse\) predicate and the additional \textsc{constancy} rule; note that the \textsc{choice} and \textsc{loop} rules become redundant: they can be derived from the \textsc{ifelse} and \textsc{while} rules because all morphisms, and all guards in particular, are deterministic.
\end{example}

\begin{example}[Expectation reasoning]%
  \label{ex:assertion-correctness-stoch}
  Consider the \kl{posetal imperative bicategory} \(\subStoch\) of sets and countably-supported partial stochastic channels and the signatures \((\esig, \gsig^{\subStoch}, \emptyset, \ssig^{\subStoch}, \emptyset)\), where \(\esig\) is as above (\Cref{ex:signature-expressions}), \(\gsig^{\subStoch}\) contains symbols for equality, inequality and binary probabilistic choices (\Cref{ex:signature-guards-stoch}), and \(\ssig^{\subStoch}\) contains constants for the integers and for the uniform distributions on \(n\) elements (\Cref{ex:signature-states-stoch}); the signatures for predicates and commands are empty.
  We obtain a probabilistic guarded command language with both probabilistic and conditional choices and loops~\cite{kozen1985probabilistic,batz2021relatively}, but no nondeterministic choice.
  In this case, a triple \(\triple{p}{c}{q}\) is valid whenever \(p(x) \cdot c(y \given x) = c(y \given x) \cdot q(y)\) for all \(x,y \in X^{N}\).
\end{example}

\subsection{Incorrectness triples}

This section considers \kl{state-incorrectness triples}.
In the category \(\Rel\) of sets and relations, these are known as \emph{reverse Hoare triples}~\cite{deVries2011reverse} or \emph{incorrectness triples}~\cite{o2019incorrectness}.

\begin{definition}
  A \intro{state-incorrectness triple}, $\triple{s}{c}{t}$, consists of a \kl{command}, $c$, and two \kl{states}, $s$ and \(t\).
  A \kl{state-incorrectness triple} is \intro[valid state-incorrectness triple]{valid} in a \kl{semantics}, \((\cat{C}, X, N, \gsemn{-})\), if \(s \ccomp c \sgeq t\).
\end{definition}

We derive the rules of incorrectness logic~\cite{o2019incorrectness} in the more general setting of \kl{imperative posetal bicategories}.
The original incorrectness rules for choices and loops are a particular case of the ones below (see \Cref{ex:state-incorrectness-rel}).

\begin{toappendix}
\begin{figure}[h!]
  \begin{gather*}
    \inferrule[skip]
      { }
      { \triple{s}{\noop}{s} }
    \qquad
    \inferrule[comp]
      { \triple{s}{c_1}{t} \\ \triple{t}{c_2}{r} }
      { \triple{s}{c_1 \ccomp c_2}{r} }
    \qquad
    \inferrule[comp (error)]
      { \triple{s}{c_1}{\sfalse} }
      { \triple{s}{c_1 \ccomp c_2}{\sfalse} }
    \\
    \inferrule[assign]
      { }
      { \triple{ s }{ \assign{x_{i}}{e} }{ \ssub{s}{x_{i}}{e} }}
    \qquad
    \inferrule[sample]
      {  }
      { \triple{s}{ \samp{(x_{i_{1}}, \dots, x_{i_{k}})}{t}}{ \smute{x_{i_{1}}, \dots, x_{i_{k}}}{s}{t} } }
    \\
    \inferrule[choice (left)]
      { \triple{\restrict{s}{\pred{b}}}{c_1}{t}
      }
      { \triple{s}{\cchoice{c_1}{b}{c_2}}{t}
        }
    \quad
    \inferrule[choice (right)]
      { \triple{\restrict{s}{\pred{(\gnot{b})}}}{c_2}{t}       }
      { \triple{s}{\cchoice{c_1}{b}{c_2}}{t}
        }
    \quad
    \inferrule[convex]
      { \triple{s_{1}}{c}{t_{1}} \\ \triple{s_{2}}{c}{t_{2}} }
      { \triple{\schoice{s_{1}}{b}{s_{2}}}{c}{\schoice{t_{1}}{b}{t_{2}}}
        }
    \\
    \inferrule[iter zero]
      {  }
      { \triple{s}{ \while{b}{c} }{\restrict{s}{\pred{(\gnot{b})}}} }
    \qquad
    \inferrule[iter]
      { \triple{\restrict{s}{\pred{b}}}{ c \ccomp \while{b}{c} }{ t } }
      { \triple{s}{ \while{b}{c} }{ t } }
    \qquad
    \inferrule[assert]
      {  }
      { \triple{s}{\assert{p}}{\restrict{s}{p}} }
  \end{gather*}

  \begin{gather*}
    \inferrule[monotone]
      { \semn{s_1} \pgeq \semn{s_2} \\ \triple{ s_{2}}{c}{t_{2}} \\ \semn{t_2} \pgeq \semn{t_1} }
      { \triple{s_{1}}{c}{t_{1}} }
    \qquad
    \inferrule[fail]
      { }
      { \triple{s}{\abort}{\sfalse} }
    \qquad
    \inferrule[bot]
      { }
      { \triple{s}{c}{\sfalse} }
  \end{gather*}
  \begin{gather*}
    \inferrule[constancy]
    { \triple{s}{c}{t} \\ \modv{c} \intersection \freev{p} = \emptyset }
    { \triple{ \restrict{s}{p} }{c}{ \restrict{t}{p} } }
    \qquad
    \inferrule[auxiliary variable]
    { \triple{s}{c}{t} \\ x_{i_{1}}, \dots, x_{i_{k}} \notin \freev{c} }
    { \triple{ \smute{x_{i_{1}}, \dots, x_{i_{k}}}{s}{u} }{c}{ \smute{x_{i_{1}}, \dots, x_{i_{k}}}{t}{u} } }
  \end{gather*}
  \caption{Derivation rules for \kl{state-incorrectness triples} (\Cref{thm:incorrectnesslogic}).}%
  \label{fig:state-incorrectness-rules}
\end{figure}
\end{toappendix}

\begin{theoremrep}%
  \label{thm:incorrectnesslogic}
  The rules in \Cref{fig:state-incorrectness-rules} derive \kl{valid state-incorrectness triples} in any \kl{imperative posetal bicategory} where $\zeromap_{A,B} \pleq f$ for all morphisms $f \colon A \to B$.
\end{theoremrep}
\begin{proof}
  The \textsc{skip}, \textsc{comp} and \textsc{comp (error)} rules follow from \Cref{prop:predicate-state-actions}.
  From now on, we will use \Cref{prop:predicate-state-actions} implicitly.
  \begin{align*}
    s \ccomp \noop &\sequiv s & s \ccomp c_{1} \ccomp c_{2} &\sgeq t \ccomp c_{2} \sgeq u & s \ccomp c_{1} \ccomp c_{2} &\sgeq \bot \ccomp c_{2} \sequiv \bot
  \end{align*}
  \textsc{assign} and \textsc{sample}: use \Cref{prop:predicate-state-actions}.
  \begin{align*}
    s \ccomp (\assign{x_i}{e}) & \sequiv \ssub{s}{x_i}{e} & s \ccomp (\samp{x_{i_1}, \dots, x_{i_k}}{t}) & \sequiv \smute{x_{i_1}, \dots, x_{i_k}}{s}{t}
  \end{align*}
  \textsc{choice (left)} and \textsc{choice (right)} \begin{align*}
    &s \ccomp (\cchoice{c_{1}}{b}{c_{2}}) & &s \ccomp (\cchoice{c_{1}}{b}{c_{2}})  && \\
    & \sgeq s \ccomp (\cchoice{c_{1}}{b}{\abort}) & & \sgeq s \ccomp (\cchoice{\abort}{b}{c_{2}}) && \text{(Hypothesis)} \\
    & \sequiv s \ccomp \assert{\pred{b}} \ccomp c_{1} & & \sequiv s \ccomp \assert{\pred{(\gnot{b})}} \ccomp c_{2} && \text{(\Cref{cor:assert-guards})} \\
    & \sequiv (\restrict{s}{\pred{b}}) \ccomp c_{1} & & \sequiv (\restrict{s}{\pred{(\gnot{b})}}) \ccomp c_{2} && \text{(\Cref{lem:states-restrict-assert})} \\
    & \sgeq t & & \sgeq t && \text{(Assumption)}
  \end{align*}
  \textsc{convex}
  \begin{align*}
    & (\schoice{s_{1}}{b}{s_{2}}) \ccomp c && \\
    & \sequiv \schoice{(s_{1} \ccomp c)}{b}{(s_{2} \ccomp c)} && \text{(\Cref{prop:predicate-state-actions})}\\
    & \sgeq \schoice{t_{1}}{b}{t_{2}} && \text{(Assumption)}
  \end{align*}
  \textsc{iter zero} and \textsc{iter}
  \begin{align*}
    & s \ccomp \while{b}{c} && s \ccomp \while{b}{c} && \\
    & \sequiv s \ccomp \cchoice{(c \ccomp \while{b}{c})}{b}{\noop} && \sequiv s \ccomp \cchoice{(c \ccomp \while{b}{c})}{b}{\noop} && \text{(\Cref{cor:while-unrolling})} \\
    & \sgeq  s \ccomp \cchoice{\abort}{b}{\noop} && \sgeq s \ccomp \cchoice{(c \ccomp \while{b}{c})}{b}{\abort} && \text{(Hypothesis)} \\
    & \sequiv s \ccomp \assert{\pred{(\gnot b)}} && \sequiv s \ccomp \assert{\pred{b}} \ccomp c \ccomp \while{b}{c} && \text{(\Cref{cor:assert-guards})} \\
    & \sequiv \restrict{s}{\assert{\pred{(\gnot b)}}} && \sequiv \restrict{s}{\assert{\pred{b}}} \ccomp c \ccomp \while{b}{c} && \text{(\Cref{lem:states-restrict-assert})} \\
    & && \sgeq t && \text{(Assumption)}
  \end{align*}
  \textsc{assert}: \Cref{lem:states-restrict-assert}.
  \[s \ccomp \assert{p} \sequiv \restrict{s}{p}\]
  \textsc{monotone}: by assumption.
  \[s_1 \ccomp c \sgeq s_2 \ccomp c \sgeq t_2 \sgeq t_1 \]
  \textsc{fail}: by \Cref{prop:predicate-state-actions}.
  \[s \ccomp \abort \sequiv \sfalse\]
  \textsc{bot}: by hypothesis.
  \[s \ccomp c \sgeq \sfalse\]
  \textsc{constancy}: use \Cref{cor:assert-sample-interchange}.
  \begin{align*}
    & (\restrict{s}{p}) \ccomp c && \\
    & \sequiv s \ccomp \assert{p} \ccomp c && \text{(\Cref{prop:predicate-state-actions})}\\
    & \sequiv s \ccomp c \ccomp \assert{p} && \text{(\Cref{cor:assert-sample-interchange})}\\
    & \sgeq t \assert{p} && \text{(Assumption)}\\
    & \sequiv \restrict{t}{p} && \text{(\Cref{prop:predicate-state-actions})}
  \end{align*}
  \textsc{auxiliary variable}
  \begin{align*}
    & (\smute{x_{i_{1}}, \dots, x_{i_{k}}}{s}{u}) \ccomp c && \\
    & \sequiv s \ccomp (\samp{x_{i_{1}}, \dots, x_{i_{k}}}{u}) \ccomp c && \text{(\Cref{prop:predicate-state-actions})} \\
    & \sequiv s \ccomp c \ccomp (\samp{x_{i_{1}}, \dots, x_{i_{k}}}{u}) && \text{(\Cref{cor:assert-sample-interchange})} \\
    & \sgeq t \ccomp (\samp{x_{i_{1}}, \dots, x_{i_{k}}}{u}) && \text{(Assumption)} \\
    & \sequiv \smute{x_{i_{1}}, \dots, x_{i_{k}}}{t}{u} && \text{(\Cref{prop:predicate-state-actions})}
  \end{align*}
\end{proof}

\begin{example}[Incorrectness logic]%
  \label{ex:state-incorrectness-rel}
  Consider the \kl{imperative posetal bicategory} \(\Rel\) of sets and relations with the set \(\integers\) of integers as chosen object and fix a number \(N\) of variables.
  Recall that states of type \(1 \to X^{N}\) in \(\Rel\) correspond to subsets of \(X^{N}\).
  In this case, a \kl{state-incorrectness triple} \(\triple{s}{c}{t}\) is valid if and only if for all \(y \in \semn{t}\) there exist \(x \in \semn{s}\) such that \(y \in \semn{c}(x)\); this is the definition of validity of incorrectness triples~\cite[Definition~1]{o2019incorrectness}.

  Consider the signatures \((\esig, \gsig^{\Rel}, \emptyset, \ssig^{\Rel}, \emptyset)\), where \(\esig\) contains symbols for the arithmetic operations and constants for the integers (\Cref{ex:signature-expressions}), \(\gsig^{\Rel}\) contains symbols for equality, inequality and nondeterministic branching (\Cref{ex:signature-guards-par-rel}), \(\ssig^{\Rel}\) contains constants for the integers and a constant for nondeterministic assignment (\Cref{ex:signature-states-par-rel}); the signatures for predicates and commands are empty.
  We can derive a unary predicate, \(\mathsf{nat}(x) \defn \pred{(0 \geq x)}\), that tests if an integer is a natural number, and we can derive an operation on commands, \(\mathsf{local}\ x. c\), that makes the variable \(x\) local in the command \(c\) (as defined in \Cref{ex:local-command}).

  The derivation rules that we obtain are those of incorrectness logic~\cite[Figure~2]{o2019incorrectness}.
  The empty under-approximates, consequence, unit, sequencing, iterate zero, iterate nonzero, error and assume rules are clear instantiations of the corresponding rules in \Cref{thm:incorrectnesslogic}.
  We explain the remaining rules in more detail.

  The \textsc{constancy} rule particularises to the constancy rule because \(\semn{\restrict{s}{p}} = \semn{s} \intersection \semn{p}\).
  Thanks to the nondeterministic choice generator (\(\nondetchoice\)) the \textsc{convex} rule particularises to the disjunction rule, and the \textsc{choice (left)} and \textsc{choice (right)} rules particularise to the nondeterministic choice rules because \(\semn{s} = \semn{\restrict{s}{\ptrue}} = \semn{\restrict{s}{\pred{\nondetchoice}}} = \semn{\restrict{s}{\pred{(\gnot \nondetchoice)}}}\).
  The \textsc{sample} rule with the state \(\nondetvar\) particularises to nondeterministic assignment: \(\semn{\smute{x_{i}}{s}{\nondetvar}} = \{x_{1}, \dots, x_{i-1}, x_{i+1}, \dots, x_{N} \in \integers \st \exists x_{i} \in \integers\ .\ (x_{1}, \dots, x_{N}) \in \semn{s}\}\).
  The \textsc{assign} rule instantiates to the assignment rule because \(\semn{\ssub{s}{x_{i}}{e}} = \{x_{1}, \dots, x_{i-1}, y, x_{i+1}, \dots, x_{N} \in \integers \st \exists x_{i} \in \integers\ .\ \semn{e}(x_{1}, \dots, x_{N}) = y \land (x_{1}, \dots, x_{N}) \in \semn{s}\}\).

  For the local variable rule, suppose that \(x_{i} \notin \freev{s}\) and that \(\triple{s}{c}{t}\) is valid.
  Then \(\semn{\smute{x_{i}}{s}{\nondetvar}} = \semn{s}\) and, as a consequence,
  \[s \ccomp (\mathsf{local}\ x_{i}. c) = s \ccomp (\samp{x_{i}}{\nondetvar}) \ccomp c \ccomp (\samp{x_{i}}{\nondetvar}) = (\smute{x_{i}}{s}{\nondetvar}) \ccomp c \ccomp (\samp{x_{i}}{\nondetvar}) = s \ccomp c \ccomp (\samp{x_{i}}{\nondetvar})\]
  Using the second assumption, and applying the \textsc{consequence} rule, we obtain that \(\semn{s \ccomp (\mathsf{local}\ x_{i}. c)} \pgeq \semn{t \ccomp (\samp{x_{i}}{\nondetvar})} = \semn{\smute{x_{i}}{t}{\nondetvar}} = \{x_{1}, \dots, x_{i-1}, x_{i+1}, \dots, x_{N} \in \integers \st \exists x_{i} \in \integers\ .\ (x_{1}, \dots, x_{N}) \in \semn{t}\}\).

\end{example}

\begin{example}%
  \label{ex:state-incorrectness-stoch}
  In the setting of \Cref{ex:assertion-correctness-stoch}, the pre- and postconditions are interpreted as subdistributions.
  A triple \(\triple{s}{c}{t}\) is valid iff \(\sum_{x \in X^{N}} \semn{s}(x) \cdot \semn{c}(y \given x) \geq \semn{t}(y)\) for all \(y \in X^{N}\).
  The interpretation of such a triple mirrors the incorrectness reasoning for nondeterministic semantics: if the inputs are distributed according to \(\semn{s}\), then the probability that \(c\) outputs a `bad' state \(y\) is at least \(\semn{t}(y)\).
\end{example}

\subsection{Weakest-precondition triples}
This section considers \kl{predicate-correctness triples}.

\begin{definition}
  A \intro{predicate-correctness triple}, $\triple{p}{c}{q}$, consists of a \kl{command} $c$, and two \kl{predicates}, $p$ and $q$.
  A \kl{predicate-correctness triple} is \intro[valid state-incorrectness triple]{valid} in a semantics, \((\cat{C}, X, N, \gsemn{-})\), if \(p \sleq c \ccomp q\).
\end{definition}

\begin{toappendix}
\begin{figure}[h!]
  \begin{gather*}
    \inferrule[skip]
      { }
      { \triple{p}{\noop}{p} }
    \qquad
    \inferrule[comp]
      { \triple{p}{c_1}{q} \\ \triple{q}{c_2}{r} }
      { \triple{p}{c_1 \ccomp c_2}{r} }
    \\
    \inferrule[assign]
      { }
      { \triple{ \psub{p}{x_{i}}{e} }{ \assign{x_{i}}{e} }{ p }}
    \qquad
    \inferrule[sample]
      { }
      { \triple{ \pmute{x_{i_{1}}, \dots, x_{i_{k}}}{p}{t} }{ \samp{(x_{i_{1}}, \dots, x_{i_{k}})}{t} }{ p }}
    \\
    \inferrule[unroll]
      { \triple{p}{ \cchoice{(c \ccomp \while{b}{c})}{b}{\noop} }{ q } }
        { \triple{p}{ \while{b}{c} }{ q } }
    \qquad
    \inferrule[choice]
      { \triple{p_1}{c_1}{q} \\
        \triple{p_2}{c_2}{q}
      }
      { \triple{\pchoice{p_1}{b}{p_2} }{\cchoice{c_1}{b}{c_2}}{q}
      }
    \\
    \inferrule[choice (tot)]
      { \triple{p}{c_1}{q} \\
        \triple{p}{c_2}{q} \\ b \text{ total}
      }
      { \triple{p}{\cchoice{c_1}{b}{c_2}}{q}
      }
    \qquad
    \inferrule[ifelse]
      { \triple{ \pred{b} \pand p_1 }{ c_{1} }{ q } \\
        \triple{ \pred{(\lnot b)} \pand p_2 }{ c_{2} }{ q } \\
        b \mbox{ deterministic}}
      { \triple{ \pchoice{p_1}{b}{p_2} }{\cchoice{c_{1}}{b}{c_{2}}}{q} }
    \\
    \inferrule[ifelse (tot)]
      { \triple{ \pred{b} \pand p }{ c_{1} }{ q } \\
        \triple{ \pred{(\lnot b)} \pand p }{ c_{2} }{ q} \\
        b \mbox{ deterministic and total}}
      { \triple{p}{\cchoice{c_{1}}{b}{c_{2}}}{q} }
  \end{gather*}
  \begin{gather*}
    \inferrule[assert]
      { \semn{\pred{(\gnot b)} \pand q} \pleq \semn{\pfalse} \\ \semn{b} \text{ deterministic } }
      { \triple{\pchoice{p}{b}{q}}{\assert{\pred{b}}}{ p } }
    \qquad
    \inferrule[convex]
      { \triple{p_1}{c}{q_{1}} \\ \triple{p_2}{c}{q_2} \\ b \text{ constant}}
      { \triple{\pchoice{p_{1}}{b}{p_{2}}}{c}{\pchoice{q_{1}}{b}{q_{2}}} }
    \\
    \inferrule[monotone]
      { \semn{p_1} \pleq \semn{p_2} \\ \triple{ p_{2}}{c}{q_{2}} \\ \semn{q_2} \pleq \semn{q_1} }
      { \triple{p_{1}}{c}{q_{1}} }
    \qquad
    \inferrule[bot]
      { }
      { \triple{\pfalse}{c}{q} }
  \end{gather*}
  \begin{gather*}
    \inferrule[constancy]
    { \triple{p}{c}{q} \\ \modv{c} \intersection \freev{r} = \emptyset }
    { \triple{ r \pand p }{c}{ r \pand q } }
    \qquad
    \inferrule[auxiliary variable]
    { \triple{p}{c}{q} \\ x_{i_{1}}, \dots, x_{i_{k}} \notin \freev{c} }
    { \triple{ \pmute{x_{i_{1}}, \dots, x_{i_{k}}}{p}{t} }{c}{ \pmute{x_{i_{1}}, \dots, x_{i_{k}}}{q}{t} } }
  \end{gather*}
  \caption{Derivation rules for \kl{predicate-correctness triples} (\Cref{thm:outcomelikelogic}).}%
  \label{fig:predicate-correctness-rules}
\end{figure}
\end{toappendix}

\begin{theoremrep}%
  \label{thm:outcomelikelogic}
  The rules in \Cref{fig:predicate-correctness-rules} derive \kl{valid predicate-correctness triples} in any \kl{imperative posetal bicategory} where \(\zeromap_{A,B} \pleq f\) for all morphisms \(f \colon A \to B\).
\end{theoremrep}
\begin{proof}
  \textsc{skip} and \textsc{comp}: apply \Cref{prop:predicate-state-actions}.
  \begin{align*}
    p & \sequiv \noop \ccomp p & p \sleq c_{1} \ccomp q &\sleq c_{1} \ccomp c_{2} \ccomp r
  \end{align*}
  \textsc{assign} and \textsc{sample}: use \Cref{prop:predicate-state-actions}.
  \begin{align*}
    \psub{p}{x_i}{e} &\sequiv (\assign{x_i}{e}) \ccomp p & \pmute{x_{i_1}, \dots, x_{i_k}}{p}{t} & \sequiv (\samp{x_{i_1}, \dots, x_{i_k}}{t}) \ccomp p
  \end{align*}
  \textsc{unroll}
  \begin{align*}
    p & \sleq (\cchoice{(c \ccomp \while{b}{c})}{b}{\noop}) \ccomp q && \text{(Assumption)} \\
    & \sequiv (\while{b}{c}) \ccomp q && \text{(\Cref{cor:while-unrolling})} 
  \end{align*}
  \textsc{choice}
  \begin{align*}
    \pchoice{p_1}{b}{p_2}
    & \sleq \pchoice{(c_1 \ccomp q)}{b}{(c_2 \ccomp q)} && \text{(Assumption)} \\
    & \sequiv (\cchoice{c_1}{b}{c_2}) \ccomp q && \text{(\Cref{prop:predicate-state-actions})}
  \end{align*}
  \textsc{choice (tot)}
  \begin{align*}
    p
    & \sequiv \pchoice{p}{b}{p} && \text{(\(b\) total)} \\
    & \sleq \pchoice{(c_1 \ccomp q)}{b}{(c_2 \ccomp q)} && \text{(Assumption)} \\
    & \sequiv (\cchoice{c_1}{b}{c_2}) \ccomp q && \text{(\Cref{prop:predicate-state-actions})}
  \end{align*}
  \textsc{ifelse}
  \begin{align*}
    \pchoice{p_1}{b}{p_2}
    & \sequiv \pchoice{(\pred{b} \pand p_1)}{b}{(\pred{(\gnot b)} \pand p_2)} && \text{(\(b\) deterministic)} \\
    & \sleq \pchoice{(c_1 \ccomp q)}{b}{(c_2 \ccomp q)} && \text{(Assumption)} \\
    & \sequiv (\cchoice{c_1}{b}{c_2}) \ccomp q && \text{(\Cref{prop:predicate-state-actions})}
  \end{align*}
  \textsc{ifelse (tot)}
  \begin{align*}
    p
    & \sequiv \pchoice{p}{b}{p} && \text{(\(b\) total)} \\
    & \sequiv \pchoice{(\pred{b} \pand p)}{b}{(\pred{(\gnot b)} \pand p)} && \text{(\(b\) deterministic)} \\
    & \sleq \pchoice{(c_1 \ccomp q)}{b}{(c_2 \ccomp q)} && \text{(Assumption)} \\
    & \sequiv (\cchoice{c_1}{b}{c_2}) \ccomp q && \text{(\Cref{prop:predicate-state-actions})}    
  \end{align*}
  \textsc{assert}
  \begin{align*}
    \pchoice{p}{b}{q}
    & \sequiv \pchoice{p}{b}{(\pred{(\gnot b)} \pand q)} && \text{(\(b\) deterministic)} \\
    & \sleq \pchoice{p}{b}{\pfalse} && \text{(Assumption)} \\
    & \sequiv \pred{b} \pand p && \text{(\Cref{cor:morph-guard-to-predicates})} \\
    & \sequiv \assert{\pred{b}} \ccomp p && \text{(\Cref{prop:predicate-state-actions})}
  \end{align*}
  \textsc{convex}: use the assumption and \Cref{cor:constant-guards-interchange}.
  \[\pchoice{p_{1}}{b}{p_{2}} \sleq \pchoice{(c \ccomp q_{1})}{b}{(c \ccomp q_{2})} \sequiv c \ccomp (\pchoice{q_{1}}{b}{q_{2}})\]
  \textsc{monotone}: apply \Cref{prop:predicate-state-actions}.
  \[p_{1} \sleq p_{2} \sleq c \ccomp q_{2} \sleq c \ccomp q_{1}\]
  \textsc{bot}: use the hypothesis.
  \[\bot \sleq c \ccomp q\]
  \textsc{constancy}: use \Cref{cor:assert-sample-interchange}.
  \begin{align*}
    & r \pand p && \\
    & \sequiv \assert{r} \ccomp p && \text{(\Cref{prop:predicate-state-actions})}\\
    & \sleq \assert{r} \ccomp c \ccomp q && \text{(Assumption)}\\
    & \sequiv c \ccomp \assert{r} \ccomp q && \text{(\Cref{cor:assert-sample-interchange})}\\
    & \sequiv c \ccomp (q \pand r) && \text{(\Cref{prop:predicate-state-actions})}
  \end{align*}
  \textsc{auxiliary variable}.
  \begin{align*}
    & (\pmute{x_{i_{1}}, \dots, x_{i_{k}}}{p}{t}) && \\
    & \sequiv (\samp{x_{i_{1}}, \dots, x_{i_{k}}}{t}) \ccomp p && \text{(\Cref{prop:predicate-state-actions})} \\
    & \sleq (\samp{x_{i_{1}}, \dots, x_{i_{k}}}{t}) \ccomp c \ccomp q  &&  \text{(Assumption)} \\
    & \sequiv c \ccomp (\samp{x_{i_{1}}, \dots, x_{i_{k}}}{t}) \ccomp q &&  \text{(\Cref{cor:assert-sample-interchange})} \\
    & \sequiv \pmute{x_{i_{1}}, \dots, x_{i_{k}}}{p}{t} && \text{(\Cref{prop:predicate-state-actions})}
  \end{align*}
\end{proof}

\begin{example}[Weakest precondition reasoning]
  In the setting of \Cref{ex:state-incorrectness-rel}, a triple \(\triple{p}{c}{q}\) is valid whenever, if \(x \in \semn{p}\), then \(\semn{c}(x) \in \semn{q}\).
\end{example}

\begin{example}[Weakest pre-expectation reasoning]
  In the setting of \Cref{ex:assertion-correctness-stoch}, a triple \(\triple{p}{c}{q}\) is valid whenever \(\semn{p}(x) \leq \sum_{y \in X^{N}} \semn{c}(y \given x) \cdot q(y)\) for all \(x \in X^{N}\).
  The logic that we obtain reasons about weakest pre-expectations~\cite{batz2021relatively}.
\end{example}
\begin{toappendix}
\section{Relational program logics}%
\label{sec:relational-program-logics}

Relational program triples compare pairs of programs in a shared context.
They are a tuple of two commands, a precondition on the product of the input types and a postcondition on the product of the output types.
As for (not relational) program triples, the validity of relational program triples can be defined in terms of any of the inequalities in \Cref{fig:relational-program-triples}.
This time, \(p\) and \(q\) are predicates on a product type, \(s\) and \(t\) are states on a product type, and the commands need to be replaced by couplings of commands as one cannot assume that their effects are independent~\cite{barthe2009formal,barthe2019relational,avanzini2025quantitative}.
\begin{figure}[h!]
\begin{align*}
  & &&\text{State} && \text{Predicate} && \text{Assertion}\\
  & \text{Relational correctness} && \semn{s} \dcomp \synch{h} \pleq \semn{t} && \semn{p} \pleq \synch{h} \dcomp \semn{q} && \semn{\assert{p}} \dcomp \synch{h} \pleq \synch{h} \dcomp \semn{\assert{q}} \\
  & \text{Relational incorrectness} && \semn{s} \dcomp \synch{h} \pgeq \semn{t} && \semn{p} \pgeq \synch{h} \dcomp \semn{q} && \semn{\assert{p}} \dcomp \synch{h} \pgeq \synch{h} \dcomp \semn{\assert{q}}
\end{align*}
\caption{Inequalities that define validity of relational program triples \(\triple{p}{c \sim d}{q}\) or \(\triple{s}{c \sim d}{t}\), where \(\couple{h}{\semn{c}}{\semn{d}}\) is a coupling of the interpretations of the commands \(c\) and \(d\), and \(\synch{h} \defn h \dcomp \cproj_{X \tensor Y}\).}\label{fig:relational-program-triples}
\end{figure}

\begin{definition}\AP%
  \label{def:coupling}
  A \intro{coupling} of two morphisms, \(f_{1} \colon X_{1} \to Y_{1}\) and \(f_{2} \colon X_{2} \to Y_{2}\) in an \kl{imperative category}, is a morphism \(h \colon X_{1} \tensor X_{2} \to Y_{1} \tensor Y_{2} + Y_{1} + Y_{2}\) such that \(h \dcomp [\proj_{1}, \id, \zeromap] = \proj_{1} \dcomp f_{1}\) and \(h \dcomp [\proj_{2}, \zeromap, \id] = \proj_{2} \dcomp f_{2}\), where \([\proj_{1}, \id, \zeromap]\) indicates the copairing of the first projection \(\proj_{1} \colon X_{1} \tensor X_{2} \to X_{1}\), the identity \(\id_{X_{1}}\) and the zero morphism \(\zeromap \colon X_{2} \to X_{1}\).
  A \emph{strong coupling} is a coupling of the form \(h \dcomp \inject_{Y_{1} \tensor Y_{2}}\), where \(\inject_{Y_{1} \tensor Y_{2}} \colon Y_{1} \tensor Y_{2} \to Y_{1} \tensor Y_{2} + Y_{1} + Y_{2}\) denotes the coproduct injection.

  We write that \(h\) is a coupling of \(f_{1}\) and \(f_{2}\) as \(\couple{h}{f_{1}}{f_{2}}\).
  Given a coupling \(h \colon X_{1} \tensor X_{2} \to Y_{1} \tensor Y_{2} + Y_{1} + Y_{2}\), define \(\synch{h} \colon X_{1} \tensor X_{2} \to Y_{1} \tensor Y_{2} \) by postcomposing with the maps to the zero object, \intro[coupling synchronization]{\(\synch{h} = h \dcomp \cproj_{Y_{1} \tensor Y_{2}}\)}.
\end{definition}

\begin{remark}\label{rem:star-couplings}
  We spell out the definition of coupling for states in \(\subStoch\) to show that, in this case, our definition of coupling coincides with the definition of \(\star\)-coupling for subdistributions~\cite{avanzini2025quantitative}.
  Two states \(s \colon 1 \to X\) and \(t \colon 1 \to Y\) in \(\subStoch\) are two subdistributions \(s \in \subdistr(X)\) and \(t \in \subdistr(Y)\).
  A coupling of \(s\) and \(t\) is a subdistribution \(u \colon 1 \to X \times Y + X + Y\) such that \(s(x) = \sum_{y \in Y} u(x,y) + u(x,\star)\) and \(t(x) = \sum_{x \in X} u(x,y) + u(\star,y)\), where \((x,\star)\) denotes the element \(x\) in the second component of the coproduct, and \((\star,y)\) denotes the element \(y\) in the third component of the coproduct.
  A subdistribution on \(X \times Y + X + Y\) is the same as a distribution on \(X \times Y + X + Y + 1\), thus couplings of states in \(\subStoch\) coincide with \(\star\)-couplings of subdistributions~\cite{avanzini2025quantitative}.
  Similarly, strong couplings coincide with (total) couplings of subdistributions~\cite{barthe2009formal,avanzini2025quantitative}.
\end{remark}

Strong couplings enforce the same termination behaviour as total couplings of subdistributions do~\cite{avanzini2025quantitative}.
If \(\couple{h}{f_{1}}{f_{2}}\) is a strong coupling, \((f_{1} \dcomp \discard) \tensor \discard = h \dcomp (\discard \tensor \discard) = \discard \tensor (f_{2} \dcomp \discard)\), where \(f_{i} \dcomp \discard\) gives the termination predicate of \(f_{i}\).

\begin{remark}\label{rem:deterministic-couplings}
  When all morphisms are deterministic, then strong couplings trivialise: all strong couplings of \(f\) and \(g\) need to be \(f \tensor g\).
  This is the case of the category \(\Par\) of sets and partial functions.
  For this reason, couplings were only introduced with probabilistic examples of relational program logics~\cite{barthe2009formal,avanzini2025quantitative}.
\end{remark}

Sometimes couplings can be constructed only if the inputs satisfy some property.
This is also done in~\cite{avanzini2025quantitative}: the triples are of the form \(\triple{P \mid \phi}{c \sim d}{Q \mid \psi}\), whose meaning is that, when the inputs satisfy \(P\), then there is a coupling \(\couple{h}{c}{d}\) and the triple \(\triple{P \pand \phi}{\synch{h}}{Q \pand \psi}\) is valid.
The next definition generalises these couplings to our setting.

\begin{definition}
  A \intro{coupling with precondition}, \(p \colon X_{1} \tensor X_{2} \to 1\), of two morphisms, \(f_{1} \colon X_{1} \to Y_{1}\) and \(f_{2} \colon X_{2} \to Y_{2}\) in an \kl{imperative category}, is a morphism \(h \colon X_{1} \tensor X_{2} \to Y_{1} \tensor Y_{2} + Y_{1} + Y_{2}\) such that \(\assert{p} \dcomp h \dcomp [\proj_{1}, \id, \zeromap] = \assert{p} \dcomp \proj_{1} \dcomp f_{1}\) and \(\assert{p} \dcomp h \dcomp [\proj_{2}, \zeromap, \id] = \assert{p} \dcomp \proj_{2} \dcomp f_{2}\).
  In this case, we write \(\pcouple{p}{h}{f_{1}}{f_{2}}\).
\end{definition}

Note that \kl{couplings} are \kl{couplings with precondition} \(\ptrue\), the \emph{true} predicate.

\begin{definition}
  A \intro{relational semantics} for the monoidal guarded command language over the signatures \((\esig, \gsig, \psig, \ssig, \csig)\) is a tuple, \((\cat{C}, X_{1},x_{2}, N_{1},N_{2}, \gsemn{-})\), of an \kl{imperative posetal bicategory}, \(\cat{C}\), two objects in it, \(X_{1}\) and \(X_{2}\), two natural numbers, \(N_{1}, N_{2} \in \naturals\), and a function, \(\gsemn{-}\) as described in \(\Cref{sec:monoidal-gcl}\).
\end{definition}

Predicates and states in relational program logics need to express properties about pairs of programs.
Thus, they need a few extra constructs to express properties of joint executions: for two guards, \(b_{1} \colon X_{1} \to 1 +1\) and \(b_{2} \colon X_{2} \to 1 +1\), we denote with \(b_{1} = b_{2}\) the predicate on \(X_{1} \tensor X_{2}\) that succeeds when \(b_{1}\) and \(b_{2}\) are both true or both false, and fails otherwise;
we use \(\pred{b_{1}} \tensor \pred{b_{2}}\) to denote the predicate on \(X_{1} \tensor X_{2}\) obtained as the monoidal product of \(\pred{b_{1}} \colon X_{1} \to 1\) and \(\pred{b_{2}} \colon X_{2} \to 1\);
for a predicate \(p \colon X_{1} \tensor X_{2} \to 1\), we indicate with \(\swap \ccomp p \colon X_{2} \tensor X_{1} \to 1\) the predicate obtained by permuting its inputs;
the simultaneous substitution of variables in a predicate \(p \colon X_{1} \tensor X_{2} \to 1\) is indicated with \(\psub{p}{(x_{i_{1}}, x_{i_{2}})}{(e_{1},e_{2})}\), while substitution on one side only with \(\psub{p}{(x_{i}, -)}{(e, -)}\), \(\swap \ccomp p\);
an analogous notation is used for states.

\subsection{The algebra of couplings}

We study the algebra of couplings so that we can more easily derive the rules of relational program logics.
Couplings can be composed.

\begin{lemma}%
  \label{lem:coupling-tensor-totals}
  For two total commands, \(c\) and \(d\), there is always a \kl{coupling} of them, \(\couple{(\semn{c} \tensor \semn{d}) + \initmap + \initmap}{c}{d}\).
  We denote this operation as\intro[coupling tensor]{} \(c \couptensor d \defn (\semn{c} \tensor \semn{d}) + \initmap + \initmap\).
  In particular, there is always a \kl{coupling} of identities, \(\couple{(\id \couptensor \id)}{\noop}{\noop}\), and there is always a coupling of assignments, \(\couple{(\semn{\assign{x_{i_1}}{e_1}} \couptensor \semn{\assign{x_{i_2}}{e_2}})}{\assign{x_{i_1}}{e_1}}{\assign{x_{i_2}}{e_2}}\).
\end{lemma}
\begin{proof}
  By naturality of the morphisms \(\initmap\) and by totality of \(d\), we obtain:
  \begin{align*}
    & (c \couptensor d) \dcomp [\proj_{1}, \id, \zeromap] \defn ((\semn{c} \tensor \semn{d}) + \initmap + \initmap) \dcomp [\proj_{1}, \id, \zeromap] = (\semn{c} \tensor \semn{d}) \dcomp \proj_{1} = \proj_{1} \dcomp \semn{c}.
  \end{align*}
  Symmetrically, we obtain that \((c \couptensor d) \dcomp [\proj_{2}, \zeromap, \id] = \proj_{2} \dcomp \semn{d}\).
\end{proof}

\begin{lemma}%
  \label{lem:coupling-composition}
  Suppose that there are two \kl{couplings} of commands, \(\couple{h_1}{c_1}{d_1}\) and \(\couple{h_2}{c_2}{d_2}\); then there is a \kl{coupling} of the compositions, \(\couple{(h_1 \dcomp \copairing{h_2}{\initmap + d_1 + d_2})}{(c_1 \ccomp c_2)}{(d_1 \ccomp d_2)}\).
  We denote this operation as\intro[coupling composition]{} \(h_{1} \coupcomp h_{2} \defn h_1 \dcomp \copairing{h_2}{\initmap + d_1 + d_2}\).
\end{lemma}
\begin{proof}
  Let \(\couple{u}{f_{1}}{f_{2}}\) and \(\couple{v}{g_{1}}{g_{2}}\).
  Their composition is defined below.
  \begin{align*}
    u \coupcomp v & \defn \figCouplingCompositionDef
  \end{align*}
  We check that this defines a coupling.
  \begin{align*}
    & \figCouplingCompositionWellDefProofOne = \figCouplingCompositionWellDefProofTwo \\
    & = \figCouplingCompositionWellDefProofThree = \figCouplingCompositionWellDefProofFour \\
    & = \figCouplingCompositionWellDefProofFive
  \end{align*}
\end{proof}

\begin{proposition}%
  \label{prop:coupling-category}
  For two \kl{couplings} of commands, \(\couple{h_1}{c_1}{d_1}\) and \(\couple{h_2}{c_2}{d_2}\), there is a coupling of their composition \(\couple{(h_{1} \coupcomp h_{2})}{(c_{1} \ccomp c_{2})}{(d_{1} \ccomp c_{2})}\) that satisfies \(\synch{(h_{1} \coupcomp h_{2})} = \synch{h_{1}} \dcomp \synch{h_{2}}\).
  The operation \((\coupcomp)\) is associative and unital; we indicate its unit with\intro[identity coupling]{} \(\idcoup\), which satisfies \(\synch{(\idcoup)} = \id\).
\end{proposition}
\begin{proof}
  \Cref{lem:coupling-composition} gives the definition of \kl{coupling composition}; \Cref{lem:coupling-tensor-totals} gives the definition of identity coupling, \(\idcoup \defn \id \couptensor \id\).
  \begin{align*}
    & \synch{(h_{1} \coupcomp h_{2})} && \synch{(\idcoup)} \\
    & \defn h_1 \dcomp \copairing{h_2}{\initmap + d_1 + d_2} \dcomp \cproj_{1} && \defn ((\id \tensor \id) + \initmap + \initmap) \dcomp \cproj_{1} \\
    & = h_1 \dcomp \copairing{h_2 \dcomp \cproj_{1}}{\zeromap} && = \id \tensor \id \\
    & = h_{1} \dcomp \cproj_{1} \dcomp h_{2} \dcomp \cproj_{1} && \\
    & \codefn \synch{h_{1}} \dcomp \synch{h_{2}} &&
  \end{align*}
  We check associativity of composition.
  \begin{align*}
    \figCouplingCompositionAssociativeLeftProof
    & = \figCouplingCompositionAssociativeMidProof \\
    & = \figCouplingCompositionAssociativeRightProof
  \end{align*}
  We check unitality of composition.
  \begin{align*}
    \figCouplingCompositionUnitalLeftProof & = \figCouplingCompositionUnitalLeftProofOne = \figCouplingCompositionUnitalRightProof
  \end{align*}
\end{proof}

Couplings of conditional statements can be assembled from couplings of the two branches.

\begin{proposition}%
  \label{prop:symmetry-coupling}
  Suppose that there is a \kl{coupling} of two commands, \(\couple{h}{c}{d}\).
  Then, there is a coupling of the same commands in reverse order, \(\couple{\swap \dcomp h \dcomp (\swap + \cswap)}{d}{c}\).
\end{proposition}
\begin{proof}
  This is easily checked with diagrams.
\end{proof}

\begin{lemma}%
  \label{lem:coupling-choice}
  Suppose that there are four \kl{couplings} of commands, \(\couple{h}{c_1}{c_{2}}\), \(\couple{g}{d_{1}}{d_{2}}\), \(\couple{f_{1}}{c_{1}}{d_{2}}\) and \(\couple{f_2}{d_{1}}{c_2}\); suppose moreover that there is a \kl{coupling} of guards, \(\couple{a}{b_{1}}{b_{2}}\).
  Then, there is a \kl{coupling} of the choice statements,
  \[\couple{a \dcomp [h, f_{1}, f_{2}, g, (\initmap + c_{1} + \initmap), (\initmap + d_{1} + \initmap), (\initmap + \initmap + c_{2}), (\initmap + \initmap + d_{2})]}{(\cchoice{c_{1}}{b_{1}}{d_{1}})}{(\cchoice{c_{2}}{b_{2}}{d_{2}})}.\]
\end{lemma}
\begin{proof}
  The equations defining couplings follow from the assumptions and from naturality of the monoid structure of coproducts.
\end{proof}

\begin{proposition}%
  \label{prop:coupling-total-choice}
  Suppose that there are \kl{couplings} of commands \(\couple{h}{c_1}{c_{2}}\), \(\couple{g}{d_{1}}{d_{2}}\), \(\couple{f_{1}}{c_{1}}{d_{2}}\) and \(\couple{f_2}{d_{1}}{c_2}\); suppose moreover that \(b_{1}\) and \(b_{2}\) are \kl{total} guards.
  Then, there is a \kl{coupling} of the choice statements, \(\couple{(b_{1} \tensor b_{2}) \dcomp [h, f_{1}, f_{2}, g]}{(\cchoice{c_{1}}{b_{1}}{d_{1}})}{(\cchoice{c_{2}}{b_{2}}{d_{2}})}\).
  This \kl{coupling} satisfies \(\synch{((b_{1} \tensor b_{2}) \dcomp [h, f_{1}, f_{2}, g])} = (b_{1} \tensor b_{2}) \dcomp [\synch{h}, \synch{f_{1}}, \synch{f_{2}}, \synch{g}]\).
\end{proposition}
\begin{proof}
  Apply \Cref{lem:coupling-choice} with the coupling of total morphisms in \Cref{lem:coupling-tensor-totals}.
\end{proof}

\begin{proposition}%
  \label{prop:precondition-coupling-total-choice}
  Suppose that there are \kl{couplings with precondition} of commands \(\pcouple{\pred{b_{1}} \tensor \pred{b_{2}}}{h}{c_1}{c_{2}}\) and \(\pcouple{\pred{(\gnot b_{1})} \tensor \pred{(\gnot b_{2})}}{g}{d_{1}}{d_{2}}\); suppose moreover that \(b_{1}\) and \(b_{2}\) are \kl{total} and \kl{deterministic} guards.
  Then, there is a \kl{coupling with precondition} \((b_{1} = b_{2})\) of the choice statements, \(\pcouple{b_{1} = b_{2}}{(b_{1} \tensor b_{2}) \dcomp [h, \zeromap, \zeromap, g]}{(\cchoice{c_{1}}{b_{1}}{d_{1}})}{(\cchoice{c_{2}}{b_{2}}{d_{2}})}\).
  This \kl{coupling with precondition} satisfies \(\synch{((b_{1} \tensor b_{2}) \dcomp [h, \zeromap, \zeromap, g])} = (b_{1} \tensor b_{2}) \dcomp [\synch{h}, \zeromap, \zeromap, \synch{g}]\).
\end{proposition}
\begin{proof}
  The equations defining couplings follow from the assumptions and from naturality of the monoid structure of coproducts.
  By determinism of the guards, we may disregard the middle branches where \(b_{1}\) and \(b_{2}\) disagree.
\end{proof}

Similarly, couplings of while statements can be assembled from couplings of the loop body.

\begin{proposition}%
  \label{prop:coupling-while}
  Suppose that there is a \kl{coupling with precondition} of commands \(\pcouple{\pred{b_{1}} \tensor \pred{b_{2}}}{h}{c_1}{c_{2}}\); suppose moreover that \(b_{1}\) and \(b_{2}\) are \kl{total} and \kl{deterministic} guards.
  Then, there is a \kl{coupling with precondition} \((b_{1} = b_{2})\) of the while statements, \(\pcouple{b_{1} = b_{2}}{\trace(\join \dcomp (b_{1} \tensor b_{2}) \dcomp (h + \finmap + \finmap + \id)) \dcomp \swap_{(1,2,3)} }{(\while{b_{1}}{c_{1}})}{(\while{b_{2}}{c_{2}})}\).
  This \kl{coupling with precondition} satisfies \(\synch{(\trace(\join \dcomp (b_{1} \tensor b_{2}) \dcomp (h + \finmap + \finmap + \id)) \dcomp \swap_{(1,2,3)})} = \trace(\join \dcomp (b_{1} \tensor b_{2}) \dcomp (\synch{h} + \finmap + \finmap + \id))\).
\end{proposition}
\begin{proof}
  The proof relies on uniformity of the trace and, crucially, on determinism of the guards.
\end{proof}

\begin{proposition}%
  \label{prop:coupling-free-modified-variables}
  Couplings preserve free and modified variables: for two commands, \(c\) and \(d\), there is always a coupling, \(\couple{h}{c}{d}\), satisfying \(\modv{h} = \modv{c} \union \modv{d}\) and \(\freev{h} = \freev{c} \union \freev{d}\).
\end{proposition}
\begin{proof}
  The proof uses the distributive structure; the free and modified variables give a factorization of the commands \(c\) and \(d\), which carries over to the coupling.
\end{proof}

\subsection{Relational correctness triples}
This section considers \kl{relational assertion-correctness triples}.
In the category \(\Par\) of sets and partial functions, these correspond to relational Hoare triples~\cite{benton2004relational}.

\begin{definition}
  A \intro{relational assertion-correctness triple}, \(\triple{p}{c \sim c'}{q}\), consists of two \kl{commands}, \(c\) and \(c'\), and two \kl{predicates} on \(N_{1} + N_{2}\) variables, \(p\) and \(q\).
  A \kl{relational assertion-correctness triple} is \intro[valid relational assertion-correctness triple]{valid} in a \kl{relational semantics}, \((\cat{C}, X_{1},X_{2}, N_{1}, N_{2}, \gsemn{-})\), if there exists a coupling, \(\couple{h}{\semn{c}}{\semn{c'}}\), satisfying \(\semn{\assert p} \dcomp \synch{h} \pleq \synch{h} \dcomp \semn{\assert{q}}\).
\end{definition}

\begin{theorem}%
  [Soundness]%
  \label{thm:relationalcorrectness}
  The following are \kl{valid relational assertion-correctness triples} in any \kl{imperative posetal bicategory} where \(\zeromap_{A,B} \pleq f\) for all morphisms \(f \colon A \to B\).
  \begin{gather*}
    \inferrule[skip]
      { }
      { \triple{p}{\noop \sim \noop}{p} }
    \qquad
    \inferrule[comp]
      { \triple{p}{c_1 \sim d_{1}}{q} \\ \triple{q}{c_2 \sim d_{2}}{r} }
      { \triple{p}{(c_1 \ccomp c_2) \sim (d_{1} \ccomp d_{2})}{r} }
    \\
    \inferrule[assign]
      { }
      { \triple{ \psub{p}{(x_{i_{1}}, x_{i_{2}})}{(e_{1}, e_{2})}}{ (\assign{x_{i_{1}}}{e_{1}}) \sim (\assign{x_{i_{2}}}{e_{2}})}{ p }}
    \\
    \inferrule[choice]
      { \triple{p}{c_1 \sim c_{2}}{q} \\
        \triple{p}{c_1 \sim d_{2}}{q} \\
        \triple{p}{d_1 \sim c_{2}}{q} \\
        \triple{p}{d_1 \sim d_{2}}{q} \\
        \semn{b_{1}}, \semn{b_{2}} \mbox{ total}
      }
      { \triple{p}{(\cchoice{c_1}{b_{1}}{d_1}) \sim (\cchoice{c_2}{b_{2}}{d_2})}{q}
      }
    \\
    \inferrule[ifelse]
      { \semn{b_{1}}, \semn{b_{2}} \mbox{ deterministic and total} \\
        \triple{ \pred{b_{1}} \tensor \pred{b_{2}} \mid p }{ c_{1} \sim c_{2} }{ q } \\
        \triple{ \pred{(\gnot b_{1})} \tensor \pred{(\gnot b_{2})} \mid p  }{ d_{1} \sim d_{2} }{ q }}
      { \triple{b_{1} = b_{2} \mid p}{(\cchoice{c_{1}}{b_{1}}{d_{1}}) \sim (\cchoice{c_{2}}{b_{2}}{d_{2}})}{q} }
      \\
    \inferrule[while]
      { \triple{ \pred{b_{1}} \tensor \pred{b_{2}} \mid p }{c_{1} \sim c_{2}}{ b_{1} = b_{2} \mid p} \\ \semn{b_{1}}, \semn{b_{2}} \text{ deterministic and total}}
      { \triple{b_{1} = b_{2} \mid p}{ (\while{b_{1}}{c_{1}}) \sim (\while{b_{2}}{c_{2}}) }{ \pred{(\gnot b_{1})} \tensor \pred{(\gnot b_{2})} \mid p } }
  \end{gather*}
  \begin{gather*}
    \inferrule[monotone]
      { \semn{p_1} \pleq \semn{p_2} \\ \triple{ p_{2}}{c \sim d}{q_{2}} \\ \semn{q_2} \pleq \semn{q_1} }
      { \triple{p_{1}}{c \sim d}{q_{1}} }
    \qquad
    \inferrule[symm]
      { \triple{ p }{c \sim d}{ q } }
      { \triple{ \swap \ccomp p }{d \sim c}{ \swap \ccomp q } }
  \end{gather*}
  \begin{gather*}
    \inferrule[assign-L]
      {  }
      { \triple{ \psub{p}{(x_{i}, -)}{(e, -)} }{(\assign{x_{i}}{e}) \sim \noop}{p} }
    \qquad
    \inferrule[choice-L]
      { \triple{p}{c \sim \noop}{q} \\
        \triple{p}{d \sim \noop}{q} \\
        \semn{b} \mbox{ total}
      }
      { \triple{p}{(\cchoice{c}{b}{d}) \sim \noop}{q}
      }
    \\
    \inferrule[ifelse-L]
    { \semn{b_{1}}, \semn{b_{2}} \mbox{ deterministic and total}\\
      \triple{ \pred{b_{1}} \tensor \pred{b_{2}} \mid p }{ c \sim \noop }{ q } \\
        \triple{ \pred{(\gnot b_{1})} \tensor \pred{(\gnot b_{2})} \mid p  }{ d \sim \noop }{ q }
        }
      { \triple{p}{(\cchoice{c}{b_{1}}{d}) \sim \noop}{q} }
    \\
    \inferrule[constancy]
    { \triple{p}{c \sim d}{q} \\ (\modv{c} \union \modv{d}) \intersection \freev{r} = \emptyset }
    { \triple{ p \pand r }{c \sim d}{ q \pand r } }
  \end{gather*}
\end{theorem}
\begin{proof}
  \textsc{skip}: by \Cref{prop:coupling-category}, there is an identity coupling \(\idcoup\) satisfying \(\synch{(\idcoup)} = \id\).
  \begin{align*}
    & \semn{\assert{p}} \dcomp \synch{(\idcoup)} \\
    & = \semn{\assert{p}} \dcomp \id \\
    & = \id \dcomp \semn{\assert{p}} \\
    & = \synch{(\idcoup)} \dcomp \semn{\assert{p}}
  \end{align*}
  \textsc{comp}: by the assumptions, there are couplings \(\couple{g_{1}}{c_{1}}{d_{1}}\) and \(\couple{g_{2}}{c_{2}}{d_{2}}\) satisfying \(\semn{\assert{p}} \dcomp \synch{g_{1}} \leq \synch{g_{1}} \dcomp \semn{\assert{q}}\) and \(\semn{\assert{q}} \dcomp \synch{g_{2}} \leq \synch{g_{2}} \dcomp \semn{\assert{r}}\).
  By \Cref{prop:coupling-category}, there is a coupling \(\couple{(g_{1} \coupcomp g_{2})}{(c_{1} \dcomp c_{2})}{(d_{1} \dcomp d_{2})}\) satisfying \(\synch{(g_{1} \coupcomp g_{2})} = \synch{g_{1}} \dcomp \synch{g_{2}}\).
  \begin{align*}
    & \semn{\assert{p}} \dcomp \synch{(g_{1} \coupcomp g_{2})} \\
    & = \semn{\assert{p}} \dcomp \synch{g_{1}} \dcomp \synch{g_{2}} \\
    & \leq \synch{g_{1}} \dcomp \semn{\assert{q}} \dcomp  \synch{g_{2}} \\
    & \leq \synch{g_{1}} \dcomp \synch{g_{2}} \dcomp \semn{\assert{r}} \\
    & = \synch{(g_{1} \coupcomp g_{2})} \dcomp \semn{\assert{r}}
  \end{align*}

  \textsc{assign}: by \Cref{lem:coupling-tensor-totals}, there is a \kl{coupling} of assignments, \(\couple{(\assign{x_{i_1}}{e_{1}} \couptensor \assign{x_{i_2}}{e_{2}})}{(\assign{x_{i_1}}{e_{1}})}{(\assign{x_{i_2}}{e_{2}})}\).
  This coupling satisfies the triple by determinism of \(e_{1}\) and \(e_{2}\).
  \begin{align*}
    & \semn{ \assert{ (\psub{p}{(x_{i_{1}}, x_{i_{2}})}{(e_{1}, e_{2})})}} \dcomp \synch{(\assign{x_{i_1}}{e_{1}} \couptensor \assign{x_{i_2}}{e_{2}})} \\
    & = \semn{ \assert{ (\psub{p}{(x_{i_{1}}, x_{i_{2}})}{(e_{1}, e_{2})})}} \dcomp (\semn{\assign{x_{i_1}}{e_{1}}} \tensor \semn{\assign{x_{i_2}}{e_{2}}})\\
    & = \cp \dcomp ((\semn{\assign{x_{i_1}}{e_{1}}} \tensor \semn{\assign{x_{i_2}}{e_{2}}}) \tensor ((\semn{\assign{x_{i_1}}{e_{1}}} \tensor \semn{\assign{x_{i_2}}{e_{2}}}) \dcomp \semn{p}))\\
    & = (\semn{\assign{x_{i_1}}{e_{1}}} \tensor \semn{\assign{x_{i_2}}{e_{2}}}) \dcomp \semn{\assert{p}} \\
    & = \synch{(\assign{x_{i_1}}{e_{1}} \couptensor \assign{x_{i_2}}{e_{2}})} \dcomp \semn{\assert{p}}
  \end{align*}

  \textsc{choice}.
  The assumptions give us couplings \(\couple{h}{c_1}{c_{2}}\), \(\couple{g}{d_{1}}{d_{2}}\), \(\couple{f_{1}}{c_{1}}{d_{2}}\) and \(\couple{f_2}{d_{1}}{c_2}\).
  \Cref{prop:coupling-total-choice} gives us a coupling of the conditional statements \(\couple{(b_{1} \tensor b_{2}) \dcomp [h, f_{1}, f_{2}, g]}{(\cchoice{c_{1}}{b_{1}}{d_{1}})}{(\cchoice{c_{2}}{b_{2}}{d_{2}})}\) satisfying \(\synch{((b_{1} \tensor b_{2}) \dcomp [h, f_{1}, f_{2}, g])} = (b_{1} \tensor b_{2}) \dcomp [\synch{h}, \synch{f_{1}}, \synch{f_{2}}, \synch{g}]\).
  \begin{align*}
    & \semn{\assert{p}} \dcomp \synch{(b_{1} \tensor b_{2}) \dcomp [h, f_{1}, f_{2}, g]} \\
    & = \semn{\assert{p}} \dcomp (b_{1} \tensor b_{2}) \dcomp [\synch{h}, \synch{f_{1}}, \synch{f_{2}}, \synch{g}] \\
    & = (b_{1} \tensor b_{2}) \dcomp [\semn{\assert{p}} \dcomp \synch{h}, \semn{\assert{p}} \dcomp \synch{f_{1}}, \semn{\assert{p}} \dcomp \synch{f_{2}}, \semn{\assert{p}} \dcomp \synch{g}] \\
    & \leq (b_{1} \tensor b_{2}) \dcomp [\synch{h} \dcomp \semn{\assert{q}}, \synch{f_{1}} \dcomp \semn{\assert{q}}, \synch{f_{2}} \dcomp \semn{\assert{q}}, \synch{g} \dcomp \semn{\assert{q}}] \\
    & = (b_{1} \tensor b_{2}) \dcomp [\synch{h}, \synch{f_{1}}, \synch{f_{2}}, \synch{g}] \dcomp \semn{\assert{q}} \\
    & = \synch{((b_{1} \tensor b_{2}) \dcomp [h, f_{1}, f_{2}, g])} \dcomp \semn{\assert{q}}
  \end{align*}

  \textsc{ifelse}.
  The assumptions give us \kl{couplings with preconditions} \(\pcouple{\pred{b_{1}} \tensor \pred{b_{2}}}{h}{c_1}{c_{2}}\) and \(\pcouple{\pred{(\gnot b_{1})} \tensor \pred{(\gnot b_{2})}}{g}{d_{1}}{d_{2}}\) satisfying \(\semn{\assert{((\pred{b_{1}} \tensor \pred{b_{2}}) \pand p)}} \dcomp \synch{h} \leq \synch{h} \dcomp \semn{\assert{q}}\) and \(\semn{\assert{((\pred{(\gnot b_{1})} \tensor \pred{(\gnot b_{2})}) \pand p)}} \dcomp \synch{g} \leq \synch{g} \dcomp \semn{\assert{q}}\).
  \Cref{prop:precondition-coupling-total-choice} gives us a \kl{coupling with precondition} \((b_{1} = b_{2})\) of the choice statements, \(\pcouple{b_{1} = b_{2}}{(b_{1} \tensor b_{2}) \dcomp [h, \zeromap, \zeromap, g]}{(\cchoice{c_{1}}{b_{1}}{d_{1}})}{(\cchoice{c_{2}}{b_{2}}{d_{2}})}\), satisfying \(\synch{((b_{1} \tensor b_{2}) \dcomp [h, \zeromap, \zeromap, g])} = (b_{1} \tensor b_{2}) \dcomp [\synch{h}, \zeromap, \zeromap, \synch{g}]\).
  \begin{align*}
    & \semn{\assert{((b_{1} = b_{2}) \pand p)}} \dcomp \synch{((b_{1} \tensor b_{2}) \dcomp [h, \zeromap, \zeromap, g])} \\
    & = \semn{\assert{((b_{1} = b_{2}) \pand p)}} \dcomp (b_{1} \tensor b_{2}) \dcomp [\synch{h}, \zeromap, \zeromap, \synch{g}] \\
    & = (b_{1} \tensor b_{2}) \dcomp [\semn{\assert{((\pred{b_{1}} \tensor \pred{b_{2}}) \pand p)}} \dcomp \synch{h}, \zeromap, \zeromap, \semn{\assert{((\pred{(\gnot b_{1})} \tensor \pred{(\gnot b_{2})}) \pand p)}} \dcomp \synch{g}] \\
    & \leq (b_{1} \tensor b_{2}) \dcomp [\synch{h} \dcomp \semn{\assert{q}}, \zeromap, \zeromap, \synch{g} \dcomp \semn{\assert{q}}] \\
    & = (b_{1} \tensor b_{2}) \dcomp [\synch{h}, \zeromap, \zeromap, \synch{g}] \dcomp \semn{\assert{q}} \\
    & = \synch{((b_{1} \tensor b_{2}) \dcomp [h, \zeromap, \zeromap, g])} \dcomp \semn{\assert{q}}
  \end{align*}

  \textsc{while}: apply \Cref{prop:coupling-while}.

  \textsc{monotone}.
  Let \(\couple{h}{c}{d}\) be the coupling given by the assumption.
  \[\semn{\assert{p_{1}}} \dcomp \synch{h} \pleq \semn{\assert{p_{2}}} \dcomp \synch{h} \pleq \synch{h} \dcomp \semn{\assert{q_{2}}} \pleq \synch{h} \dcomp \semn{\assert{q_{1}}}\]

  \textsc{symm}.
  Let \(\couple{h}{c}{d}\) be the coupling given by the assumption.
  By \Cref{prop:symmetry-coupling}, \(\couple{(\swap \dcomp h \dcomp (\swap + \cswap))}{d}{c}\) and this satisfies the desired inequality.
  \begin{align*}
    & \semn{\assert{(\swap \ccomp p)}} \dcomp \synch{(\swap \dcomp h \dcomp (\swap + \cswap))} && \\
    & = \semn{\assert{(\swap \ccomp p)}} \dcomp \swap \dcomp \synch{h} \dcomp \swap && \text{(\Cref{def:coupling})}\\
    & = \swap \dcomp \semn{\assert{p}} \dcomp \synch{h} \dcomp \swap && \text{(Deterministic symmetries)} \\
    & \pleq \swap \dcomp \synch{h} \dcomp \semn{\assert{q}} \dcomp \swap && \text{(Assumption)} \\
    & = \swap \dcomp \synch{h} \dcomp \swap \dcomp \semn{\assert{(\swap \ccomp q)}} && \text{(Deterministic symmetries)} \\
    &\synch{(\swap \ccomp h \ccomp (\swap + \cswap))} \dcomp \semn{\assert{(\swap \ccomp q)}} &&
  \end{align*}

  The one-sided rules are particular instances of the two sided rules: (\textsc{assign-L}) take the expression \(e_{2}\) to be the variable \(x_{i_2}\); (\textsc{choice-L}, \textsc{ifelse-L}) take the commands \(c_{2}\) and \(d_{2}\) to be \(\noop\).

  \textsc{constancy}.
  Let \(\couple{h}{c}{d}\) be the coupling given by the assumption.
  By \Cref{prop:coupling-free-modified-variables}, we may assume that \(\modv{h} = \modv{c} \union \modv{d}\).
  \begin{align*}
    & \semn{\assert{(p \pand r)}} \dcomp \synch{h} && \\
    & = \semn{\assert{p}} \dcomp \semn{\assert{r}} \dcomp \synch{h} && \text{(\Cref{lem:assert-monoid-morph})}\\
    & = \semn{\assert{p}} \dcomp \synch{h} \dcomp \semn{\assert{r}}  && \text{(\Cref{cor:assert-sample-interchange})} \\
    & \pleq \synch{h} \dcomp \semn{\assert{q}} \semn{\assert{r}}  && \text{(Assumption)} \\
    & = \synch{h} \dcomp \semn{\assert{(q \pand r)}}  && \text{(\Cref{lem:assert-monoid-morph})}
  \end{align*}
\end{proof}

\begin{example}
  Benton's work~\cite{benton2004relational} restricts to strong couplings, which simplify in the case of partial functions (\Cref{rem:deterministic-couplings}).
  The validity condition of a triple \(\triple{p}{c \sim c'}{q}\), thus, simplifies to \(\assert{p} \dcomp (c \tensor c') \leq (c \tensor c') \dcomp \assert{q}\).
  We present the rules in the general case to allow semantics different from partial functions.
\end{example}

\subsection{Relational incorrectness triples}

This section considers \kl{relational predicate-incorrectness triples}.
In the category \(\subStoch\) of sets and partial stochastic functions, these correspond to quantitative probabilistic relational Hoare triples~\cite{avanzini2025quantitative}.

\begin{definition}
  A \intro{relational predicate-incorrectness triple}, \(\triple{p}{c \sim c'}{q}\), consists of two \kl{commands}, \(c\) and \(c'\), and two \kl{predicates} on \(N_{1}+N_{2}\) variables, \(p\) and \(q\).
  A \kl{relational predicate-incorrectness triple} is \intro[valid relational predicate-incorrectness triple]{valid} in a relational semantics, \((\cat{C}, X_{1},X_{2}, N_{1}, N_{2}, \gsemn{-})\), if there exists a coupling, \(\couple{h}{\semn{c}}{\semn{c'}}\), satisfying \(\semn{p} \pgeq \synch{h} \dcomp \semn{q}\).
\end{definition}

We derive the rules of relational predicate-incorrectness logic.
Compared to the rules of quantitative probabilistic relational Hoare logic~\cite{avanzini2025quantitative}, we do not assume that guards are deterministic, so we derive additional rules for nondeterministic choice and iteration.
The \textsc{strassen} rule of quantitative probabilistic relational Hoare logic~\cite{avanzini2025quantitative} is missing as it is a consequence of Strassen's theorem, a characterisation of couplings particular to subdistributions.

\begin{theorem}%
  [Soundness]%
  \label{thm:relationalincorrectness}
  The following are \kl{valid relational predicate-incorrectness triples} in any \kl{imperative posetal bicategory} where \(\zeromap_{A,B} \pleq f\) and \(f \dcomp \discard_{B} \pleq \discard_{A}\) for all morphisms \(f \colon A \to B\).
  \begin{gather*}
    \inferrule[skip]
      { }
      { \triple{p}{\noop \sim \noop}{p} }
    \qquad
    \inferrule[comp]
      { \triple{p}{c_1 \sim d_{1}}{q} \\ \triple{q}{c_2 \sim d_{2}}{r} }
      { \triple{p}{(c_1 \ccomp c_2) \sim (d_{1} \sim d_{2})}{r} }
    \\
    \inferrule[assign]
      { }
      { \triple{ \psub{p}{(x_{i_{1}}, x_{i_{2}})}{(e_{1}, e_{2})} }{ (\assign{x_{i_{1}}}{e_{1}}) \sim (\assign{x_{i_{2}}}{e_{2}})}{ p }}
    \\
    \inferrule[sample]
      { \couple{h}{s_{1}}{s_{2}} }
      { \triple{ \pmute{(x_{i_{1}}, \dots, x_{i_{k}}; y_{j_{1}}, \dots, y_{j_{l}})}{p}{\synch{h}} }{ (\samp{(x_{i_{1}}, \dots, x_{i_{k}})}{s_{1}}) \sim (\samp{(y_{j_{1}}, \dots, y_{j_{l}})}{s_{2}})}{ p }}
    \\
    \inferrule[choice]
      { \semn{b_{1}}, \semn{b_{2}} \text{ total}\\
        \triple{p}{c_1 \sim c_{2}}{q} \\
        \triple{p}{c_1 \sim d_{2}}{q} \\
        \triple{p}{d_1 \sim c_{2}}{q} \\
        \triple{p}{d_1 \sim d_{2}}{q}
      }
      { \triple{p}{(\cchoice{c_1}{b_{1}}{d_1}) \sim (\cchoice{c_2}{b_{2}}{d_2})}{q}
      }
    \\
    \inferrule[ifelse]
      { \semn{b_{1}}, \semn{b_{2}} \text{ deterministic and total}\\
        \triple{ \pred{b_{1}} \tensor \pred{b_{2}} \mid p }{ c_{1} \sim c_{2} }{ q } \\
        \triple{ \pred{(\gnot b_{1})} \tensor \pred{(\gnot b_{2})} \mid p  }{ d_{1} \sim d_{2} }{ q }}
      { \triple{b_{1} = b_{2} \mid p}{(\cchoice{c_{1}}{b_{1}}{d_{1}}) \sim (\cchoice{c_{2}}{b_{2}}{d_{2}})}{q} }
    \\
    \inferrule[while]
      { \triple{ \pred{b_{1}} \tensor \pred{b_{2}} \mid p }{c_{1} \sim c_{2}}{b_{1} = b_{2} \mid p} \\ \semn{b_{1}}, \semn{b_{2}} \text{ deterministic and total}}
      { \triple{b_{1} = b_{2} \mid p}{ (\while{b_{1}}{c_{1}}) \sim (\while{b_{2}}{c_{2}}) }{ \pred{(\gnot b_{1})} \tensor \pred{(\gnot b_{2})} \mid p } }
  \end{gather*}
  \begin{gather*}
    \inferrule[monotone]
      { \semn{p_1} \pgeq \semn{p_2} \\ \triple{ p_{2}}{c \sim d}{q_{2}} \\ \semn{q_2} \pgeq \semn{q_1} }
      { \triple{p_{1}}{c \sim d}{q_{1}} }
    \qquad
    \inferrule[symm]
      { \triple{ p }{c \sim d}{ q } }
      { \triple{ \swap \ccomp p }{d \sim c}{ \swap \ccomp q } }
    \\
    \inferrule[assign-L]
      {  }
      { \triple{ \psub{p}{(x_{i},-)}{(e,-)} }{(\assign{x_{i}}{e}) \sim \noop}{p} }
    \qquad
    \inferrule[sample-L]
      { \semn{t} \text{ total} }
      { \triple{ \pmute{(x_{i_{1}}, \dots, x_{i_{k}};-)}{p}{t} }{ (\samp{(x_{i_{1}}, \dots, x_{i_{k}})}{t}) \sim \noop}{ p }}
    \\
    \inferrule[choice-L]
      { \triple{p}{c \sim \noop}{q} \\
        \triple{p}{d \sim \noop}{q} \\
        \semn{b} \mbox{ total}
      }
      { \triple{p}{(\cchoice{c}{b}{d}) \sim \noop}{q}
      }
    \\
    \inferrule[ifelse-L]
    { \semn{b_{1}}, \semn{b_{2}} \mbox{ deterministic and total} \\
      \triple{ \pred{b_{1}} \tensor \pred{b_{1}} \mid p }{ c \sim \noop }{ q } \\
        \triple{ \pred{(\gnot b_{1})} \tensor \pred{(\gnot b_{2})} \mid p  }{ d \sim \noop }{ q }
        }
      { \triple{ b_{1} = b_{2} \mid p}{(\cchoice{c}{b}{d}) \sim \noop}{q} }
  \end{gather*}
  \begin{gather*}
    \inferrule[constancy]
    { \triple{p}{c \sim d}{q} \\ (\modv{c} \union \modv{d}) \intersection \freev{r} = \emptyset }
    { \triple{ r \pand p }{c \sim d}{ r \pand q } }
    \qquad
    \inferrule[auxiliary variable]
    { \triple{p}{c \sim d}{q} \\ x_{i_{1}}, \dots, x_{i_{k}} \notin \freev{c} \union \freev{d} }
    { \triple{ \pmute{x_{i_{1}}, \dots, x_{i_{k}}}{p}{t} }{c \sim d}{ \pmute{x_{i_{1}}, \dots, x_{i_{k}}}{q}{t} } }
  \end{gather*}
\end{theorem}
\begin{proof}
  \textsc{skip}.
  By \Cref{lem:coupling-tensor-totals}, the monoidal product gives a coupling: \(\couple{(\noop \couptensor \noop)}{\noop}{\noop}\).
  By unitality and \Cref{prop:coupling-category}, we obtain the rule.
  \[\semn{p} = (\id \tensor \id) \dcomp \semn{p} = \synch{(\noop \couptensor \noop)} \dcomp \semn{p} \]

  \textsc{comp}.
  Suppose there are couplings \(\couple{g_{1}}{c_{1}}{d_{1}}\) and \(\couple{g_{2}}{c_{2}}{d_{2}}\) satisfying \(p \geq \synch{g_{1}} \ccomp q\) and \(q \geq \synch{g_{2}} \ccomp r\).
  By \Cref{prop:coupling-category}, there is a coupling \(\couple{(g_{1} \coupcomp g_{2})}{(c_{1} \ccomp c_{2})}{(d_{1} \ccomp d_{2})}\) satisfying \(\synch{(g_{1} \coupcomp g_{2})} = \synch{g_{1}} \dcomp \synch{g_{2}}\).
  \begin{align*}
    & \synch{(g_{1} \coupcomp g_{2})} \dcomp \semn{r} && \\
    & = \synch{g_{1}} \dcomp \synch{g_{2}} \dcomp \semn{r} && \text{(\Cref{prop:coupling-category})} \\
    & \pleq \synch{g_{1}} \dcomp \semn{q} && \text{(Assumption)}\\
    & \pleq \semn{p} && \text{(Assumption)}
  \end{align*}

  \textsc{assign}.
  By \Cref{lem:coupling-tensor-totals}, the monoidal product gives a coupling: \(\couple{((\assign{x_{i_1}}{e_{1}}) \couptensor (\assign{x_{i_2}}{e_{2}}))}{(\assign{x_{i_1}}{e_{1}})}{(\assign{x_{i_2}}{e_{2}})}\).
  This coupling satisfies the triple by definition.
  \[\semn{\psub{(x_{i_1},x_{i_2})}{(e_{1},e_{2})}{p}} = \semn{(\assign{x_{i_1}}{e_{1}}) \tensor (\assign{x_{i_2}}{e_{2}})} \dcomp \semn{p} = \synch{((\assign{x_{i_1}}{e_{1}}) \couptensor (\assign{x_{i_2}}{e_{2}}))} \dcomp \semn{p}\]

  \textsc{sample}.
  Given a coupling \(\couple{h}{s_{1}}{s_{2}}\), the triple is satisfied by the definition of interpretation of sampling.
  \[\semn{\pmute{(x_{i_{1}}, \dots, x_{i_{k}}; y_{j_{1}}, \dots, y_{j_{l}})}{p}{\synch{h}}} = \synch{h} \dcomp \semn{p} \]

  \textsc{choice}.
  The assumptions give us couplings \(\couple{h}{c_1}{c_{2}}\), \(\couple{g}{d_{1}}{d_{2}}\), \(\couple{f_{1}}{c_{1}}{d_{2}}\) and \(\couple{f_2}{d_{1}}{c_2}\).
  \Cref{prop:coupling-total-choice} gives us a coupling of the conditional statements \(\couple{(b_{1} \tensor b_{2}) \dcomp [h, f_{1}, f_{2}, g]}{(\cchoice{c_{1}}{b_{1}}{d_{1}})}{(\cchoice{c_{2}}{b_{2}}{d_{2}})}\) satisfying \(\synch{((b_{1} \tensor b_{2}) \dcomp [h, f_{1}, f_{2}, g])} = (b_{1} \tensor b_{2}) \dcomp [\synch{h}, \synch{f_{1}}, \synch{f_{2}}, \synch{g}]\).
  \begin{align*}
    & \semn{p} \\
    & = (b_{1} \tensor b_{2}) \dcomp [\semn{p}, \semn{p}, \semn{p}, \semn{p}] \\
    & \pgeq (b_{1} \tensor b_{2}) \dcomp [\synch{h} \dcomp \semn{q}, \synch{f_{1}} \dcomp \semn{q}, \synch{f_{2}} \dcomp \semn{q}, \synch{g} \dcomp \semn{q}] \\
    & = (b_{1} \tensor b_{2}) \dcomp [\synch{h}, \synch{f_{1}}, \synch{f_{2}}, \synch{g}] \dcomp \semn{q} \\
    & = \synch{((b_{1} \tensor b_{2}) \dcomp [h, f_{1}, f_{2}, g])} \dcomp \semn{q}
  \end{align*}

  \textsc{ifelse}.
  The assumptions give us \kl{couplings with preconditions} \(\pcouple{\pred{b_{1}} \tensor \pred{b_{2}}}{h}{c_1}{c_{2}}\) and \(\pcouple{\pred{(\gnot b_{1})} \tensor \pred{(\gnot b_{2})}}{g}{d_{1}}{d_{2}}\) satisfying \(\semn{\assert{((\pred{b_{1}} \tensor \pred{b_{2}}) \pand p)}} \dcomp \synch{h} \leq \synch{h} \dcomp \semn{\assert{q}}\) and \(\semn{\assert{((\pred{(\gnot b_{1})} \tensor \pred{(\gnot b_{2})}) \pand p)}} \dcomp \synch{g} \leq \synch{g} \dcomp \semn{\assert{q}}\).
  \Cref{prop:precondition-coupling-total-choice} gives us a \kl{coupling with precondition} \((b_{1} = b_{2})\) of the choice statements, \(\pcouple{b_{1} = b_{2}}{(b_{1} \tensor b_{2}) \dcomp [h, \zeromap, \zeromap, g]}{(\cchoice{c_{1}}{b_{1}}{d_{1}})}{(\cchoice{c_{2}}{b_{2}}{d_{2}})}\), satisfying \(\synch{((b_{1} \tensor b_{2}) \dcomp [h, \zeromap, \zeromap, g])} = (b_{1} \tensor b_{2}) \dcomp [\synch{h}, \zeromap, \zeromap, \synch{g}]\).
  \begin{align*}
    & \semn{(b_{1} = b_{2}) \pand p} \\
    & = (b_{1} \tensor b_{2}) \dcomp [\semn{(b_{1} = b_{2}) \pand p}, \zeromap, \zeromap, \semn{(b_{1} = b_{2}) \pand p}] \\
    & \pgeq (b_{1} \tensor b_{2}) \dcomp [\synch{h} \dcomp \semn{q}, \zeromap, \zeromap, \synch{g} \dcomp \semn{q}] \\
    & = (b_{1} \tensor b_{2}) \dcomp [\synch{h}, \zeromap, \zeromap, \synch{g}] \dcomp \semn{q} \\
    & = \synch{((b_{1} \tensor b_{2}) \dcomp [h, \zeromap, \zeromap, g])} \dcomp \semn{q}
  \end{align*}

  \textsc{while}: apply \Cref{prop:coupling-while}.

  \textsc{monotone}.
  Let \(\couple{h}{c}{d}\) be the coupling given by the assumption.
  \[\semn{p_{1}} \pgeq \semn{p_{2}} \pgeq \synch{h} \dcomp \semn{q_{2}} \pgeq \synch{h} \dcomp \semn{q_{1}}\]

  \textsc{symm}.
  Let \(\couple{h}{c}{d}\) be the coupling given by the assumption.
  By \Cref{prop:symmetry-coupling}, \(\couple{(\swap \dcomp h \dcomp (\swap + \cswap))}{d}{c}\) and this satisfies the desired inequality.
  \begin{align*}
    & \synch{(\swap \dcomp h \dcomp (\swap + \cswap))} \dcomp \semn{\swap \ccomp q} && \\
    & = \swap \dcomp \synch{h} \dcomp \swap \dcomp \swap \dcomp \semn{q} && \text{(\Cref{def:coupling})}\\
    & = \swap \dcomp \synch{h} \dcomp \semn{q} && \text{(Symmetries)}\\
    & \pleq \swap \dcomp \semn{p} && \text{(Assumption)}\\
    & = \semn{\swap \ccomp p}
  \end{align*}

  The one-sided rules are particular instances of the two sided rules: (\textsc{assign-L}) take the expression \(e_{2}\) to be the variable \(x_{i_2}\); (\textsc{sample-L}) the number of sampled variables \(l = 0\); (\textsc{choice-L}, \textsc{ifelse-L}) the commands \(c_{2}\) and \(d_{2}\) to be \(\noop\).

  \textsc{constancy}.
  Let \(\couple{h}{c}{d}\) be the coupling given by the assumption.
  By \Cref{prop:coupling-free-modified-variables}, we may assume that \(\modv{h} = \modv{c} \union \modv{d}\).
  \begin{align*}
    & \semn{r \pand p} && \\
    & = \semn{\assert{r}} \dcomp \semn{p} && \text{(\Cref{prop:predicate-state-actions})}\\
    & \pgeq \semn{\assert{r}} \dcomp \synch{h} \dcomp \semn{q}  && \text{(Assumption)} \\
    & = \synch{h} \dcomp \semn{\assert{r}} \semn{q}  && \text{(\Cref{cor:assert-sample-interchange})} \\
    & = \synch{h} \dcomp \semn{\assert{(r \pand q)}}  && \text{(\Cref{prop:predicate-state-actions})}
  \end{align*}

  \textsc{auxiliary variable}.
  Let \(\couple{h}{c}{d}\) be the coupling given by the assumption.
  By \Cref{prop:coupling-free-modified-variables}, we may assume that \(\freev{h} = \freev{c} \union \freev{d}\).
  \begin{align*}
    & \semn{\pmute{x_{i_{1}}, \dots, x_{i_{k}}}{p}{t}} && \\
    & = \semn{\samp{x_{i_{1}}, \dots, x_{i_{k}}}{t}} \dcomp \semn{p} && \text{(\Cref{prop:predicate-state-actions})} \\
    & \pgeq \semn{\samp{x_{i_{1}}, \dots, x_{i_{k}}}{t}} \dcomp \synch{h} \dcomp \semn{q} && \text{(Assumption)} \\
    & = \synch{h} \dcomp \semn{\samp{x_{i_{1}}, \dots, x_{i_{k}}}{t}} \dcomp \semn{q} && \text{(\Cref{cor:assert-sample-interchange})} \\
    & = \synch{h} \dcomp \semn{\pmute{x_{i_{1}}, \dots, x_{i_{k}}}{q}{t}} && \text{(\Cref{prop:predicate-state-actions})}
  \end{align*}
\end{proof}

\begin{example}
  In the setting of \Cref{ex:assertion-correctness-stoch}, \kl{relational predicate-incorrectness rules} above can reason about probabilistic sensitivity~\cite{aguirre2021pre} by considering triples where the two commands coincide.
\end{example}
\end{toappendix}
\section{Conclusions and future work}
\label{sec:conclusions}

We have introduced \kl{imperative categories} as a principled approach to program logics (\Cref{sec:imperative-categories}).
We have defined generalisation of the guarded command language that accommodates monoidal effects (\Cref{sec:monoidal-gcl}), which takes semantics in \kl{imperative categories}.
The categorical structure derives an algebra of guarded commands that generalises guarded Kleene algebras with tests; it also derives the rules of various existing program logics and relational program logics (\Cref{sec:program-logics} and \Cref{sec:relational-program-logics}).

\paragraph{External logic, fibrations, and enrichment}
While we focused on the logics given by the internal structure of the category, we could derive more variants if we accept the logic to be external (e.g.~the extra operation \(\oplus\) of \emph{outcome logic}).
In particular, a \emph{fibration} would structure the use of two different categories: one for predicates and one for commands.
We considered poset-enriched categories to express program triples.
We could extend the treatment to metric-enriched categories to express \emph{quantitative} properties of program behaviour.

\paragraph{Separation logic and premonoidal semantics}
The \emph{logic of bunched implications} has semantics in categories that are both cartesian closed and monoidal closed with a second tensor; additional distributivity with coproducts is admissible~\cite{o1999logic}.
We believe a careful adaptation of our techniques could derive separation logic from categorical first principles: this could account for its probabilistic
versions~\cite{barthe2019probabilistic}, or be extended to higher-order versions~\cite{birkedal2006semantics}.
The condition that modules have restricted access to some parts of memory~\cite{o2004separation} may be modelled with \kl{premonoidal categories} and their internal language~\cite{jeffrey1997:premonoidal}.

\bibliographystyle{alpha}
\bibliography{main.bib}

\end{document}